\documentclass[aps,prx,noshowpacs,twocolumn,superscriptaddress,nobibnotes]{revtex4-1}
\usepackage{amssymb}
\usepackage{graphicx}
\usepackage{amsmath,amsfonts}
\usepackage{hyperref}
\usepackage[utf8] {inputenc}
\usepackage{xcolor}  % !!!!! To be removed prior to submission. !!!!!
\usepackage[normalem]{ulem}
\usepackage[T1]{fontenc} % Required for accented characters
\usepackage{multirow}
\usepackage{booktabs}
\usepackage{threeparttable}
\usepackage{tabularx}
\begin{document}
	\title{Scaling up real networks by geometric branching growth}
	\date{\today}
	\author{Muhua Zheng}
	\affiliation{Departament de F\'isica de la Mat\`eria Condensada, Universitat de Barcelona, Mart\'i i Franqu\`es 1, E-08028 Barcelona, Spain}
	\affiliation{Universitat de Barcelona Institute of Complex Systems (UBICS), Universitat de Barcelona, Barcelona, Spain}
	
	\author{Guillermo Garc\'ia-P\'erez}
	\affiliation{QTF Centre of Excellence, Turku Centre for Quantum Physics, Department of Physics and Astronomy, University of Turku, FI-20014 Turun Yliopisto, Finland }
	\affiliation{Complex Systems Research Group, Department of Mathematics and Statistics, University
		of Turku, FI-20014 Turun Yliopisto, Finland}
	
	\author{Mari\'an Bogu\~n\'a}
	\affiliation{Departament de F\'isica de la Mat\`eria Condensada, Universitat de Barcelona, Mart\'i i Franqu\`es 1, E-08028 Barcelona, Spain}
	\affiliation{Universitat de Barcelona Institute of Complex Systems (UBICS), Universitat de Barcelona, Barcelona, Spain}
	
	\author{M. {\'A}ngeles Serrano}
	\email[]{marian.serrano@ub.edu}
	\affiliation{Departament de F\'isica de la Mat\`eria Condensada, Universitat de Barcelona, Mart\'i i Franqu\`es 1, E-08028 Barcelona, Spain}
	\affiliation{Universitat de Barcelona Institute of Complex Systems (UBICS), Universitat de Barcelona, Barcelona, Spain}
	\affiliation{ICREA, Passeig Llu\'is Companys 23, E-08010 Barcelona, Spain}
			
\begin{abstract}
	Real networks often grow through the sequential addition of new nodes that connect to older ones in the graph. However, many real systems evolve through the branching of fundamental units, whether those be scientific fields, countries, or species. Here, we provide empirical evidence for self-similar growth of network structure in the evolution of real systems and present the Geometric Branching Growth model, which predicts this evolution and explains the symmetries observed. The model produces multiscale unfolding of a network in a sequence of scaled-up replicas preserving network features, including clustering and community structure, at all scales. Practical applications in real instances include the tuning of network size for best response to external influence and finite-size scaling to assess critical behavior under random link failures.
\end{abstract}
\maketitle
\let\oldaddcontentsline\addcontentsline% Store \addcontentsline
\renewcommand{\addcontentsline}[3]{}% Make \addcontentsline a no-op

%One-sentence summaries, maximum 125 words, typically very short, like 15-20 words.
%{\bf One-sentence summary:} Over large time spans, real networks are found to evolve in a self-similar way that can be explained by geometric branching growth: a process that produces a multiscale unfolding of a real network in a sequence of scaled-up replicas.\\

\section{Introduction}

In the context of network science, growth is most often modeled through the sequential addition of new nodes that connect to older ones in a graph by different attachment mechanisms~\cite{Barabasi1999,Krapivsky:2001nr}, including models of hidden variables where nodes are characterized by intrinsic properties \cite{Bianconi_2001,Papadopoulos2012}. Other growth processes have also been considered, such as duplication to explain large-scale proteome evolution~\cite{Sole:2002,Pastor-Satorras:2003}. Here, we take an alternative approach and explore the relation between branching growth~\cite{Kolmogorov:1947} and geometric renormalization~\cite{Garcia2018} to explain self-similar network evolution. Renormalization in networks, based on the ideas of the renormalization group in statistical physics~\cite{Wilson1975,Wilson1983,Kadanoff2000}, acts as a sort of inverse branching process by coarse-graining nodes and rescaling interactions. Thus, branching growth can be seen as an inverse renormalization transformation: an idea that was introduced in~\cite{Song2006} using a purely topological approach to reproduce the structure of fractal networks, where fractality was interpreted as an evolutionary drive towards robustness. However, topological distances in networks are seriously constrained by the small-world property; while the characterization of fractality in real networks disregards fundamental features of their structure, including clustering and community organization. 

Geometric renormalization~\cite{Garcia2018} (GR) is an alternative technique that can be performed by virtue of the discovery that the structure of real networks is underlain by a latent hyperbolic geometry~\cite{Serrano2008,Krioukov2009}. Thus, the likelihood of interactions between nodes depends on their distances in the underlying space, via a universal connectivity law that operates at all scales and simultaneously encodes short- and long-range connections. This approach has been able to explain many features of the structure of real networks, including the small-world property, scale-free degree distributions, and clustering, as well as fundamental mechanisms such as preferential attachment in growing networks~\cite{Papadopoulos2012} and the emergence of communities~\cite{Garcia2018aa,Zuev2015aa}. Given a network map, GR produces a multiscale unfolding of the network in scaled-down replicas over progressively longer length scales. This transformation has revealed self-similarity to be a ubiquitous symmetry in real networks, whose structural properties remain scale-invariant as the observational resolution is decreased~\cite{Garcia2018}. This poses the question of whether this self-similarity could be related to the mechanisms driving the growth of real networks and, therefore, whether their evolution could be conceptualized within the framework of the geometric renormalization group. 

In this work, we show that real networks---citations between scientific journals~\cite{Hric2018,Fortunato2018} and international trade~\cite{Garcia2016}---have evolved in a self-similar way over time spans of more than $100$ years, meaning that their local, mesoscale, and global topological properties remain in a steady state as time goes by, with a moderate increase of the average degree. We demonstrate that the observations can be modeled by a self-similar metric expansion produced by a geometric branching growth (GBG) process. Beyond the capacity of the model to explain and predict the self-similar evolution of real networks effectively, the technique is flexible and allows us to produce scaled-up network replicas that, when combined with scaled-down network replicas~\cite{Garcia2018}, provide a full up-and-down self-similar multiscale unfolding of complex networks that covers both large and small scales. We illustrate the use of GBG multiscale unfolding in real network instances via the tuning of network size for optimal response to an external influence, referred to here as {\it the optimal mass}, and a finite-size scaling analysis of critical behavior under random link failures.

%%%%%%%%%%%%%%%%%%%%%%%%%%%%%%%%%%%%%
\section{Self-similar evolution of real networks}
We consider the evolution of the journal citation network~\cite{Hric2018} (JCN) and of the world trade web~\cite{Garcia2016} (WTW) over time spans of more than 100 years. 

The evolution of journal citation networks offers a quantitative proxy for the development of contemporary science and the emergence of a vast number of new scientific fields and subfields, driven by diversification and specialization~\cite{Hric2018,Fortunato2018,Milojevic2015,Radicchi2008,Wang2013}. Here, we analyze data from~\cite{Hric2018}, where the time period 1900-2013 is divided into time windows of ten years before $1970$ and of five years thereafter. One citation network is reconstructed for each time window, where journals are represented as nodes that are linked whenever citations between their publications exist. 

An increase of the number of actors is also a hallmark of the evolution of the international trade system. The number of states in the world increased from $42$ in 1900 to $195$ in 2016~\cite{War2017}, mainly due to processes such as decolonization, the dismantlement of large or multicultural states such as the USSR and Yugoslavia (1991) into a number of smaller sates, the parliamentary split of an existing state into two as happened in Czechoslovakia (1993), and independence processes after civil wars, like that of the Republic of South Sudan and the Republic of the Sudan (2011). Here, we use networks in the World Trade Atlas~\cite{Garcia2016}: a collection of annual world trade network maps in hyperbolic geometry, which provide information on the long-term evolution of the international trade system from 1870 to 2013, where nodes represent countries linked by bilateral trade relationships. The maps revealed that globalization, hierarchization, and localization are main forces shaping the trade space, which far from being flat is hyperbolic, as a reflection of its complex architecture. More details of the two datasets are available in Appendix~\ref{A} and the main statistical properties are in Tables~S1 and S2 in Supplemental Material (SM). 

\begin{figure*}[!t]
	\centering
	\includegraphics[width=1\linewidth]{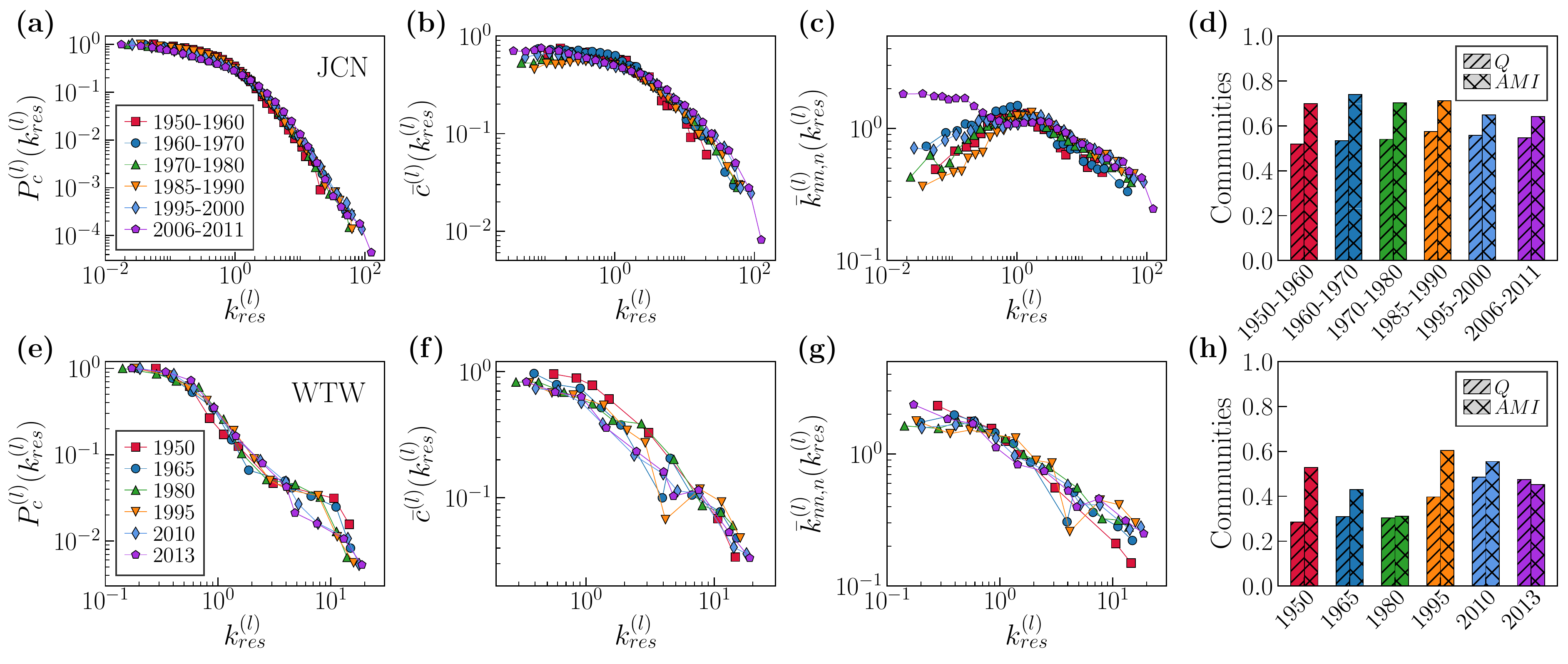}
	\caption{ {\bf Self-similar evolution of real networks.} 
		(a), (e), Complementary cumulative distribution $P_c^{(l)}(k_{res}^{(l)})$ of rescaled degrees $k^{(l)}_{res} = k^{(l)}/ \langle k^{(l)}\rangle$, where the superindex $l$ indicates different time snapshots of the corresponding system.
		(b), (f), Degree dependent clustering coefficient $\bar{c}^{(l)}(k_{res}^{(l)})$ over rescaled-degree classes. (c), (g), Degree--degree correlations, as measured by the normalized average nearest-neighbor degree $\bar{k}_{nn,n}^{(l)} (k_{res}^{(l)}) = \bar{k}_{nn}^{(l)} (k_{res}^{(l)}) \langle  k^{(l)}\rangle/\langle(k^{(l)})^2\rangle$.
		(d), (h), Modularity $Q$ and adjusted mutual information $\textrm{AMI}$ between the community partitions of two consecutive snapshots by considering nodes existing in both snapshots in the JCN and the WTW, respectively. A few representative snapshots are shown here. Results for all networks in the period analyzed are in SM, Figs.~S2-S4. 
	}
	\label{fig:evolution}
\end{figure*}

The size $N$ of the two evolving networks increases over time ranging from $118$ journals in 1900-1910 to $21460$ in 2008-2013, and from $24$ countries in $1870$ to $189$ in $2013$, Fig.~S1(a)-(b) in SM. After World War II, the average degree $\langle k \rangle$ only shows a moderate increase in the JCN and almost flat behavior in the WTW, Fig.~S1(c)-(d). Degree distributions, clustering spectra, degree--degree correlations and the community structure of some snapshots are shown for the JCN in Fig.~\ref{fig:evolution}(a)-(d) and for the WTW in Fig.~\ref{fig:evolution}(e)-(h) (results for all snapshots are in SM, Figs.~S2-S4). We observe clear-cut self-similar behavior with the curves for different networks overlapping when the degrees of the nodes are rescaled by the average degree of the corresponding network. Fig.~\ref{fig:evolution}(d) and (h) shows the modularity, $Q$, of the optimal partitions detected by the Louvain method~\cite{Blondel2008}, and the adjusted mutual information $\textrm{AMI}$~\cite{Vinh2010} between the optimal partitions of two consecutive snapshots, in which we only considered the nodes that exist in both. The level of modularity remains stable throughout the evolution of the systems and the overlap between communities in the consecutive snapshots is consistently very high. This indicates that the community structure is mostly preserved as time goes by. Hence, the empirical evidence presented so far indicates that these real networks grow in a self-similar fashion.

%%%%%%%%%%%%%%%%%%%%%%%%%%%%%%%%%%%%%%%%%%%%%%%%%%%%%%%
\section{Geometric branching growth}
\begin{figure}[!t]
	\centering
	\includegraphics[width=1\linewidth]{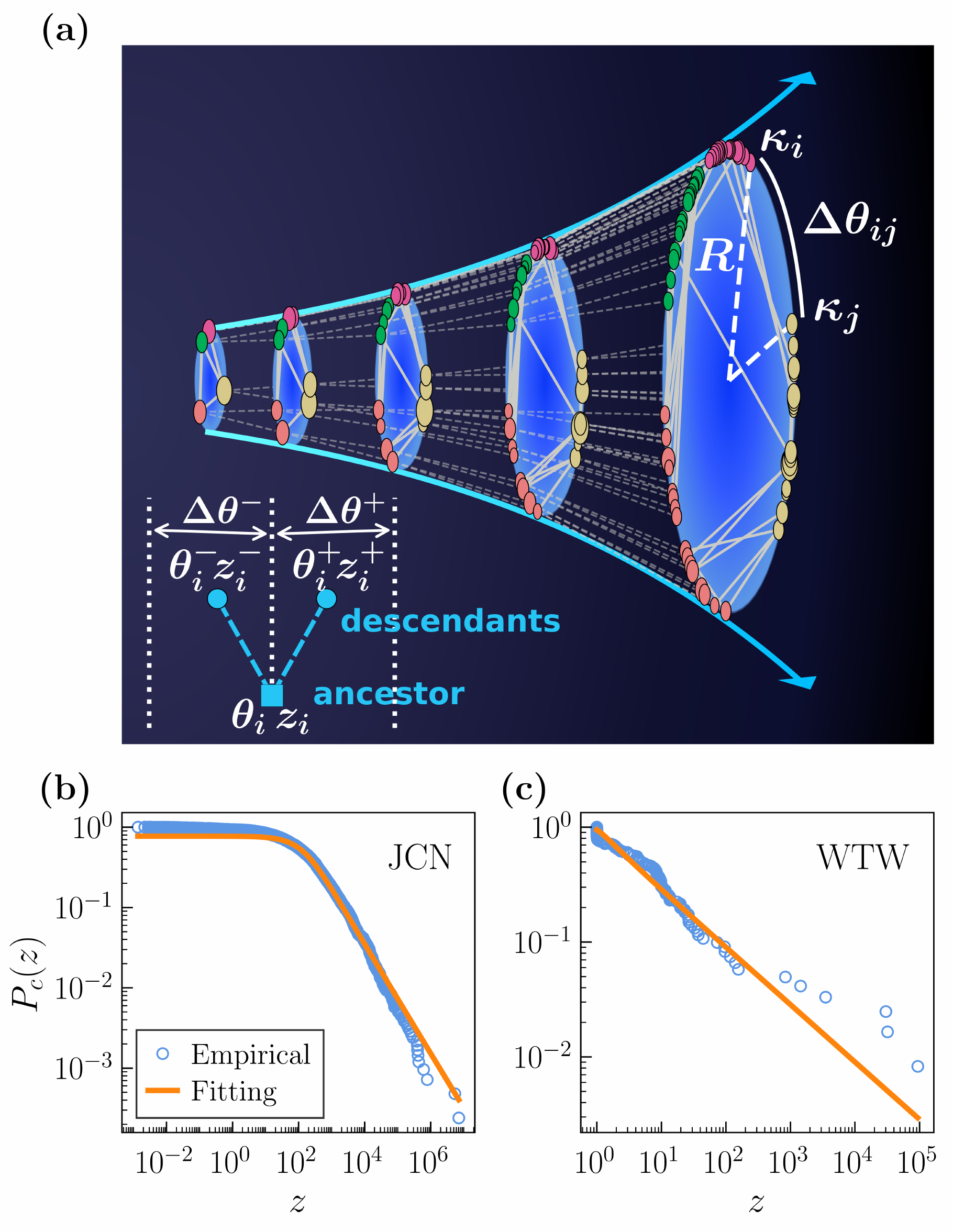}
	\caption{ {\bf Sketch of the GBG model.} 
		(a) In each layer of the self-similar upwards multiscale unfolding, the size of each node is proportional to the
		logarithm of its hidden degree, $\kappa$. Different colors represent different geometric communities. Dashed lines connect ancestors to their descendants along the flow (blue arrows). A pair of nodes, $i$ and $j$, with hidden degrees $\kappa_i$ and $\kappa_j$ has been highlighted, for which the angular separation, $\Delta \theta_{ij}$, represents their similarity distance. In the bottom left corner we show a sketch of the branching process from an ancestor to its pair of descendants. The complementary cumulative distribution of $z$ values together with their corresponding stable distribution fittings are in (b) for JCN 1965-1975, and in (c) for WTW $1965$. }
	\label{fig:sketch}
\end{figure}

To model the observed self-similar evolution of real networks, we propose the GBG transformation that produces self-similar multiscale unfolding of a network in a shell of scaled-up replicas of progressively increasing size, as illustrated in Fig.~\ref{fig:sketch}(a). The GBG transformation acts on network maps: geometric representations that reveal the manifest latent hyperbolic geometry of network structure~\cite{Serrano2008,Krioukov2010}. To describe network maps, we employ the $\mathbb{S}^1$ model~\cite{Serrano2008} which, in contrast to the isomorphic hyperbolic version $\mathbb{H}^2$~\cite{Krioukov2010}, makes the similarity dimension explicit. In the $\mathbb{S}^1$ model, each node, $i$, is assigned a hidden degree, $\kappa_i$, or popularity, and an angular position, $\theta_i$, or similarity, in a one-dimensional sphere representing the similarity space; and every pair of nodes, $i$ and $j$, is connected with probability:
\begin{align} \label{eq:con_pro}
p_{ij} = \frac{1}{1+\chi_{ij}^\beta}=\frac{1}{1+\left(\frac{R\Delta \theta_{ij}}{\mu \kappa_i \kappa_j}\right)^\beta}\ 
\end{align}
so that more popular (larger $\kappa$) or more similar (lower $\Delta \theta$) nodes are more likely to form connections. The similarity circle has radius $R$, adjusted to maintain a constant density of nodes, equal to one, without loss of generality. The choice for the connection probability Eq.~\eqref{eq:con_pro} ensures that graphs generated by the model belong to the maximum entropy ensemble of random geometric graphs that are simultaneously sparse, small-world, highly clustered, and with heterogeneous degree distributions, among other properties observed in real networks~\cite{boguna2019small}. Likewise, the $\mathbb{S}^1/\mathbb{H}^2$ model is particularly interesting because a body of analytic results for the most relevant topological properties have already been derived, including degree distribution~\cite{Serrano2008,Krioukov2010,gugelmann2012random}, clustering~\cite{Krioukov2010,gugelmann2012random,candellero2016clustering}, diameter~\cite{abdullah2017typical,friedrich2018diameter,muller2019diameter}, percolation~\cite{Serrano2011,fountoulakis2018law}, self-similarity~\cite{Serrano2008}, or spectral properties~\cite{kiwi2018spectral}.

A network map, that is, the set of hidden variables $\{ \kappa_i, \theta_i \}$ together with the parameters $\beta$ and $\mu$ controlling the local clustering coefficient $\left<c\right>$ and the average degree $\left<k\right>$, can be obtained by finding the coordinates that maximize the probability for the observed network to be generated as an instance of the $\mathbb{S}^1$ model. Due to the aforementioned isomorphism between $\mathbb{S}^1$ and $\mathbb{H}^2$, these model parameters and hidden variables also define the corresponding hyperbolic map~\cite{Boguna2010}. To produce the maps we used the embedding tool Mercator~\cite{Garcia2019}, which infers the coordinates of the nodes and parameters $R$, $\beta$, and $\mu$ from the topology of the network. More details can be found in Appendix~\ref{B}. 

The GBG transformation can be controlled to adjust the growth in the number of nodes and also the flow of the average degree, embodying a family of models that includes non-inflationary and inflationary growth. Non-inflationary growth produces a sequence of progressively magnified layers with decreasing average degree that comply with GR. Inflationary growth means that scaled-up shell layers are produced with an average degree that does not decrease very fast, or even increases. The GBG transformation is compliant with GR if, when GR is applied to the layer obtained after the GBG transformation, the result is the original network. In brief, the GR transformation~\cite{Garcia2018} proceeds by defining non-overlapping blocks of consecutive nodes of equal size $r$ around the similarity circle, which are then coarse grained into a single node in the renormalized lower-resolution map, where pairs of nodes are connected with a link if any of their precursor nodes were connected in the original layer. As a result, multiscale unfolding of self-similar scaled-down network replicas is obtained, except for the average degree of the renormalized layers, which typically grows exponentially in real networks (more details in SM). 

The fist step to generate a GBG scaled-up map is to split every node in the original layer into $r$ descendants with probability $p$, so that the population increases as $N'=N(1+p(r-1))=bN$ with branching rate $b$. For mathematical convenience, we will continue the description with $r=2$ ($b=1+p$). We can use parameter $b$ in combination with the number of layers in the multiscale unfolding to adjust the growth of the number of nodes over the evolution of the network. Every branching node produces a pair of descendants that require the assignment of similarity coordinates in the $\mathbb{S}^1$ circle and of hidden degrees, whereas nodes that do not split remain with the same coordinates. The radius of the circle is rescaled as $R'=bR$, so that the density of nodes remains equal to one. 

\noindent\textbf{Assigning coordinates to descendants.} One of the requirements for self-similar growth is the preservation of the ordering of nodes in the circle and their concentration across specific angular sectors, which defines the geometric community organization of networks~\cite{Garcia2018aa,Zuev2015aa}. To this end, the simplest means to model growth is to place the descendants at angular coordinates $\theta_i^{+}$ and $\theta_i^{-}$ to the left and right of the angular position of their corresponding ancestor, $i$, with uniform probability within a small angular separation $\Delta \theta^\pm$. The values $\Delta \theta^\pm$ are bounded by the total number of nodes in the descendant layer and by the proximity (to the left or right) of consecutive nodes to the ancestor in the similarity circle. We set $\Delta \theta^\pm=\textrm{min}\{\frac{2\pi}{N'}, \frac{\Delta \theta_{ij}}{2}\}$, where $\Delta \theta_{ij}=\pi-|\pi-|\theta_i-\theta_j||$ is the angular distance between the branching node $i$ and its consecutive neighboring node $j$ (to left or right) in the ancestor layer. This choice of $\Delta\theta^{\pm}$ ensures the preservation of large gaps between consecutive nodes by limiting the angular distance between branching node and descendant, while it prevents crossings between descendants of neighbouring branching nodes even in densely populated angular regions.

To assign the hidden degrees $\kappa^{+}$ and $\kappa^{-}$, we impose two conditions. First, the hidden degrees of ancestors and descendants need to comply with GR. This implies that the relation between hidden degrees of ancestors and descendants should be compliant with the GR transformation~\cite{Garcia2018} $z = z^{+} + z^{-}$, where $z=\kappa^\beta$. Second, the hidden degrees of descendants must be independent and identically distributed random variables with a distribution of hidden degrees that preserves that of the ancestor layer, $\rho(\kappa)$ (equivalently $\rho(z)$). Taking the two conditions together, the transformed hidden degrees $z^{\pm}$ of descendants should satisfy:
\begin{equation}\label{eq:global_constraint}
\int \int \mathrm{d} z^{+}\mathrm{d} z^{-} \rho(z^{+}) \rho(z^{-}) \delta \left( z - \left( z^{+} + z^{-} \right) \right) = \rho(z).
\end{equation}

The equation above implies that $\rho(z)$ is a stable distribution~\cite{levy1925,Borak2005,Nolan2018}, meaning that the linear combination of two independent variables with probability distribution $\rho(z)$ has the same distribution, up to scaling and location factors. Stable distributions admit multiple parametrizations but are always defined by four parameters $f(z; \alpha,\eta, c,d)$: the tail exponent $\alpha \in (0,2]$ and skewness $\eta \in [-1,1]$ which control the shape; and $c$ and $d$ for scale and location (see Appendix~\ref{C0}). Stable distributions conform a rich family of models including Gaussian ($\alpha=2$), Cauchy ($\alpha=1$ and $\eta=0$), L\'evy ($\alpha=1/2$ and $\eta=1$), and Landau ($\alpha=1$ and $\eta=1$) distributions. Stable distributions are infinitely divisible and are the only possible limit distributions for properly normalized and centered sums of {\it iid} random variables (generalized Central Limit Theorem)~\cite{Gnedenko:1954}. In addition, they can accommodate fat tails and asymmetry, and therefore often offer a very good fit for empirical data~\cite{Mandelbrot1963,Fama1965,Nolan2018,Kateregga2017}.

We proceed by fitting a stable distribution to the distribution of transformed hidden degrees in the original layer~\cite{Nolan:1997,Royuela2017} (see Appendix~\ref{C}). Fig.~\ref{fig:sketch}(b) and (c) show a very good fit for JCN and WTW (see Fig.~S5 for more empirical networks, and the corresponding fitting parameters in Table~S3 in SM). If $b=2$, meaning that all nodes split, the distribution for descendants, $f(z^{\pm}; \alpha^{\pm},\eta^{\pm}, c^{\pm},d^{\pm})=f(z^{\pm}; \alpha,\eta, c/2^{1/\alpha}, d/2)$, follows immediately from Eq.~(\ref{eq:global_constraint}) and basic properties of the stable distribution, with the shape parameters remaining invariant, and scale and location being adjusted so that the stable distribution of the ancestor layer is recovered when we sum the hidden variables $z^{\pm}$ of the descendants. These functions and Bayes rule can be used to generate numerically the values of $z^{+}$ from the probability of hidden degrees of descendants, conditional on the degree of the ancestor $\rho(z^{+}|z)^{\mathrm{nor}}$, normalized to ensure that the hidden degrees of descendants are non-negative. Finally, $z^{-}$ is calculated deterministically using $z^{-}=z -z^{+}$, and the variables $z^{\pm}$ are transformed back into $\kappa^{\pm}$ using $\kappa = z^{1/\beta}$. In the case of fractionary $b$, we produce the hidden variables $z^{\pm}$ of the descendants of branching nodes using $f(z; \alpha,\eta, c/b^{1/\alpha}, d/b)$, and assume that the stable distribution in the new layer is $f(z; \alpha,\eta, c/b^{1/\alpha}, d/b)$. This gives a good approximation to the mixture of stable distributions that result from nodes with different branching behavior. Fig.~S5 in SM demonstrates that the distribution of hidden variables $z$ of descendants has the same shape as that of the ancestor layer in different real networks.\\ 

\begin{figure*}[t]
	\centering	
	\includegraphics[width=0.9\linewidth]{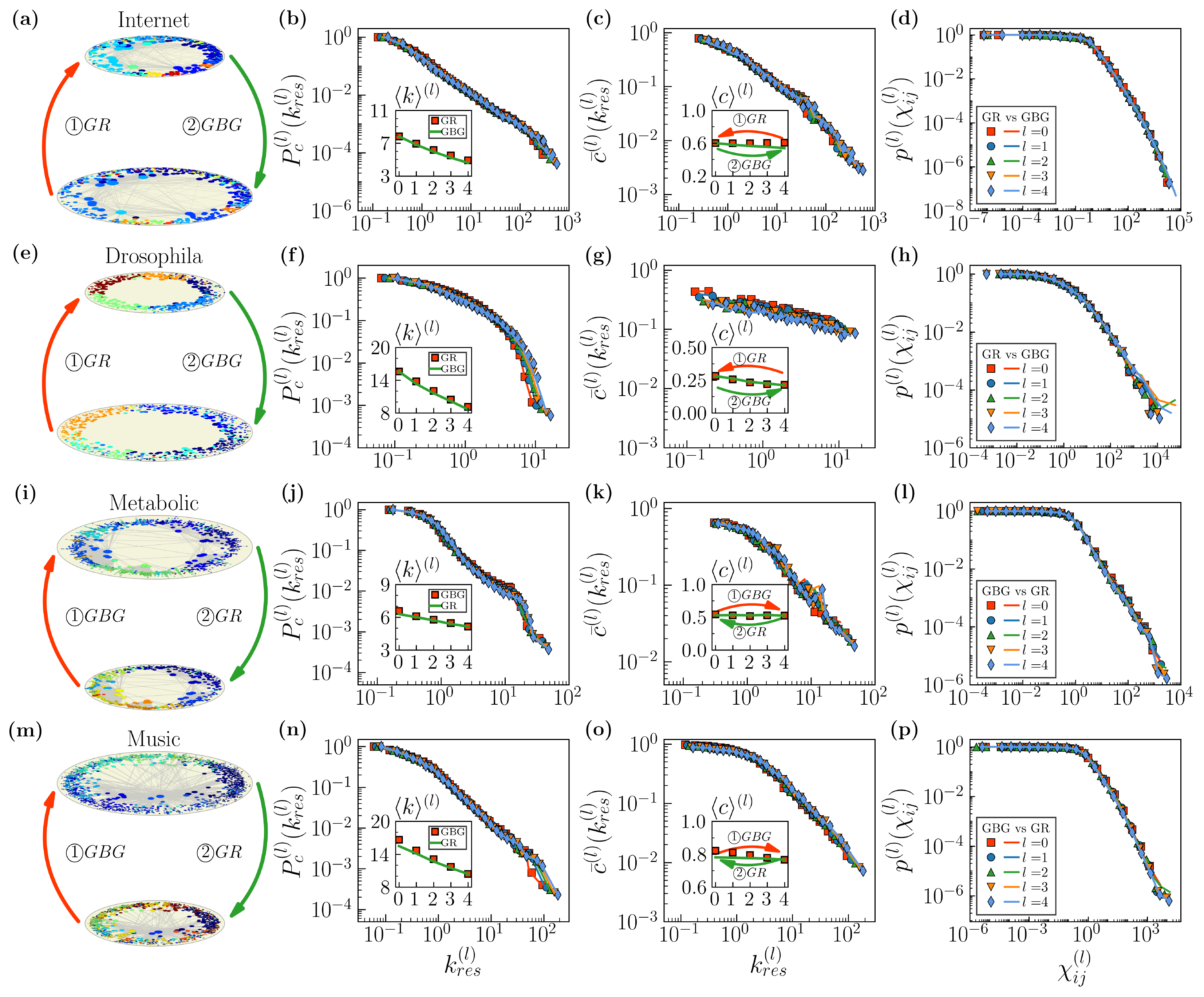}
	\caption{ {\bf GBG is compliant with GR in the non-inflationary limit.} (a)-(d) GR-GBG transformation in the Internet. We renormalized the original network from layer $4$ to layer $0$, and applied GBG to layer $0$ with $b=1.2$ to go back to layer $4$. Colors of nodes  in sketch (a) indicate community structure as detected by the Louvain algorithm. (b) Flow of the average degree. (c) Flow of the average clustering. (d)  Flow of the connection probability $p^{(l)}(\chi_{ij}^{(l)})$ as a function of the effective distance $\chi_{ij}^{(l)}$ in layer $l$. (e)-(h) The same for Drosophila. (i) and (l), GBG-GR transformation in the human metabolic network. The original network as transformed by applying GBG with $b = 1.2$ to produce 4 layers, and applied GR to layer 4 to go back to layer 0. (m)-(p) The same for Music.}
	\label{fig:up2down}
\end{figure*}

\noindent\textbf{Connecting nodes in the descendant layer.} Once coordinates have been assigned to nodes, connections between descendants in the new layer are implemented such that the resulting network belongs to the $\mathbb{S}^1$ ensemble. In what we call the {\em non-inflationary} limit, we also require that the new network is compliant with GR, that is, GR applied to the descendant layer should result in the ancestor layer. We use the probability of connection $p_{ij}$ Eq.~\eqref{eq:con_pro} as in the $\mathbb{S}^1$ model, rescaling $\mu$ in the new layer to control the flow of the average degree, and with $\beta$ remaining invariant as in the GR transformation. We use $\mu'=b\mu$ and connect descendants branching from the same ancestor with probability $p_{ij}(\mu')$. Then, for every pair of connected ancestors, we establish potential links among their descendants with the same probability $p_{ij}(\mu')$, but making sure that at least one link is formed between them (see Appendix~\ref{D}). \\

\noindent\textbf{{\em Inflationary} GBG}. We first proceed as in the non-inflationary case. Once we have a non-inflationary GBG map, we set $\mu'_a=a\mu'=a b \mu, (a\geqslant1)$ to adjust the average degree to a larger value by adding extra links between any pair of nodes that remained unconnected using probability:
\begin{equation}\label{eq:conprob_extra}
\pi_{ij}=\frac{p_{ij}(\mu'_a)-p_{ij}(\mu')}{1-p_{ij}(\mu')}. 
\end{equation}
These steps ensure that: i) all pairs of descendants in the GBG layer are connected with probability $p_{ij}(\mu'_a)$, with the original form Eq.~\eqref{eq:con_pro} in the ancestor layer, and hence the resulting network belongs to the $\mathbb{S}^1$ ensemble; ii) links exist between descendants of connected ancestors; and iii) the non-inflationary limit is recovered for $a=1$, that is, in this case, $\pi_{ij}=0$ and no extra links are formed so that GBG complies with GR and there are only connections in the descendant layer between descendants of the same ancestor or of connected ancestors.\\

\noindent\textbf{GBG is a statistical inverse of GR}. We support this claim with the results shown in Fig.~\ref{fig:up2down} and Figs.~S6-S13 in SM for different real networks and branching rates $b$, which show that the results are robust. In Fig.~\ref{fig:up2down}(a)-(h), we show that after applying GR to the Internet or Drosophila, the original network can be recovered with high fidelity (in a statistical sense) by applying the non-inflationary GBG to the renormalized layer. Conversely, if we first apply the non-inflationary GBG technique to obtain the scaled-up network and then recover the networks by geometric renormalization, the result is analogous; see Fig.~\ref{fig:up2down}(e)-(p) for Metabolic and Music (details of the datasets are in Appendix~\ref{A}). This means that non-inflationary GBG and GR flows produce the same values of average clustering, average degree and empirical connection probability, among other properties. The GBG transformation also preserves the original community structure, as detected by the Louvain algorithm~\cite{Blondel2008}. Furthermore, since the transformation also preserves the correlation between hidden angles and degrees, the self-similarity of the scaled-up networks extends to structural correlations among nodes, such as degree-degree correlations~\cite{Garcia2019}. 

\begin{figure}[!t]
	\centering	
	\includegraphics[width=1\linewidth]{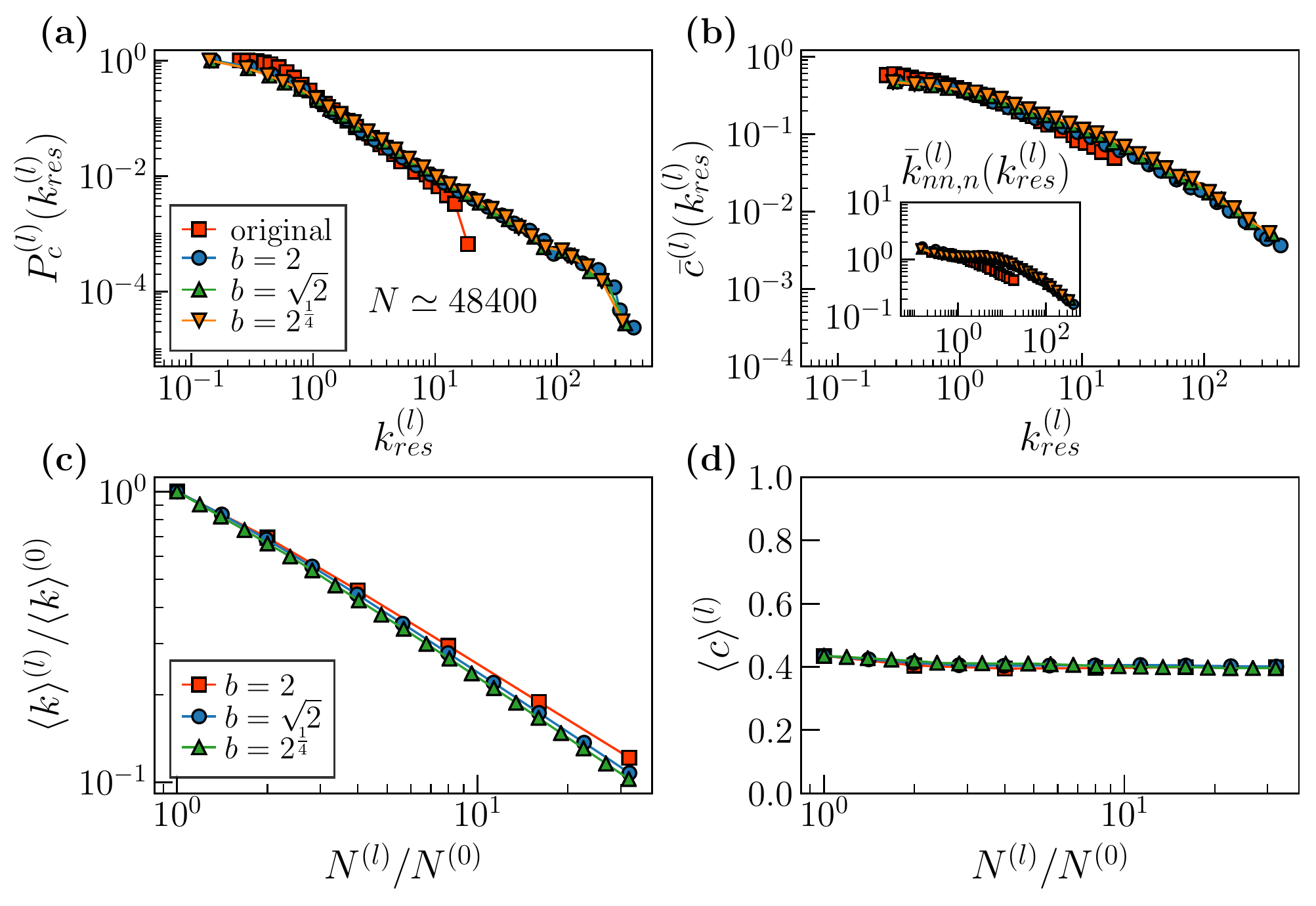}
	\caption{{\bf Non-inflationary GBG has semigroup structure.} (a)-(b) Topological properties of layer 5,10 and 20 in the GBG flow ($N\simeq 48400$) of a synthetic network with $N=1513$, $\langle k\rangle=47.08$, $\beta=1.44$, $\gamma=2.17$ for $b=2,2^{1/2},2^{1/4}$ respectively. (a) Complementary cumulative distribution $P_c^{(l)}(k_{res}^{(l)})$ of rescaled degrees $k^{(l)}_{res} = k^{(l)}/ \langle k^{(l)}\rangle$.
	(b) Degree dependent clustering coefficient $\bar{c}^{(l)}(k_{res}^{(l)})$ over rescaled-degree classes, inset shows degree--degree correlations. (c) and (d) behavior of average degree and clustering coefficient versus network size in the GBG flow, where the superindex $l=0$ indicates the original synthetic network. 
}
	\label{fig:semigroup}
\end{figure}

Notice that the semigroup property of GR also holds for GBG, meaning that two consecutive GBG transformations of scale $b$ over a network are equivalent to a single transformation of scale $b^2$. This is easy to derive when the branching rate has an integer value, but it holds even if $b<2$. Results supporting this claim are shown in Fig.~\ref{fig:semigroup} for a synthetic network produced with the $\mathbb{S}^1$ model and different values of $b$, and in Figs.~S14 and S15 in SM for different real networks.

Strictly speaking, inflationary GBG is not a statistical inverse of GR because of the new links added to increase the average degree over the value given by inverse GR. If we first apply the inflationary GBG technique to obtain a scaled-up network and subsequently apply geometric renormalization, we would recover an {\it inflated} version of the original network that we would need to {\it deflate} to recover the original network. To rebalance the average degree one needs an extra mechanism, like the pruning used in~\cite{Garcia2018} to produce scaled-down network replicas. Given that the addition of links in the inflationary step of the inflationary GBG process, as well as the pruning of links to decrease the average degree of GR layers are compliant with the $\mathbb{S}^1$ model, we say then that inflationary GBG (GBG+addition of links) is a statistical inverse of deflationary GR (GR+pruning of links), see Fig.~S16 in SM.\\ 

\noindent\textbf{Behavior of the average degree.} In the non-inflationary GBG model ($a=1$), we can use the inverse of the GR relation between the average degrees in a descendant layer and in the ancestor layer~\cite{Garcia2018}, using $\mu'= b\mu$ to obtain $\langle k\rangle^{(l)}=(b^{-\nu})^l {\langle k \rangle^{(0)}}$, where the scaling factor $\nu$ depends on the connectivity structure of the original network, and $\langle k\rangle^{(l)}$ (the mean degree of layer $l$) refers to the original network when $l=0$. Typically, as the scaling factor, $\nu$, is positive in real networks~\cite{Garcia2018}, the average degree of the descendant layers decreases exponentially. 

In the inflationary regime, $\mu'= a b\mu$ and following the same derivations as in~\cite{Garcia2018}, we find:
\begin{eqnarray} \label{eq:k_Max2}
\langle k\rangle_{a}^{(l)}\!=\!a^l {\langle k \rangle^{(l)}}\!=\!(a b^{-\nu})^l {\langle k \rangle^{(0)}}\!=\!\left [\frac{N^{(l)}}{N^{(0)}}\right ] ^{-\nu+\frac{\ln a}{\ln b}} \mkern-18mu {\langle k \rangle^{(0)}},
\end{eqnarray}
where we have used $l=\frac{\ln \frac{N^{(l)}}{N^{(0)}}}{\ln b}$, with $N^{(l)}$ and $N^{(0)}$ being the network sizes on layer $l$ and $0$, respectively; and $N^{(l)}=bN^{(l-1)}$. Notice that the inflationary process was applied here to every layer in the flow. If, instead, it is applied in a single step to the last layer produced in a non-inflationary GBG transformation, then $\langle k\rangle_{a}^{(l)}=a {\langle k \rangle^{(l)}}=a (b^{-\nu})^l {\langle k \rangle^{(0)}}$. From Eq.~(\ref{eq:k_Max2}), the average degree $\langle k\rangle_{a}^{(l)}$ increases as a power of $N^{(l)}$. Fig.~S17 in SM shows the high degree of congruency between this theoretical prediction, the empirical data, and simulations (as explained below) of the inflationary version of GBG, in JCN and WTW.\\

\begin{figure*}[!t]
	\centering
	\includegraphics[width=1\linewidth]{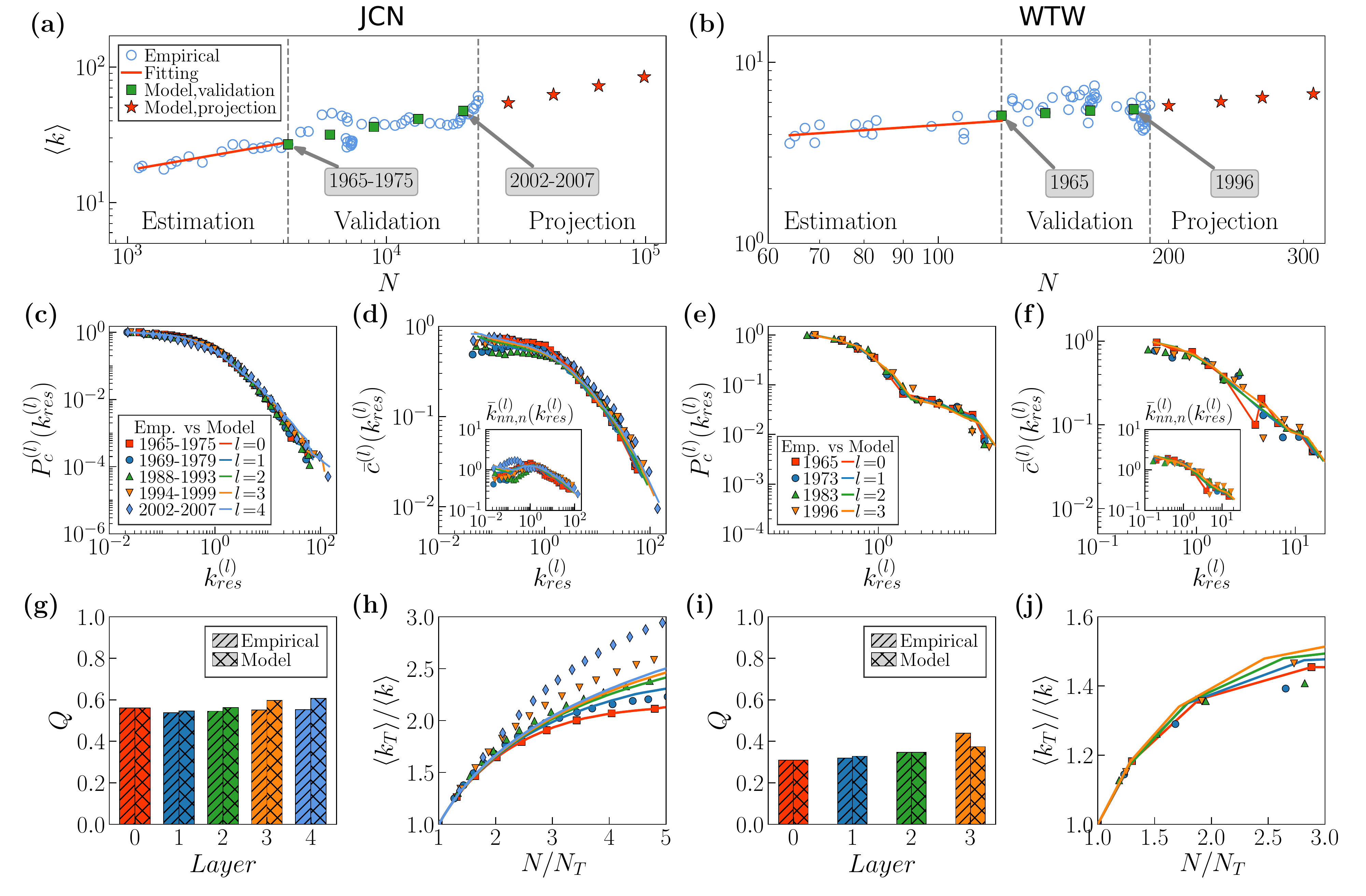}	
	\caption{
		\textbf{The GBG model predicts the self-similar evolution of real networks.} (a), (b), Evolution of the average degree $\langle k\rangle$ vs network size $N$. The estimation, validation, and projection sections are separated by vertical dashed lines. Blue circles, green squares and red stars represent empirical data, validation points, and projection from the model, respectively. The data in the estimation section are used to find values of $a$ (see details in Appendix~\ref{E}). The branching rates, $b$, are fixed to $1.5$ in JCN and $1.15$ in WTW, and the corresponding values of $a$ are $1.415$ and $1.090$ respectively, see Fig.~S18 in SM. For validation purposes, we grow the network from 1965-1975 in JCN and 1965 in WTW, using GBG and compared the resulting networks with empirical snapshots of the same size. (c)-(j), Comparison of the topological properties of simulated and empirical networks.
		(c), (e), Complementary cumulative distribution of rescaled degrees. (d), (f), Degree-dependent clustering coefficient over rescaled-degree classes, insets: Degree--degree correlations; 
		(g), (i), the modularity $Q$;
		(h), (j), local rich-club effect in the JCN and the WTW, respectively. }
	\label{fig:emp2model}
\end{figure*}

\noindent\textbf{Predicting the evolution of real networks.} The inflationary GBG model reproduces the self-similar evolution of JCN and WTW. To support this claim, we divide the empirical data into two consecutive time windows: the first for estimation purposes and the second for validation purposes. Note that JCN and WTW data from before World War II are not used due to the high fluctuations of the network properties, see Figs. S2 and S3 in SM. We fix a value of $b$ in the range $1 < b < 2$ to adjust the rate of growth in our GBG simulation in such a way that we can produce enough snapshots to compare with the real data. With this value of $b$, we estimate parameter $a$ from the empirical evolution of the average degree vs network size (see details in Appendix~\ref{E}). We find that $a$ remains stable over time (see Fig.~S18 in SM), consistent with the empirical observation that the average degree grows as a power of the system size, see Fig.~\ref{fig:emp2model}(a) and (b). Next, we use the network snapshot at the end of the estimation period as the initial layer in GBG multiscale unfolding to simulate a number of scaled-up layers that we then compare to empirical snapshots of approximately the same size in the validation set. The comparisons of degree distributions, clustering, degree--degree correlations, and modularity are shown in Fig.~\ref{fig:emp2model}(c), (d), (g) and (e), (f), (i). We also measured the local rich-club and nested self-similarity effects, reported in Fig.~\ref{fig:emp2model}(h) and (j) and Figs.~S19 and S20 in SM. We name as ``local rich-club effect'' and ``nested self-similarity effect'' the observation in real networks that the nested hierarchy of subgraphs produced by progressively thresholding the degrees of the nodes presents, respectively, an increasing internal average degree and self-similar structure~\cite{Serrano2008,Serrano2011}. This is a highly non-trivial property with crucial implications, such as the absence of a critical threshold in any phase transition whose critical point depends monotonously on the average degree, including percolation, epidemic spreading processes and the Ising model~\cite{Dorogovtsev2008}. The results show that all the networks analyzed in this paper, including JCN and WTW, present the two effects, see Figs.~S19 and S20 in SM. Notice that standard growing network models, including the Barab{\'a}si-Albert model~\cite{Barabasi1999} and the Popularity--Similarity Optimization model in hyperbolic space~\cite{Papadopoulos2012}, have a constant average degree as the network grows, and they also present a constant average degree of the subgraphs in the nested hierarchy, see Fig.~S21 in SM. Therefore, they lack the local rich-club effect. In fact, if those models were adjusted to increase the average degree over time, as happens in the real networks that we analyze in this work, the flow of the average degree in the nested hierarchy would be decreasing, see Fig.~S21b in SM. In addition, the results are robust for different values of $b$, and for different starting times, see Fig.~S22-S24 in SM. Therefore, the GBG model reproduces the self-similar evolution of the structure of the two networks with high fidelity. More comparisons between the model and empirical observations are also shown in Figs.~S25 to S28 in SM.

\section{Scaled-up real network replicas}

One of the practical applications of the GBG model is the production of magnified replicas of real networks: versions where the number of nodes is increased while preserving the statistical properties of the original network, in particular its average degree $\langle k^{(0)}\rangle$. Using GBG, the procedure is straightforward and involves adjusting the parameter $b$, the number of layers $l$, and the inflationary parameter $a$. The idea is to single out a specific scale after a certain number of non-inflationary GBG steps and to tune $a$ to increase the average degree to the target value by adding new links using Eq.~\eqref{eq:conprob_extra} (see details in Appendix~\ref{F}). Notice that this application of the GBG model can also be extended to networks that do not necessarily evolve according to the model, as it exploits the underlying geometric structure and congruency with geometric models observed in many real-world networks.

We illustrate the usefulness of scaled-up network replicas through two examples. In the first, we use the upwards self-similar multiscale unfolding of a small Facebook network to produce a sequence of scaled-up replicas. The goal is to detect the network size that produces the optimal response to external modulation in a noisy environment (see Fig.~S29 in SM for the self-similar statistical properties with respect to the original network). In the second, we combine scale-up network replicas produced by GBG with scaled-down network replicas produced by GR (as described in~\cite{Garcia2018}) to explore the critical behavior of a real network, the Internet (see the same topological properties of both scaled-up and scaled-down networks in Fig.~S29), close to the transition where the global connectivity of the network disintegrates under random link failures.

\begin{figure}[!t]
	\centering
	\includegraphics[width=1\linewidth]{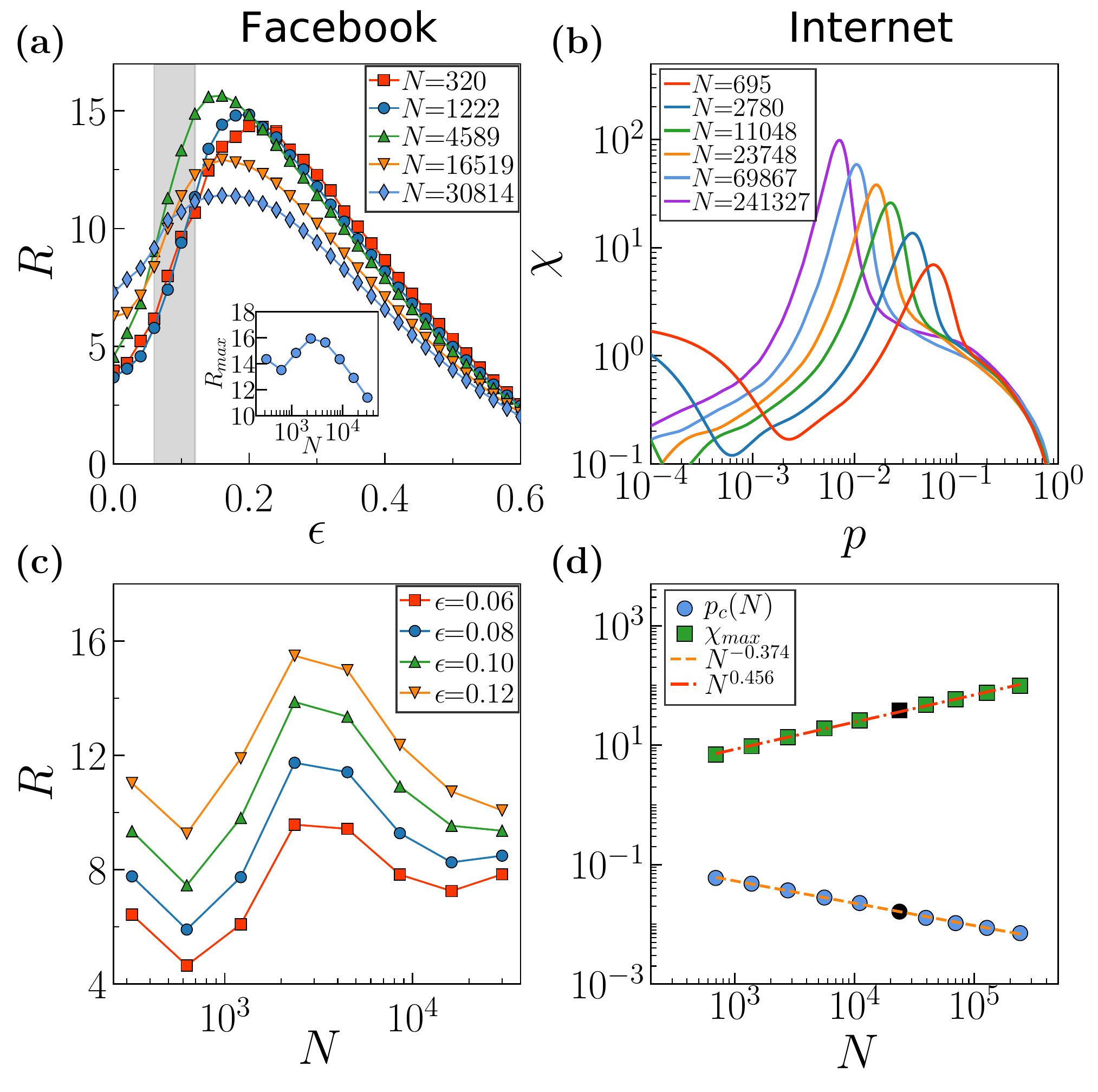}
	\caption{ {\bf Controlling network size from network snapshots.} (a) and (c), Optimal mass of a Facebook network for best response to external modulation in a noisy environment. (a), System response, $R$, as a function of the noise intensity $\epsilon$ for different network sizes. The inset shows the maximum response as a function of $N$. (b), System response, $R$, as a function of $N$ for different values of the noise intensity $\epsilon$. The range of noise values on the $x$-axis corresponds to the gray region in plot (a). (b) and (d), Scaling with system size of random link failure as a bond percolation process in the Internet. (c), Susceptibility, $\chi$, as a function of bond occupation probability, $p$, for different network sizes, $N$. (d), Critical bond occupation probability, $p_c$, and the maximum, $\chi_{\mathrm{max}}$, of the susceptibility, $\chi$, as functions of network size, $N$. The dashed lines are power law fits and the black symbols in (d) indicate the original network. The GBG and GR shells are produced with $b=2$.}		
	\label{fig:application}
\end{figure}

\noindent\textbf{Size-dependent system response to external modulation in a noisy environment.} Small size in real networks can be a limiting factor for the study of dynamical processes, especially when long-range dynamical correlations are non-negligible or when finite size effects play an important role in the final outcome of the dynamics. Here, we study the behavior of a model of opinion formation with nontrivial size dependence using a small Facebook network of 320 users, working for the same software company~\cite{Fire2016} (see data description in Appendix~\ref{A}). The opinion formation model introduced in~\cite{Kuperman2002} includes imitation following a majority rule, external influence in the form of a periodic ``fashion'' wave, and noise. This model was shown to present a noise stochastic resonance effect in small-world networks~\cite{Gammaitoni1998}, displaying an optimal response of the population to the ``fashion'' wave for some noise level. The system also displays a size stochastic resonance effect~\cite{Pikovsky2002,Toral2006}, which means there is an optimal value for the number of nodes, the optimal mass, for which the average opinion best follows the fashion, as a consequence of the coupling between noise and system size in the effective noise intensity. However, these results were for synthetic networks produced by models and not for real networks, so that the size of the networks could be controlled.

The GBG technique provides the opportunity to study system size stochastic resonance in real networks. We produce a GBG self-similar multiscale shell with the same average degree as the original Facebook network, and simulate the dynamical process described above in each layer (see Appendix~\ref{G} for details of our implementation). We model the external signal as a cosine function with amplitude $A$ and period $T$, and measure the response of the system as a function of the noise intensity $\epsilon$ for different system sizes using the spectral amplification factor~\cite{Jung1989} $R=4A^{-2}|\langle e^{i2\pi t/T} \rho(t)\rangle |$, where $\rho(t)=\frac{1}{N}\sum_i m_i(t)$ is the average opinion in the evolution, $m_i(t)$ is the dynamical state of node $i$ at time $t$, and $\langle \cdots \rangle$ denotes a time average. 

Our results are shown in Fig.~\ref{fig:application}(a). The optimal response, $R_{\mathrm{max}}$, is plotted in the inset as a function of $N$. For each size, $N$, there is a maximum response for some intermediate value of the noise and the optimal value occurs at some combination of noise and size. Interestingly, for sufficiently small values of noise, Fig.~\ref{fig:application}(b), $R$ is enhanced by increasing the noise; and for every noise intensity, $\epsilon$, the optimal response occurs at approximately the same value as in $R_{\mathrm{max}}$, $N=2379$. Hence, we conclude that there is an optimal mass for which the average opinion best follows the external influence. Moreover, we also found that there is some value of $N$ for which $R$ has a minimum, that is, the average opinion follows the external influence to the least extent.

\noindent\textbf{Critical behavior of real networks under random link failures.} The random failure of links in networks leads to a percolation transition: a continuous structural change which disaggregates the large cluster of connected network nodes into a bundle of small isolated components \cite{Serrano2011,Dorogovtsev2008}, hence disabling the system. The fraction of links removed, $p$, acts as a control parameter which can be manipulated to change the state of the system in {\it in silico} experiments, and the transition occurs at some specific value: $p_c$. Close to this critical point, the macroscopic properties of the network, such as the relative size of the largest connected cluster and the average cluster size, behave as power laws of the distance to the critical point, $(p-p_c)^{\delta}$, with some critical exponents. One way of extracting these exponents is by observing how certain quantities vary as the size of the system changes. However, the finite size scaling technique has faced serious challenges in real networks due to the lack of data beyond single snapshots.

Next, we prove that a downwards--upwards multiscale shell of replicas produced by the joint action of the GBG and GR techniques on a real network can be used to study the finite size scaling behavior of bond percolation, Fig.~\ref{fig:application}(c) and (d). In each layer, we measure the average size of the largest component, $\langle G\rangle$, and its fluctuations, i.e., susceptibility $\chi=\frac{\langle G^2\rangle-\langle G\rangle^2}{\langle G\rangle}$, for each combination of $(p, N)$ in the multiscale shell using the fast algorithm of Newman and Ziff~\cite{Newman2000}. In finite systems, a peak in the susceptibility, $\chi$, diverging with the system size indicates the presence of a continuous phase transition, and its position provides a way to estimate the percolation threshold, $p_c$: Fig.~\ref{fig:application}(c). In Fig.~\ref{fig:application}(d), we show that the critical link failure probability, $p_c$, approaches zero as a power law, $p_c(N)\sim N^{-0.374}$, and the maximum, $\chi_{\mathrm{max}}$, of the susceptibility also diverges as a power law: $\chi_{\mathrm{max}}(N)\sim N^{0.456}$. Not only do these results suggest a vanishing percolation threshold in the real Internet graph, as usually happens in scale-free networks, but they also provide a way to estimate the corresponding critical exponents numerically, thus offering a new way to study critical phenomena in single-instance real networks. 

\section{Conclusions}
Real networks are observed to evolve in a self-similar way that preserves their topology throughout the growth process over long time spans. The GBG model lays out a minimal number of simple principles that combine branching growth, one of the paradigms of evolution, and network geometry, to explain the empirical findings via a technique that generates self-similar metric expansion of a network replicating its original structure. One of the essential assumptions in the model, the preservation of the distribution of hidden degrees as the number of nodes increases, leads to the introduction of stable distributions in the context of network modeling. Stable distributions, a rich family of probability distributions with intriguing theoretical and practical properties, are widely used to model heavy-tailed data from many types of physical and economic systems, and represent an alternative to the power law paradigm in the study of complex networks. Meanwhile, the geometric branching growth model relies on a universal connectivity law that operates at all scales, simultaneously encoding short- and long-range connections, which keeps its form over time. Our results suggest that the same principles organize network connectivity at different length scales in real networks and that these principles are also sustained over time. As a result, simplicity, as one of the rationales for self-similarity, is one of the keys to understanding and predicting network evolution.

While some limitations of our model are obvious, for instance the exclusion of the birth/death processes of links and nodes, we believe that complementary hypotheses would not affect the results and our GBG model in any fundamental way. The model captures the main mechanisms that drive and predict the self-similar evolution of real networks. In parallel, and beyond the explanatory power of the model to effectively decode the self-similar evolution of real networks, GBG is also a technique to produce scaled-up replicas of networks: an effective and versatile tool facilitating analysis of the behavior of networks at different size scales. The combination of GBG with scaled-down network replicas produced by GR provides full up-and-down self-similar multiscale unfolding of complex networks that covers both large and small scales. Potential applications that require optimization or control of system size in complex systems are countless. Apart from those explained here, we can mention the assessment of scalability issues in dynamic processes in core functions of real networks, such as in Internet routing protocols.  

%%%%%%%%%%%%%%%%%%%%%%%%%%%%%%%%%%%%%%%%%%%%%%%%%%%%%%%%%%%%%%%%%%%%%
\appendix

\section{Data description}
\label{A}

\noindent\textbf{Journal Citation Network (JCN)}. The citation networks from 1900 to 2013 were reconstructed from data on citations between scientific articles extracted from the Thomson Reuters Citation Index~\cite{Hric2018}. Years were grouped in time windows of ten years before $1970$ and of five years afterwards. A node corresponds to a journal with publications in the given time period. A directed edge is connected from journal $i$ to journal $j$ if an article in journal $i$ cites an article in journal $j$, and the weight of this link is taken to be the number of such citations. Time differences between the citing and the cited articles are shorter than the length of the corresponding time window. 
In this work, we use undirected and unweighted versions of the original networks. We first discard the directions for any link and preserve the weight $\omega_{ij}$ with the sum of the citations, i.e., $\omega_{ij}=\omega_{i \to j}+\omega_{j\to i}$. The resulting networks are weighted and undirected but very dense. Next, we extract the backbone by preserving the edges that represent statistically significant deviations with respect to a null model for the local assignment of weights to edges~\cite{Serrano2009}. Eventually, we obtain the giant connected component of the undirected and unweighted networks from the backbone.

\noindent\textbf{World Trade Atlas}.  We used networks in the World Trade Atlas~\cite{Garcia2016}, a collection of annual world trade network maps in hyperbolic space that provides information for the long-term evolution of the international trade system from $1870$ to $2013$ (world war periods, 1914-1919 and 1939-1947, were not available due to the lack of reported information). The networks were reconstructed using historical aggregate import/export data. In every network, nodes are countries, undirected links represent bilateral trade relationships, and link weights correspond to the value of goods exchanged in a given year in current US millions of dollars. The original networks were extremely dense, and in this work we consider the unweighted versions of the disparity backbones~\cite{Serrano2009,Garcia2016}.

\noindent\textbf{Internet.} The AS Internet topology in June 2009 was extracted from data collected by the archipelago active measurement infrastructure
developed by Cooperative Association for Internet Data Analysis~\cite{Claffy2009}. The network contains $23748$ ASs represented as nodes, and it has average AS degree $\langle k \rangle=4.92$, and average clustering (measured over ASs of degree larger than $1$) $\bar{c}=0.61$. The exponent of the power-law degree distribution is $\gamma \approx 2.1$. 

\noindent\textbf{Metabolic.} 
This network is the one-mode projection of metabolites of the bipartite human metabolic network at the cell level~\cite{Serrano2012}. The network has $N=1436$ nodes, average degree $\langle k \rangle=6.57$, and average clustering $\bar{c}=0.54$. The exponent of the power-law degree distribution is $\gamma \approx 2.6$. 

\noindent\textbf{Facebook.} 
The Facebook dataset~\cite{Fire2016} contains $320$ nodes and $2369$ links, where nodes correspond to Facebook users who stated that they worked for a specific software development company in their Facebook profiles, and links represent the friendship relationships. The network has $N=320$ nodes, average degree $\langle k \rangle=14.81$, and average clustering $\bar{c}=0.49$. The exponent of the power-law degree distribution is $\gamma \approx 3.8$.

We also analyzed the following networks (giant connected component of the undirected and unweighted versions):
\begin{itemize}
	\item \textbf{Airports.} The airports dataset is obtained from Ref.~\cite{Airport2016,Kunegis2013}. Directed links represent flights by airlines. A undirected version is used by keeping bidirectional edges only.
	\item \textbf{Drosophila.} The dataset is from Ref.~\cite{Takemura2013}. It is a functional connectome within the Drosophila melanogaster optic medulla related to the motion detection circuit.
	\item \textbf{Enron.} The network of email communication within the Enron company is obtained from Refs.~\cite{Klimt2004,Leskovec2009}.
	\item \textbf{Music.} The dataset is obtained from Ref.~\cite{Serra2012,Garcia2018}, where nodes are chords-sets of musical notes played in a single beat-and connections represent observed transitions among them in a set of songs. Instead of the weighted, directed and dense network of original dataset, we considered the undirected, unweighted and sparser version by applied the disparity filter~\cite{Serrano2009}.
	\item \textbf{Proteome.} This proteome network is obtained from the human HI-II-14 interactome in Ref.~\cite{Rolland2014} and is removed self-loops in this work.
	\item \textbf{Words.} The Words network is the network of adjacency between words in Darwin's book “The Origin of Species”, from Ref.~\cite{Milo2004}.	
\end{itemize}
In Tables~S1 and S2 in SM, we give year by year statistics for the JCN and the WTW, respectively. Table~S3 summarizes the main topological features and fitting parameters of stable distribution for the rest of networks. 

%===================================================================================================
\section{Network embedding to produce geometric network maps}
\label{B}
 
We embed each considered network (except WTW networks) into hyperbolic space using the algorithm introduced in Ref.~\cite{Garcia2019}, named Mercator. Mercator takes the network adjacency matrix $A_{ij}$ ($A_{ij}=A_{ji}=1$ if there is a link between nodes $i$ and
$j$, and $A_{ij}=A_{ji}=0$ otherwise) as input and then returns inferred hidden degrees, angular positions of nodes and global model parameters. More precisely, the hyperbolic maps were inferred by finding the hidden degree and angular position of each node, $\{\kappa_i\}$ and $\{\theta_i\}$, that maximize the likelihood $\mathcal{L}$ that the structure of the network was generated by the $\mathbb{S}^1$ model, where
\begin{align}
\mathcal{L} = \prod_{i<j} \left[ p_{ij} \right]^{A_{ij}} \left[ 1 - p_{ij} \right]^{1 - A_{ij}} \ ,
\end{align}
and $p_{ij}$ is the connected probability with Eq.~\eqref{eq:con_pro}

For the WTW, we used the coordinates in the hyperbolic maps from our previous work~\cite{Garcia2016}, in which we used a Metropolis-Hastings algorithm to embed the sequence of networks starting from the most recent one. The embedding of each network was based on the $\mathbb{H}^2$ model, taking as initial coordinates the ones obtained in the embedding of the posterior layer. To give a more accurate estimation of parameters $\mu$ and $\beta$, we have adjusted them so that the ensemble of synthetic networks generated by Eq.~\eqref{eq:con_pro} with the set of coordinates $\{\kappa_i,\theta_i\}$ in~\cite{Garcia2016} have the same average degree and clustering coefficient as the empirical network on average. The detailed procedure to adjusted $\mu$ and $\beta$ is as follows:  
\begin{itemize}
	\item[1.] Take the coordinates $\{\kappa_i,\theta_i\}$ and parameters $\mu$ and $\beta$ from the embeddings in~\cite{Garcia2016}. 
	\item[2.] 
	To obtain a synthetic network with the average degree $\langle k\rangle_{real}$ of a real network, we correct the value of $\mu$ as $\mu_{new}=\xi \frac{\langle k\rangle_{real}}{\langle k \rangle_{syn}}\mu$, where the initial value of parameter $\mu$ is taken from the embedding in~\cite{Garcia2016}, and we generate a synthetic network by connecting pairs of nodes using Eq.~\eqref{eq:con_pro}. If $\langle k \rangle_{syn}>\langle k\rangle_{real}$, we set $\xi-0.1u \to \xi $, where $u$ is a random variable uniformly distributed between $(0,1)$. Similarly, if $\langle k \rangle_{syn}<\langle k\rangle_{real}$, $\xi+0.1u \to \xi$. The process is iterated and ends when $|\langle k \rangle_{syn}-\langle k\rangle_{real}|<0.1$. 
	
	\item[3.] To obtain a synthetic network with the average clustering coefficient $\langle c\rangle_{real}$ of the real network, we set $\beta_{max}=3$ and $\beta_{min}=1$ and $\beta_{new}=(\beta_{max}+\beta_{min})/2$, where the initial value of parameter $\beta$ is taken from the embedding in~\cite{Garcia2016}, and we generate a synthetic network connecting pairs of nodes using Eq.~\eqref{eq:con_pro}. If $\langle c \rangle_{syn}>\langle c\rangle_{real}$, we set $\beta_{max}=\beta_{new}$, else $\beta_{min}=\beta_{new}$. The process is iterated and ends when $|\langle c \rangle_{syn}-\langle c\rangle_{real}|<0.01$.
\end{itemize}
We generated $1000$ synthetic networks for each WTW snapshot and obtained the final parameters $\mu$ and $\beta$ as their average values over the generated ensembles. 

%===================================================================================================

\section{Adjustment of stable distributions and generation of $z^{\pm}$}
\label{C0}

A stable distribution requires four parameters $\{\alpha,\eta, c, d\}$ to be fully characterized. Apart from the four parameters, there are multiple parameterizations for stable laws depending on the research purpose~\cite{Nolan:1997,Nolan2018,Royuela2017}. This variety of parameterizations is caused by a combination of historical evolution, plus the numerous problems that have been analyzed using specialized forms of stable distributions. The probability density function for a general stable distribution does not have an analytic form, but its characteristic function $\Phi(t)$ (its Fourier transform) does. In the most common parametrization, the one that we use in this work with parametrization subindex $1$ as in~\cite{Royuela2017}, $\Phi(t)=\exp[\Psi(t)]$ where
\begin{eqnarray} \label{eq:CF}
\Psi(t)=
\begin{cases}
-|ct|^{\alpha}[1-i\eta\tan(\frac{\pi\alpha}{2})\text{sign}(t)]+idt, & \alpha\neq1\\
-|ct|[1+i\eta\frac{2}{\pi}\text{sign}(t)\ln(|t|)]+idt, & \alpha =1\\
\end{cases}
\end{eqnarray}

\begin{eqnarray} \label{eq:sign}
\text{sign}(t)=
\begin{cases}
1,& t>0\\
0,& t=0\\
-1,& t<0
\end{cases}
\end{eqnarray} 
and $i$ denotes the imaginary unit.

To obtain parameters $\{\alpha,\eta, c, d\}$ for the distribution of hidden degrees $z$ in the original layer we use the software ``libstable'' in Ref.~\cite{Royuela2017}. In particular, we use the modified maximum likelihood method, where the maximization search is only performed in the 2D $\alpha$-$\eta$ space, such that the procedure produces more accurate estimates with much shorter execution times. In Fig.~2 in the main text and Fig.~S5, we show that the stable distribution offers a very good fit to the hidden variables $z$ in real networks.

%===================================================================================================

\section{Adjustment of stable distributions and generation of $z^{\pm}$}
\label{C}

When $b=2$, the hidden degree $z^+$ of the descendant of an ancestor with hidden degree $z$ is obtained numerically by solving 
\begin{equation} \label{eq:generate_z_plus}
\int_{z^{\pm}_{\mathrm{cut}}}^{z^+}\rho(z'_{+}|z)^{\mathrm{nor}} \textrm{d}z'_{+}=u, 
\end{equation}
where $u$ is a random variable uniformly distributed between $0$ and $1$, and 
\begin{equation} 
\begin{aligned}
\rho(z^{+}|z)^{\mathrm{nor}} &= C(z) \rho(z^{+}|z)\\
&=  C(z) f(z^{+};\alpha^{\pm},\eta^{\pm}, c^{\pm}, d^{\pm})\\
&\times f(z-z^{+};\alpha^{\pm},\eta^{\pm}, c^{\pm}, d^{\pm}).
\end{aligned}
\label{eq:z_condition}
\end{equation}
The stable distribution for descendants $f(z^{\pm}; \alpha^{\pm},\eta^{\pm}, c^{\pm},d^{\pm})$ follows immediately from one of the basic properties of stable distributions in the parametrization used in this work~\cite{Royuela2017} (see Appendix~\ref{C0}). If $z_1\sim f(z_1;\alpha_1,\eta_1, c_1, d_1)$ and $z_2\sim f(z_2;\alpha_2,\eta_2, c_2, d_2)$ are independent variables, then $z_1+z_2\sim f(z;\alpha,\eta, c, d)$ with
\begin{equation}\label{eq:stable}
\alpha_{1}=\alpha_{2}=\alpha,\;
\eta=\frac{\eta_1c_1^\alpha+\eta_2c_2^\alpha}{c_1^\alpha+c_2^\alpha},\;
c^\alpha=c_1^\alpha+c_2^\alpha,\;
d=d_1+d_2,
\end{equation}
and then $f(z^{\pm}; \alpha^{\pm},\eta^{\pm}, c^{\pm},d^{\pm})\!=\!f(z^{\pm}; \alpha,\eta, c/2^{1/\alpha}, d/2)$. 

The purpose of factor $C(z)$ in Eq.~(\ref{eq:z_condition}) is two-fold. On the one hand, it acts as a normalization that ensures that the hidden degrees of descendants are non-negative, given that, in general, the stable distribution can have negative support. On the other hand, as we are considering real-world networks, finite-size effects play an important role. To control for these effects, we introduce the normalization factor $C(z)$ in Eq.~\eqref{eq:z_condition} defined as 
\begin{widetext}
\begin{equation} \label{eq:Cz}
C(z) =\left[\int_{z^{\pm}_{\mathrm{cut}}}^{z-z^{\pm}_{\mathrm{cut}}} \mathrm{d}z^+ f(z^{+};\alpha^{\pm},\eta^{\pm}, c^{\pm}, d^{\pm})f(z-z^{+};\alpha^{\pm},\eta^{\pm}, c^{\pm}, d^{\pm})\right]^{-1}.
\end{equation}
\end{widetext}
To ensure that $z^{\pm}$ are non-negative, we impose a minimum hidden degree cut-off to descendants that is related to the minimum hidden degree $z_0$ in the distribution of ancestors, $z^{\pm}_{\mathrm{cut}}=z_{0}/2^{1/\alpha}$.

Let us consider the following aspects:
\begin{itemize}
	\item[I.] Locally, and therefore also globally, $z^{+}$ and $z^{-}$ follow the same distribution, i.e., $f(z^+;\alpha,\eta, c^+, d^+)$=$f(z^-;\alpha,\eta, c^-, d^-)$. This can be shown straightforwardly, since
    
	\begin{equation*}
	\begin{aligned}
	\rho(z^{-}|z) &= \int \textrm{d}z^{+} \rho(z^{+}|z) \delta(z^{+}-(z-z^{-})) \\
	&= \frac{f(z-z^{-};\alpha,\eta, c^{\pm}, d^{\pm})f(z^{-};\alpha,\eta, c^{\pm}, d^{\pm})}{f(z;\alpha,\eta, c, d)}.
	\end{aligned}
	\end{equation*}
    
	\item[II.] Globally, both variables $z^{\pm}$ are distributed as $f(z^{\pm};\alpha,\eta, c^{\pm}, d^{\pm})$. To see this, we compute the resulting distribution as
	\begin{equation}\label{eq:stable_proof}
	\begin{aligned}
	\rho(z^{\pm}) &= \int\limits_{z^{\pm}}^{\infty} \textrm{d}z \rho(z^{\pm}|z)\rho(z) \\
	&= \int\limits_{z^{\pm}}^{\infty} \textrm{d}z \Big[\frac{f(z^{\pm};\alpha,\eta, c^{\pm}, d^{\pm})f(z-z^{\pm};\alpha,\eta, c^{\pm}, d^{\pm})}{f(z;\alpha,\eta, c, d)}\\
	&\times f(z;\alpha,\eta, c, d) \Big] \\
	&= f(z^{\pm};\alpha,\eta, c^{\pm}, d^{\pm}) \int \limits_{z^{\pm}}^{\infty} \textrm{d}z f(z-z^{\pm};\alpha,\eta, c^{\pm}, d^{\pm}) \\
	&= f(z^{\pm};\alpha,\eta, c^{\pm}, d^{\pm}) \int \limits_{0}^{\infty} \textrm{d}z f(z;\alpha,\eta, c, d)\\
	&= f(z^{\pm};\alpha,\eta, c^{\pm}, d^{\pm}).
	\end{aligned}
	\end{equation}
	
\end{itemize}
Eq.~(\ref{eq:stable_proof}) proves that the distribution for descendants, $f(z^{\pm}; \alpha^{\pm},\eta^{\pm}, c^{\pm},d^{\pm})$, has the basic properties of the stable distribution, with the same shape parameters and adjusted scale and location as the ancestor layer.  In Fig.~\ref{fig:sketch} and Fig.~S5, we show that the shape of the distribution of hidden variables $z^{\pm}$ for the descendants is invariant as the ancestor layer in different empirical networks, which confirms the results in Eq.~(\ref{eq:stable_proof}) very well.

In the case of  fractionary $b$,  in Eq.~(\ref{eq:z_condition}) we use $f(z^{\pm}; \alpha^{\pm},\eta^{\pm}, c^{\pm},d^{\pm})\!=\!f(z^{\pm}; \alpha,\eta, c/b^{1/\alpha}, d/b)$, which gives a better approximation to the actual distribution that is a mixture of two stable distributions, the one from nodes that split plus the one from nodes that do not split. 

%===================================================================================================
\section{Connecting descendants of connected ancestors}
\label{D}

In the non-inflationary model, after the branching process, we establish links between descendants of connected ancestors with probability $p_{ij}(\mu')$, with $\mu'=b \mu$, but fulfilling the condition that there is at least one link between them. Given a pair on connected ancestors, we characterize the state of links between descendants with different ancestor using binary variables $\sigma_e$, $e \in \{1,...,E\}$, with $\sigma_e=1$ meaning that the link exists and $\sigma_e=0$ that the link is absent and variable $E$ giving the number of all possible links. We have changed from the $ij$ notation to a one-index notation for the sake of simplicity. If we assume that nodes either remain or split in at most two descendants, $E$ can only take the values $4$, $2$, or $1$. The number of possible configurations combining absence and presence of links between descendants is $2^E-1$, and the possible states are encoded in vector $V=[\sigma_1,\sigma_2,\ldots,\sigma_E]$. Any possible configuration occurs with unbiased probability
\begin{equation}\label{eq:conn_pattern}
P(V) = \frac{\prod \limits_{e=1}^E {p_e(\mu')}^{\sigma_{e}}[1-{p_e(\mu')}]^{1-\sigma_{e}}}{1-q},
\end{equation}
where $q=\prod\limits_{e=1}^E [1-p_e(\mu')]$ is the probability that the descendants of connected ancestors are not connected. Notice that, by unbiased, we mean that the probability for any link configuration remains fully congruent with the $\mathbb{S}^1$ model, since we are not introducing any uncontrolled deviation when imposing that at least one link exist. Yet, the $\mathbb{S}^1$-model connection probabilities are capable of encoding correlations between nodes such as, for instance, degree-degree correlations (see Ref.~\cite{Garcia2019}).

We use the following simple algorithm to generate one of the possible configurations of connections between descendants satisfying the previous equation but ensuring that at least one connection is formed: 
\begin{itemize}
	\item[(1)] Compute probability $q = \prod\limits_{e=1}^E (1-p_e(\mu'))$.
	\item[(2)] For each possible link $e = 1,\ldots,E$:
	\begin{itemize}
		\item If at least one of the $E$ potential links has been already formed, connect the current link $e$ with probability $p_e(\mu')$.
		\item Else, connect the current link $e$ with probability $p_e(\mu')/(1-q)$ and then update $q$ as $q\to q/(1-p_e(\mu'))$.
	\end{itemize}
\end{itemize}
The advantage of this algorithm is that it allows the random generation of any possible configuration by assigning links sequentially only once and without discarding any history.

For the sake of completeness, let us show that the above algorithm indeed generates any sequence of links with the unbiased probability given by Eq.~\eqref{eq:conn_pattern}. To do so, consider a general configuration in which the $i$-th link is the first to be assigned, that is, $\sigma_n = 0$ for $n<i$ and $\sigma_i = 1$. Note that, at any time $n\leq i$, the value of $q$ (to which we will now refer as $q_{n}$) is
\begin{equation}
q_{n} = \prod\limits_{e=n}^{E} (1-p_{e}(\mu')).
\end{equation}
Therefore, the probability for link $n<i$ not to be assigned is
\begin{equation}
\begin{aligned}
P(\sigma_n= 0) &= 1 - \frac{p_{n}}{1-q_{n}} = \frac{1-p_{n}(\mu')-\prod\limits_{e=n}^{E} (1-p_{e}(\mu'))}{1-\prod\limits_{e=n}^{E} (1-p_{e}(\mu'))} \\
&= \frac{(1-p_{n}(\mu'))\left(1-\prod\limits_{e=n+1}^{E} (1-p_{e}(\mu'))\right)}{1-\prod\limits_{e=n}^{E} (1-p_{e}(\mu'))} \\
&= \frac{(1-p_{n})\left(1-q_{n+1}\right)}{1-q_{n}}.
\end{aligned}
\end{equation}
Now, let us take a look at the resulting probability with which the whole configuration is generated if we follow the algorithm:
\begin{equation}
\begin{aligned}
P(V) &= \left( \prod\limits_{e=1}^{i-1} \frac{(1-p_{n}(\mu'))\left(1-q_{n+1}\right)}{1-q_{n}} \right) \frac{p_{i}}{1-q_{i}} \\
&\times\prod \limits_{e=i+1}^{E} p_{e}(\mu')^{\sigma_e}(1-p_{e}(\mu'))^{1-\sigma_e} \\
&= \frac{1}{1-q_{1}}\left( \prod\limits_{e=1}^{i-1} (1-p_{e}(\mu')) \right)p_{i}\\
&\times\prod \limits_{e=i+1}^{E} p_{e}(\mu')^{\sigma_e}(1-p_{e}(\mu'))^{1-\sigma_e}\\
&= \frac{1}{1-\prod \limits_{e=1}^{E}(1-p_{e}(\mu'))}\left( \prod\limits_{e=1}^{i-1} (1-p_{e}(\mu')) \right)p_{i}\\
&\times \prod \limits_{e=i+1}^{E} p_{e}(\mu')^{\sigma_e}(1-p_{e}(\mu'))^{1-\sigma_e},
\end{aligned}
\end{equation}
which is Eq.~\eqref{eq:conn_pattern} in this particular case. Notice that, if $i=E$, $q_{i} = 1-p_{i}$ and $P(\sigma_{i}=1) = 1$, there is always one link at least.

%===================================================================================================
\section{Estimation of parameter $a$ in real networks}
\label{E}

Parameter $a$ is used to adjust the growth of the average degree as the network evolves. Based on Eq.~\eqref{eq:k_Max2}, the slope $s$ of $\langle k\rangle$ as a function of $N$ in log-log scale is 
\begin{equation}
s=\frac{\ln b^{-\nu}+\ln a}{\ln b}=\frac{\ln a \varphi}{\ln b}.
\end{equation} 

Parameter $a$ can be calculated for a certain branching ratio $b$ once $s$ and $\nu$, or alternatively $\varphi$, are estimated. We perform the following steps
\begin{itemize}
	\item[(1)] We estimate $s$ by least squares fitting the empirical data in a given time period, see Fig.~\ref{fig:emp2model}(a) and (b). 
	
	\item[(2)] We take a network snapshot and apply to it one step of non-inflationary GBG ($a=1)$, i.e., without adding extra links. Then, we measure the average degree $\langle k\rangle_{a=1}^{(1)}$ and $\langle k\rangle^{(0)}$ in the descendant and original layers and calculate $\varphi=b^{-\nu}=\langle k\rangle_{a=1}^{(1)}/\langle k\rangle^{(0)}$ using Eq.~\eqref{eq:k_Max2}. 
	
	\item[(3)] Calculate the value of $a$ using equation above.
\end{itemize}

We repeat the process for different time snapshots to get a set of values of $a$. In  Fig.~S18 and S24 in SM, we show the values of $a$ in the JCN and WTW. In both systems, the variances of $a$ are very small.

%===================================================================================================
\section{Upscaled real network replicas}
\label{F}

Depending on the desired final size, $b$ and the number of layers $l$ are fixed to some specific values. To adjust the average degree, we set $a=\xi \frac{\langle k^{(0)}\rangle}{\langle k^{(l)}\rangle}$, where $\langle k^{(0)}\rangle$ is the target average degree and $\langle k^{(l)}\rangle$ is the obtained average degree in non-inflationary GBG layer $l$. We start an iterative process with $\xi=1$ and $\langle k^{(l)}\rangle$ the initial average degree of layer $l$. In each round, we add links to the network using Eq.~\eqref{eq:conprob_extra} and calculate the average degree $\langle k_{\mathrm{new}}^{(l)}\rangle$ of the resulting network. If $\langle k_{\mathrm{new}}^{(l)}\rangle> \langle k^{(0)}\rangle$, we discard the whole realization and start the process again giving $\xi$ a new value $\xi-0.1u$, where $u$ is a random variable uniformly distributed in the interval $(0,1)$. Similarly, if $\langle k_{new}^{(l)}\rangle< \langle k^{(0)}\rangle$, $\xi$ is updated as $\xi+0.1u$. The process is repeated until $|\langle k_{new}^{(l)}\rangle- \langle k^{(0)}\rangle|$ is below a given threshold, that we set to $0.1$.

%===================================================================================================
\section{Simulation of the MEIN model}
\label{G}

Stochastic resonance was found in a very simple model for opinion formation in social systems~\cite{Kuperman2002,Tessone2005,Toral2006}. In the model, the opinion of an individual can change due to three basic ingredients: (i) the influence of connected neighbors, modeled by a simple majority rule; (ii) the influence of fashion, modeled as some external time varying signal and (iii) random events.

We implement the three effects mentioned above as follows. At time $t=0$, we assign random values $m_i=\pm1$ to each individual; then, at a given time $t$, the next three steps are applied consecutively:

\begin{itemize}
	\item[i.] One individual $i$ is randomly selected and it adopts the majority opinion among its connected neighbors, i.e., $m_i(t)=\text{sign}[\sum_{j\in n(i)} m_j(t)]$. If $\sum_{j\in n(i)} m_j(t)=0$, in case of a tie, a random value for $m_i(t)$ is selected. 
	
	\item[ii.] With probability $A|\text{cos}(2\pi t/T)|$, set $m_i(t)=\text{sign}[\text{cos}(2\pi t/T)]$ to follow the fashion. Parameter $A (0\leq A \leq 1)$ measures the strength of the fashion and $T$ its
	period. 
	
	\item[iii.] With probability $\epsilon$, let $m_i(t)$ adopt randomly a new value (independently of its present state), where $\epsilon$ represents the noise intensity in the dynamics.
\end{itemize}

After the three steps have been performed, time increases by a fixed amount $t=t+1/N$, so that after one unit of time every individual has been updated once on average. 

To measure of the response of the system to external modulation, we computed the spectral amplification factor~\cite{Jung1989}:
\begin{equation}\label{eq:R}
R=4A^{-2}|\langle e^{i2\pi t/T} \rho(t)\rangle |,
\end{equation}
where $\rho(t)=\frac{1}{N}\sum_i m_i(t)$ is the average opinion in the network at time $t$, and $\langle \cdots \rangle$ denotes a time average. The temporal output of each node was recorded for $2\times10^5$ time units, discarding the first $10^5$ time units as transient. 

In this paper, the amplitude $A=0.22$ and period $T=200$ are fixed. All points show the results averaged over 100 realizations of the dynamics. 

%%%%%%%%%%%%%%%%%%%%%%%%%%%%%%%%%%%%%%%%%%%%%%%%%%%%%%%%%%%%%%%%%%%%%

% =================================================================================================
% Acknowledgments
% =================================================================================================
%
\section*{Acknowledgments}
We acknowledge support from: a James S. McDonnell Foundation Scholar Award in Complex Systems; the ICREA Academia award, funded by the \textit{Generalitat de Catalunya}; the Spanish \textit{Ministerio de Ciencia, Innovaci\'on y Universidades} project no. FIS2016-76830-C2-2-P (AEI/FEDER, UE); project {\it Mapping Big Data Systems: embedding large complex networks in low-dimensional hidden metric spaces}, \textit{Ayudas Fundaci\'on BBVA a Equipos de Investigaci\'on Cient\'{\i}fica 2017}, and \textit{Generalitat de Catalunya} grant No.~2017SGR1064. Furthermore, G.~G.-P. acknowledges financial support from the Academy of Finland via the Centre of Excellence program (Project no.~312058 as well as Project no.~287750), and from the emmy.network foundation under the aegis of the \textit{Fondation de Luxembourg}.

\bibliography{reference}

\onecolumngrid
\appendix
\renewcommand\appendix{\setcounter{secnumdepth}{0}}
\newpage
\clearpage

\setcounter{page}{0}
\pagenumbering{arabic}

\clearpage
\begin{minipage}[h]{\textwidth}
	\begin{center}
		\large{\textbf{Supplementary Materials for\\ Scaling up real networks by geometric branching growth}}\\
		\vspace{0.5cm}
		Muhua Zheng$^{1,2}$, Guillermo Garc\'ia-P\'erez$^{3,4}$, Mari\'an Bogu\~n\'a$^{1,2}$ \& M. \'Angeles Serrano$^{1,2,5*}$\\ 
		\vspace{1cm}
		
		\small{1. Departament de F{\'\i}sica de la Mat\`eria Condensada,\\ Universitat de Barcelona, Mart\'{\i} i Franqu\`es 1, 08028 Barcelona, Spain}\\
		\small{2. Universitat de Barcelona Institute of Complex Systems (UBICS), \\Universitat de Barcelona, Barcelona, Spain}\\
		\small{3. QTF Centre of Excellence, Turku Centre for Quantum Physics, \\Department of Physics and Astronomy, \\University of Turku, FI-20014 Turun Yliopisto, Finland }
		\\
		\small{4. Complex Systems Research Group, \\Department of Mathematics and Statistics,\\ University
			of Turku, FI-20014 Turun Yliopisto, Finland}
		\\
		\small{5. ICREA, Pg. Llu\'is Companys 23, E-08010 Barcelona, Spain}

		\small{*Correspondence and requests for materials should be addressed to M.A.S. (marian.serrano@ub.edu)}
		\\

	\end{center}
	
\end{minipage}

\vspace{2cm}

%\maketitle

%\newpage

\tableofcontents
\let\addcontentsline\oldaddcontentsline% Restore \addcontentsline

\newpage
\renewcommand\thefigure{S\arabic{figure}}
\renewcommand\thetable{S\arabic{table}}
\setcounter{figure}{0}    
\setcounter{table}{0}

\renewcommand{\appendixname}{}

%%%%

%%%%%%%%%%%%%%%%%%%%%%%%%%%%%%%%%%%
\clearpage
\newpage
\section{Supplementary Figures S1 to S29}
\subsection{Self-similar evolution of real networks}
%\subsection{Self-similarity in CNS and WTW}
\begin{figure}[h]
	\centering
	\includegraphics[width=1\linewidth]{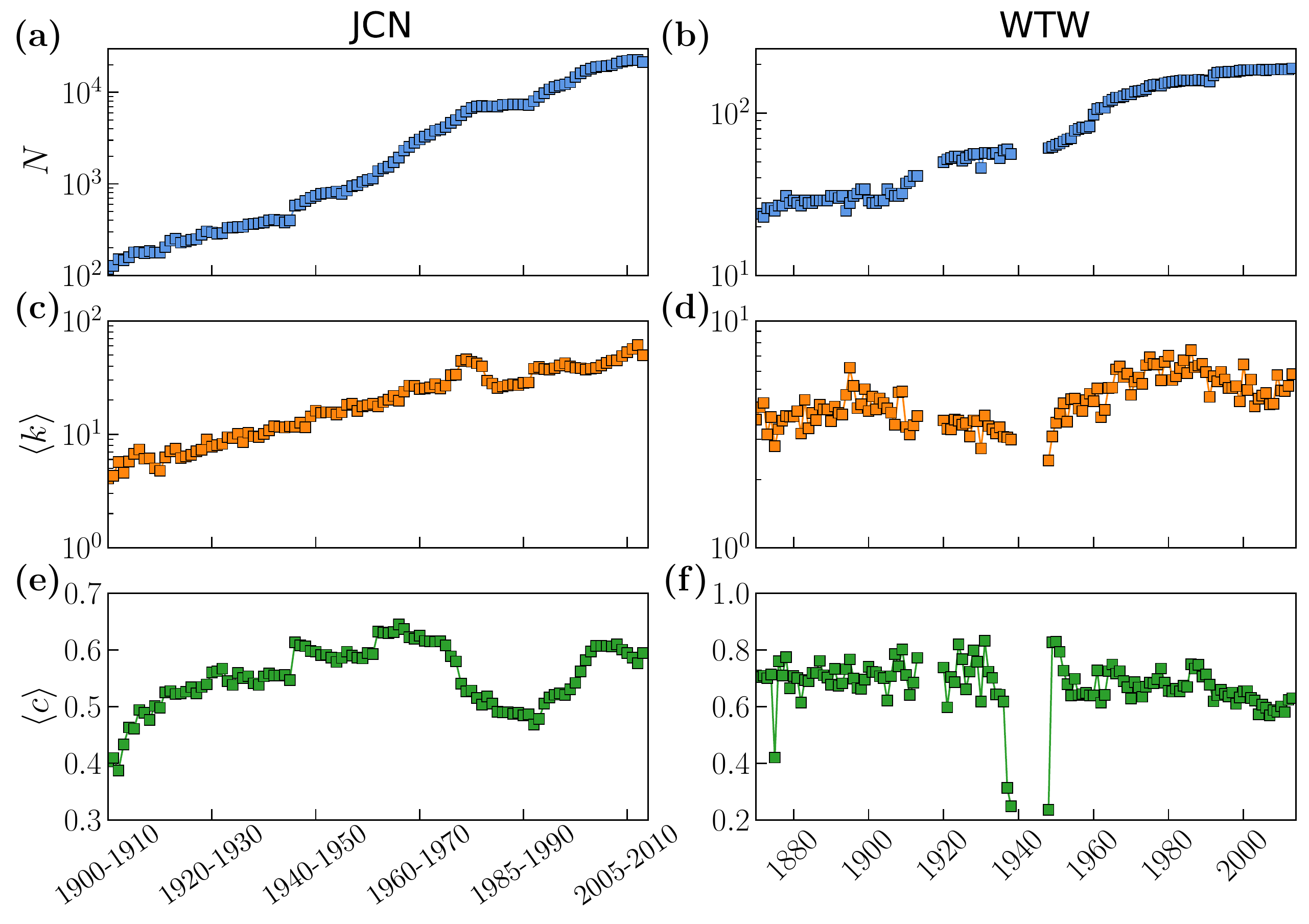}
	\caption{ {\bf Evolution of structural properties.}
		(a) and (b), the number of nodes, (c) and (d), the average degree, (e) and (f), the average clustering coefficient in the journal citation network (left column) and in the world trade web (right column). World war periods, 1914-1919 and 1939-1947, were not available in the WTW due to lack of reported information.}
	\label{figS1}
\end{figure}

\begin{figure}[htb]
	\centering
	\begin{tabular}{@{}c@{}}
		\includegraphics[width=1\linewidth]{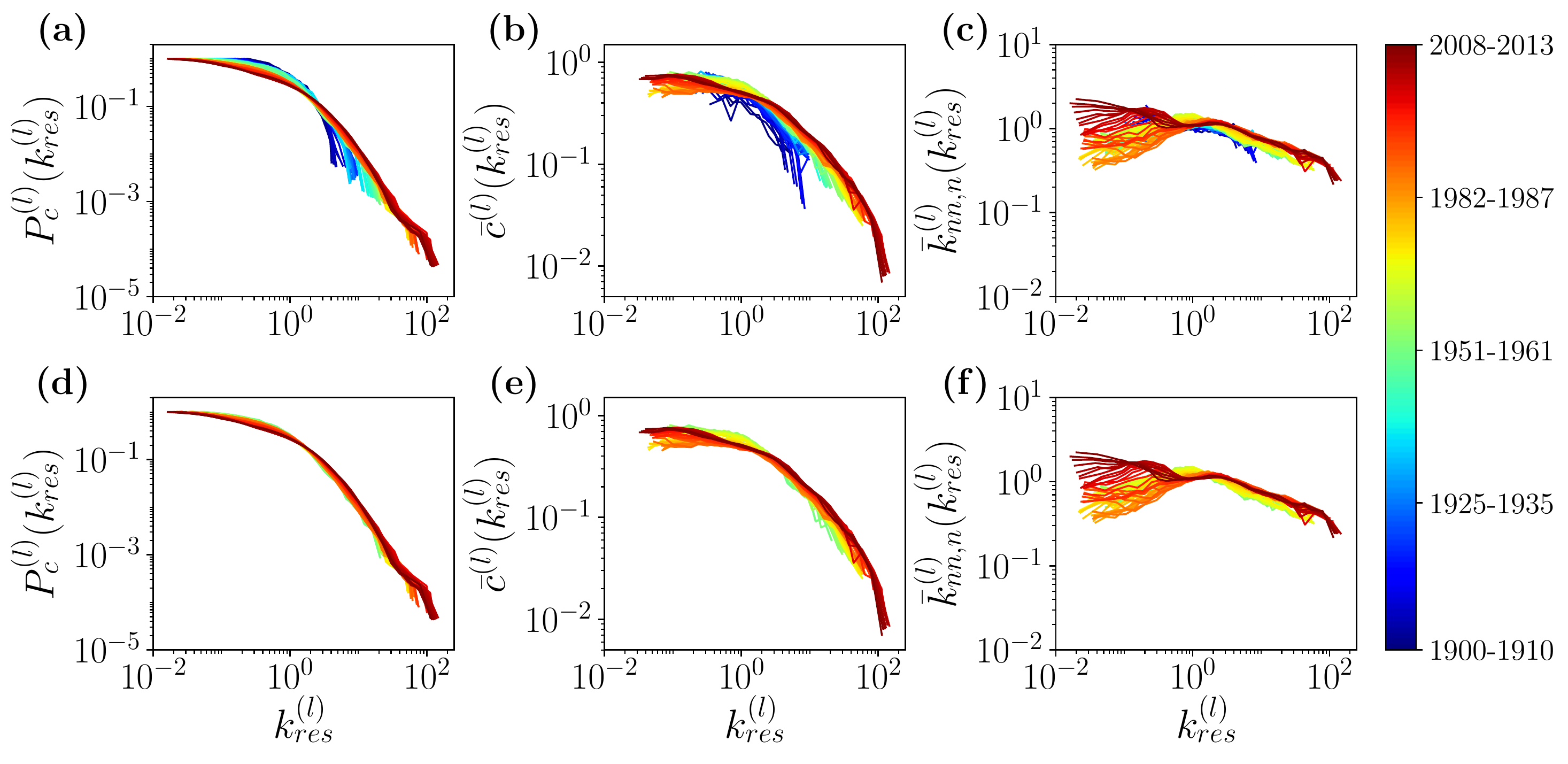}
	\end{tabular}
	\caption{(Color online). \textbf{Self-similar evolution of the journal citation network.} In (a)-(c), we show the complementary cumulative degree distributions, clustering spectra and the degree-dependent average nearest neighbors degree distributions in all the time-windows from 1900-1910 to 2008-2013. In (d)-(f), we only include snapshots from 1950-1960 to 2008-2013. Fluctuations are smaller after World War II.}
\end{figure}

\begin{figure}[htb]
	\centering
	\begin{tabular}{@{}c@{}}
		\includegraphics[width=1\linewidth]{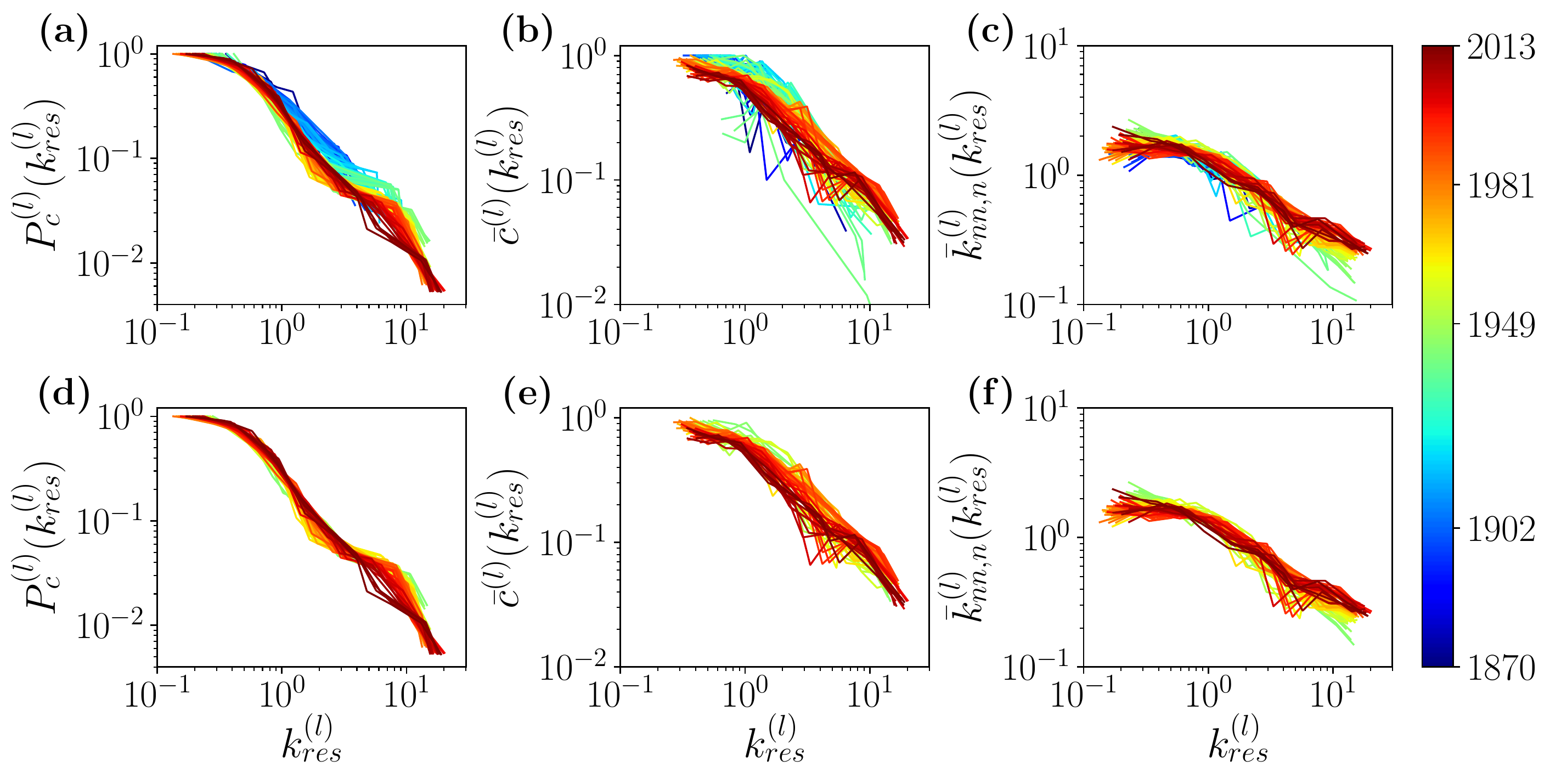}
	\end{tabular}
	\caption{(Color online). \textbf{Self-similar evolution of the world trade web.} 
		In (a)-(c), we show the complementary cumulative degree distributions, clustering spectra and the degree-dependent average nearest neighbors degree distributions in all the networks from 1870 to 2013. In (d)-(f), we only include snapshots from 1950 to 2013. Data in the world war periods, 1914-1919 and 1939-1947, was not available due to lack of reported information.}
\end{figure}

\begin{figure}[h]
	\centering
	\includegraphics[width=1\linewidth]{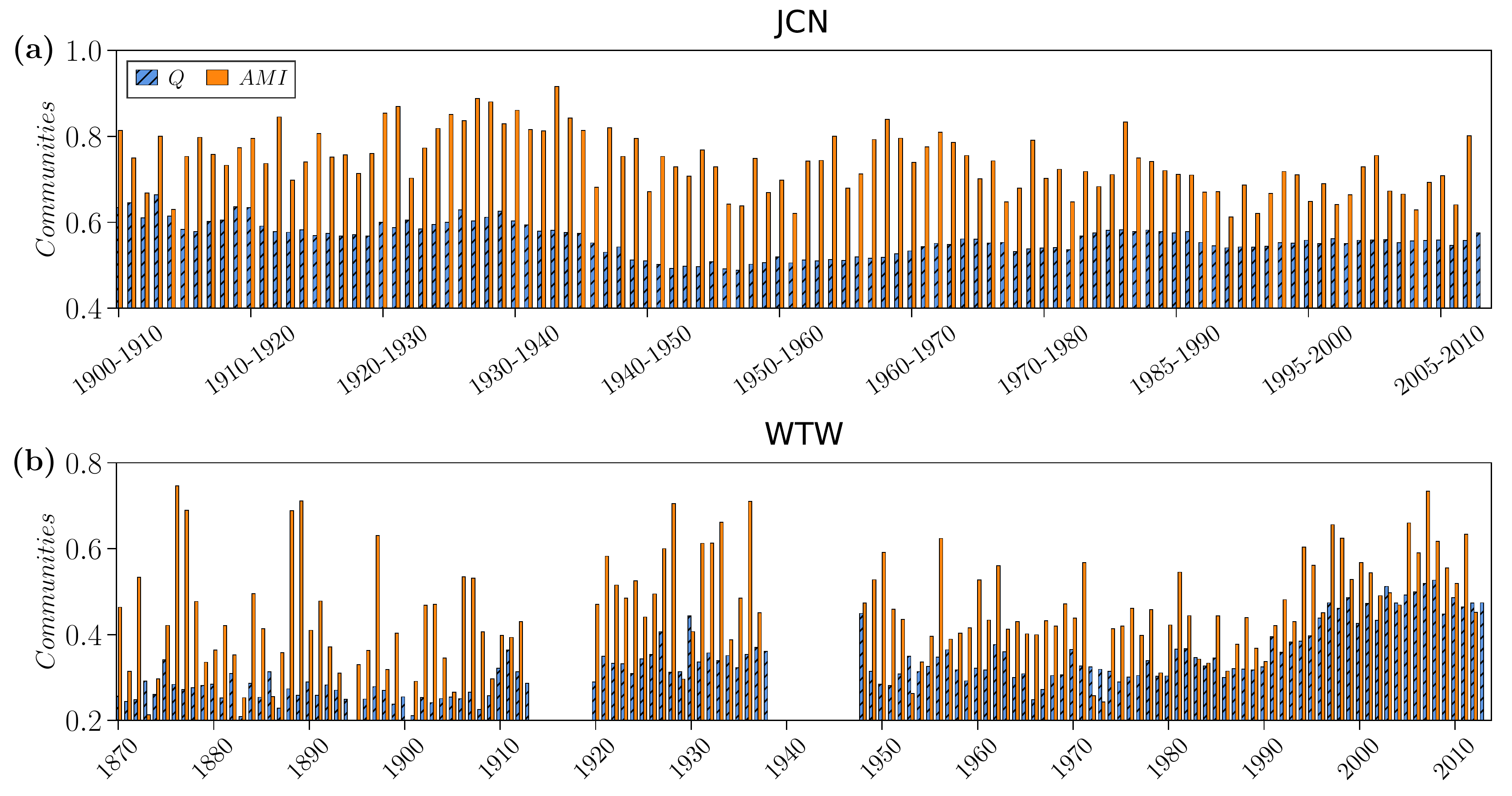}
	\caption{ {\bf Evolution of community structure.}
		Modularity $Q$ and adjusted mutual information $\textrm{AMI}$ between the community partitions of two consecutive snapshots by considering the nodes that exist in both snapshots. The JCN in (a) and the WTW in (b).}
	\label{figS1}
\end{figure}
%==================================================================================================================
%==================================================================================================================
\clearpage
\newpage
\subsection{Stable distributions in real networks}

%\begin{figure}[h]
%	\centering
%	\includegraphics[width=1\linewidth]{fig_SI/Fitting_performance.pdf}
%	\caption{ {\bf Fitting performance with the stable distribution.} The complementary cumulative distribution of $z$ on (\textbf{a}) snapshot 1965-1975 in the JCN and (\textbf{b}) $1960$ in WTW and corresponding stable distribution fitting. (\textbf{c}) and (\textbf{d}) show the distribution $P_c(z^{\pm})$ of different descendants-layers($l=0$ indicates the original network) in the JCN and WTW, respectively.
%	}	
%\end{figure}

\begin{figure}[!h]
	\centering
	\includegraphics[width=0.85\linewidth]{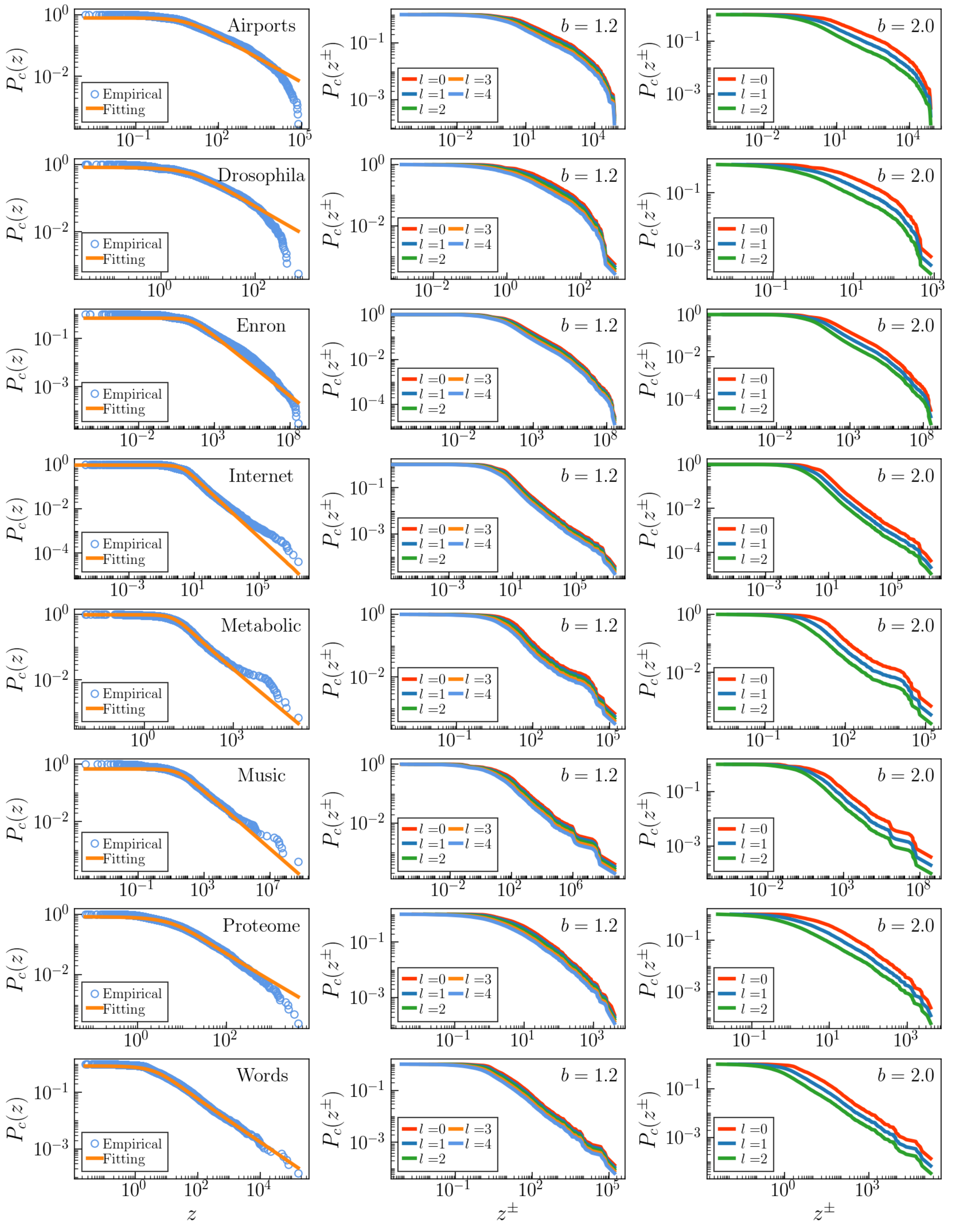}
	\caption{ {\bf Performance of the stable distribution in real networks under GBG.} The left column shows the complementary cumulative distributions of $z$ in the original networks and the stable distribution fitting. Middle and right columns show the distribution $P_c(z^{\pm})$ of different descendant layers ($l=0$ indicates the original network) with $b=1.2$ and $b=2.0$, respectively.
		Each row shows the results for a real network.	}	
\end{figure}

%==================================================================================================================
\clearpage
\newpage
\subsection{Up- and down-scaling in real networks}

\begin{figure}[htb]
	\centering
	\begin{tabular}{@{}c@{}}
		\includegraphics[width=1\linewidth]{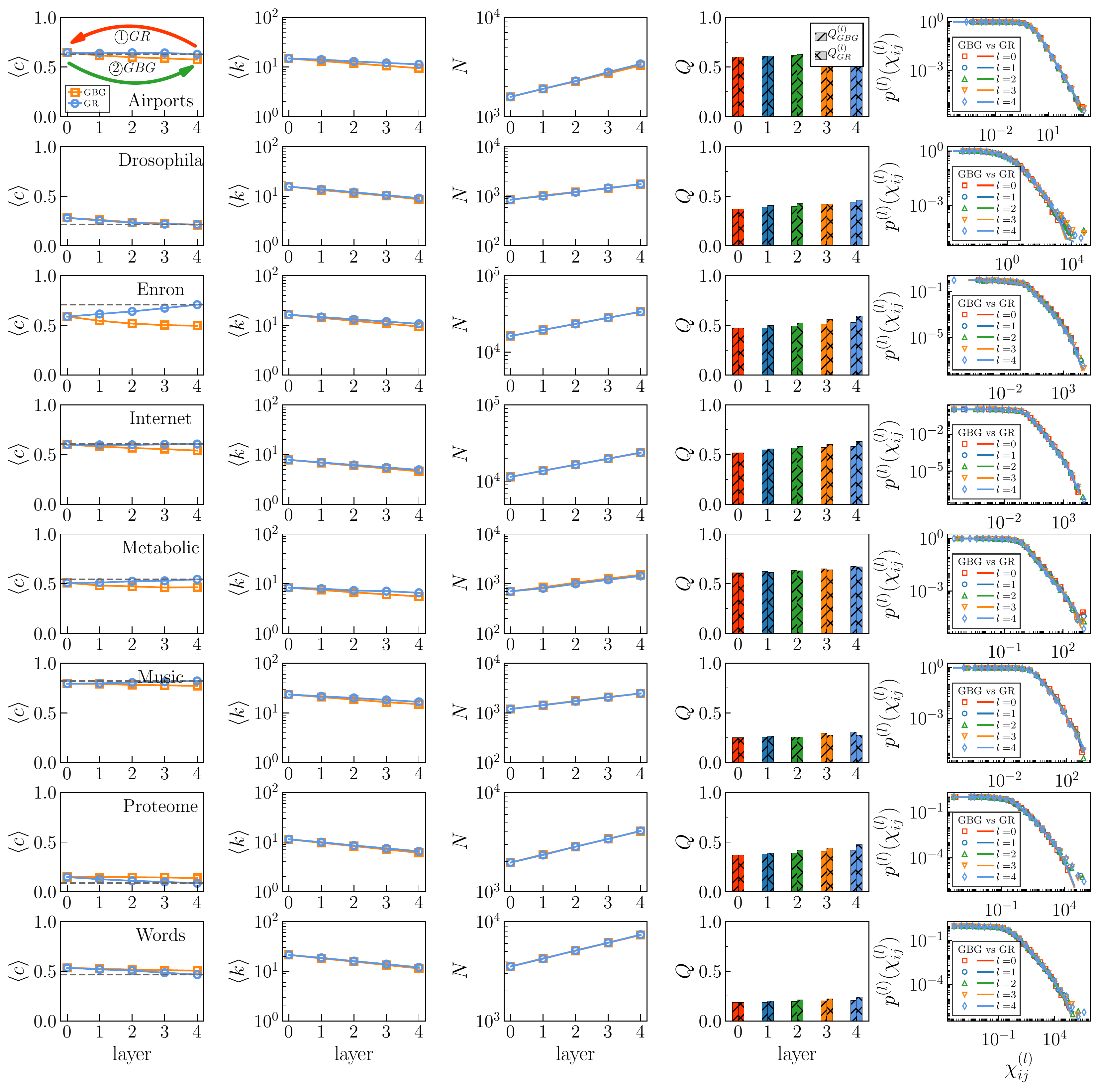}
	\end{tabular}
	\caption{\textbf{GR vs GBG on real networks with $b=1.2$.} Each column from left to right represents the mean clustering coefficient $\langle c \rangle$, average degree $\langle k \rangle$, network size $N$, modularity $Q$ and the connection probability on the GR and the GBG flows. Each row shows the results for a real network.}
\end{figure}

\begin{figure}[htb]
	\centering
	\begin{tabular}{@{}c@{}}
		\includegraphics[width=0.9\linewidth]{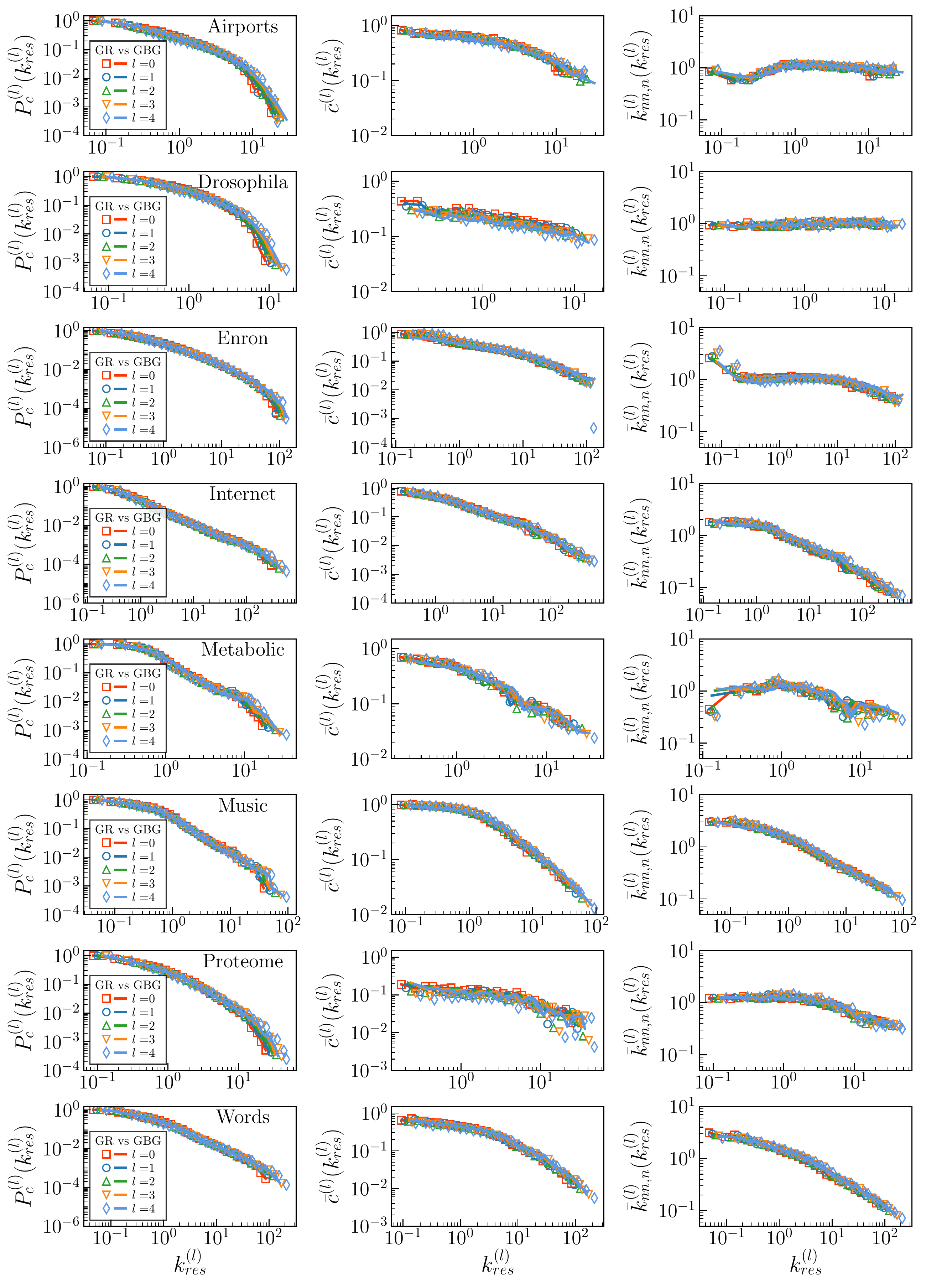}
	\end{tabular}
	\caption{\textbf{GR vs GBG on the real network with $b=1.2$.} Each column from left to right shows the complementary cumulative degree distribution, degree dependent clustering coefficient and degree-degree correlations of rescaled degrees $k^{(l)}_{res} = k^{(l)}/ \langle k^{(l)}\rangle$. Each row shows the results for a real network.}
\end{figure}

\begin{figure}[htb]
	\centering
	\begin{tabular}{@{}c@{}}
		\includegraphics[width=1\linewidth]{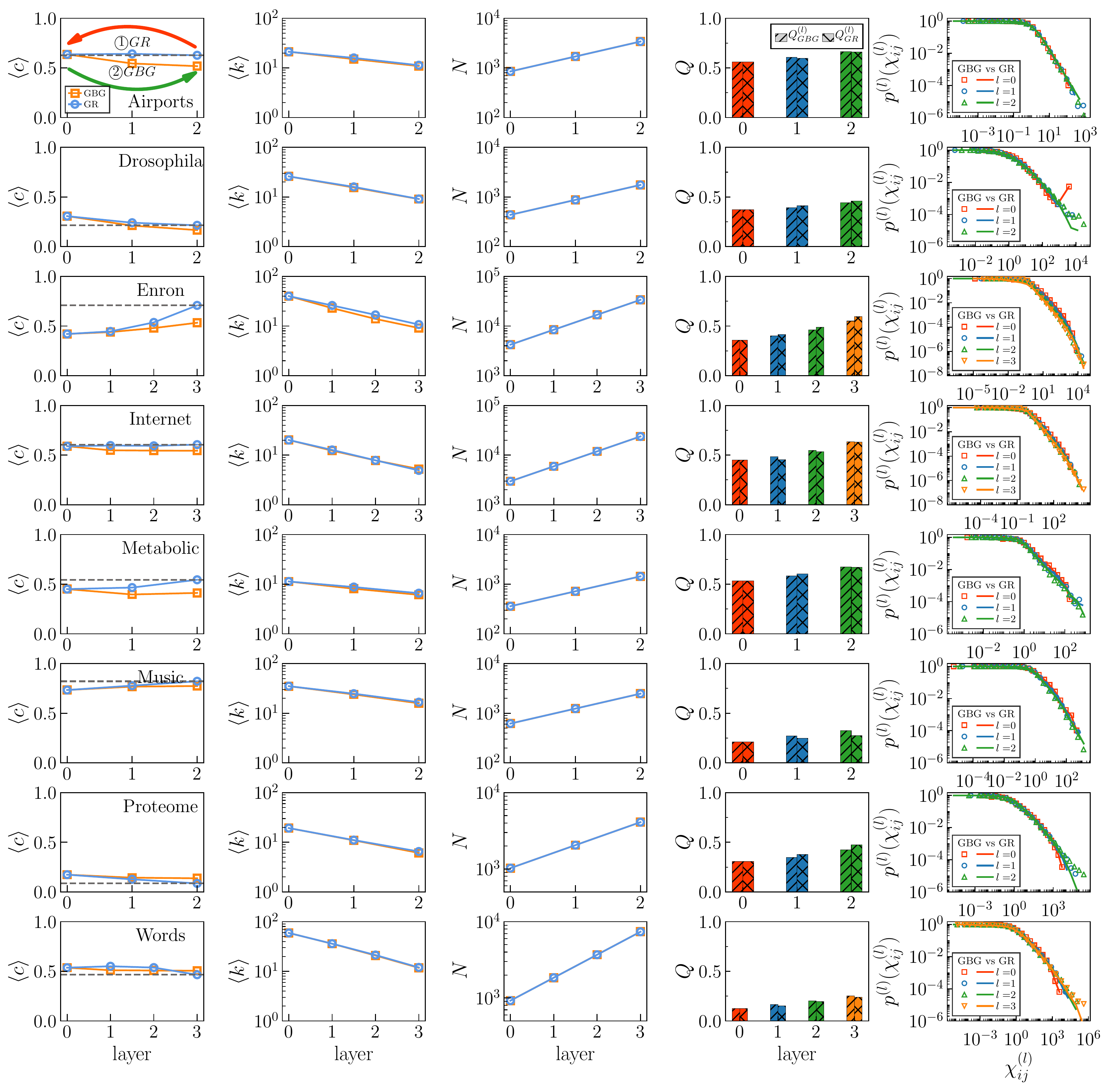}
	\end{tabular}
	\caption{\textbf{GR vs GBG on real networks with $b=2$.} Each column from left to right represents the mean clustering coefficient $\langle c \rangle$, average degree $\langle k \rangle$, network size $N$, modularity $Q$ and the connection probability on the GR and the GBG flows. Each row shows the results for a real network.}
\end{figure}

\begin{figure}[htb]
	\centering
	\begin{tabular}{@{}c@{}}
		\includegraphics[width=0.9\linewidth]{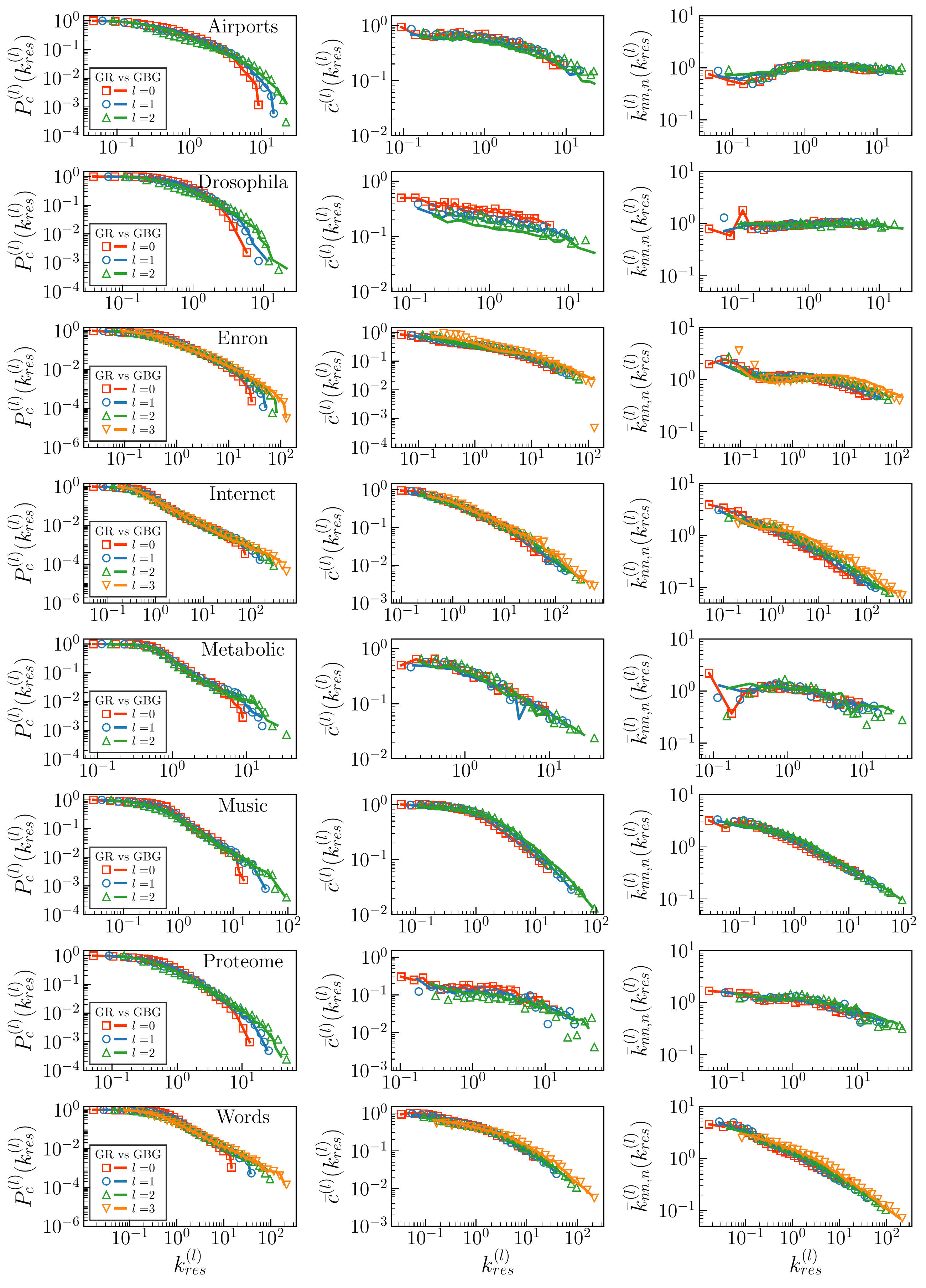}
	\end{tabular}
	\caption{\textbf{GR vs GBG on real networks with $b=2$.} Each column from left to right shows the complementary cumulative degree distribution, degree dependent clustering coefficient and degree-degree correlations of rescaled degrees $k^{(l)}_{res} = k^{(l)}/ \langle k^{(l)}\rangle$. Each row shows the results for a real network. }
\end{figure}
%----------------------------------------------------

\clearpage
\newpage
\begin{figure}[htb]
	\centering
	\begin{tabular}{@{}c@{}}
		\includegraphics[width=1\linewidth]{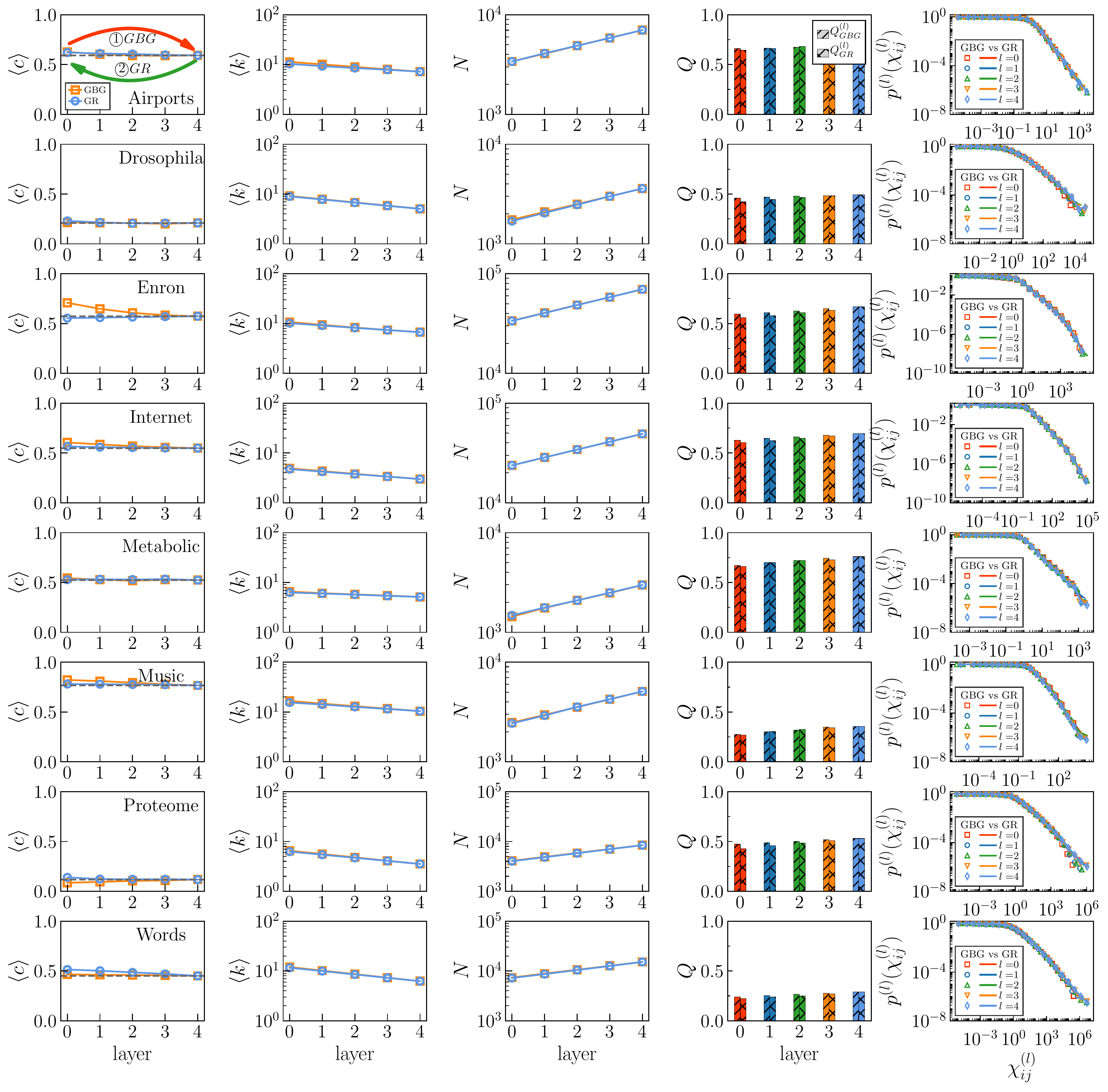}
	\end{tabular}
	\caption{\textbf{GBG vs GR on real networks with $b=1.2$.} Each column from left to right represents the mean clustering coefficient $\langle c \rangle$, average degree $\langle k \rangle$, network size $N$, modularity $Q$ and the connection probability on the GBG and GR flows. Each row shows the results for a real network.}
\end{figure}

\begin{figure}[htb]
	\centering
	\begin{tabular}{@{}c@{}}
		\includegraphics[width=0.9\linewidth]{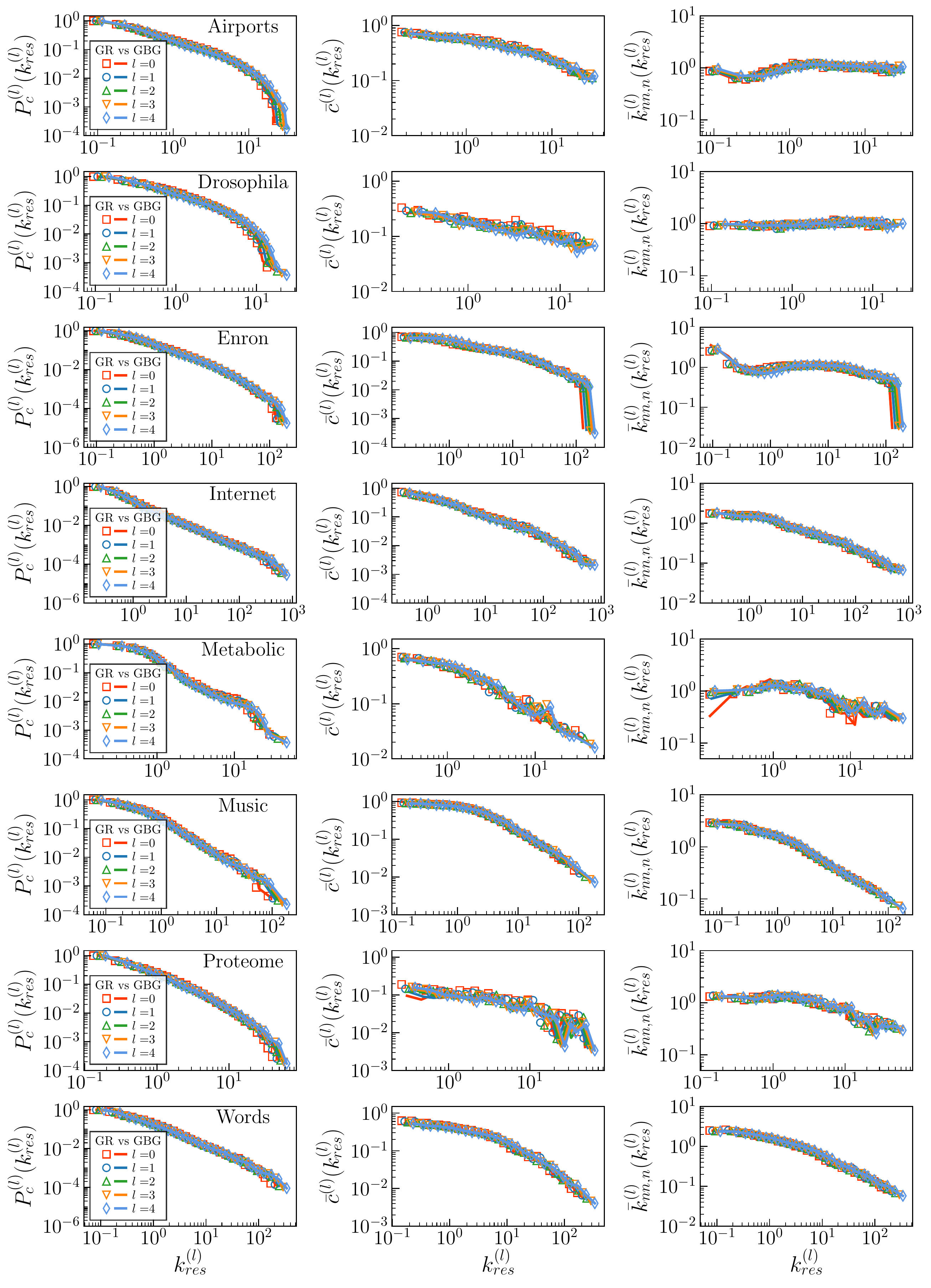}
	\end{tabular}
	\caption{\textbf{GBG vs GR on real networks with $b=1.2$.} Each column from left to right shows the complementary cumulative degree distribution, degree dependent clustering coefficient and degree-degree correlations of rescaled degrees $k^{(l)}_{res} = k^{(l)}/ \langle k^{(l)}\rangle$. Each row shows the results for a real network. }
\end{figure}

\begin{figure}[htb]
	\centering
	\begin{tabular}{@{}c@{}}
		\includegraphics[width=1\linewidth]{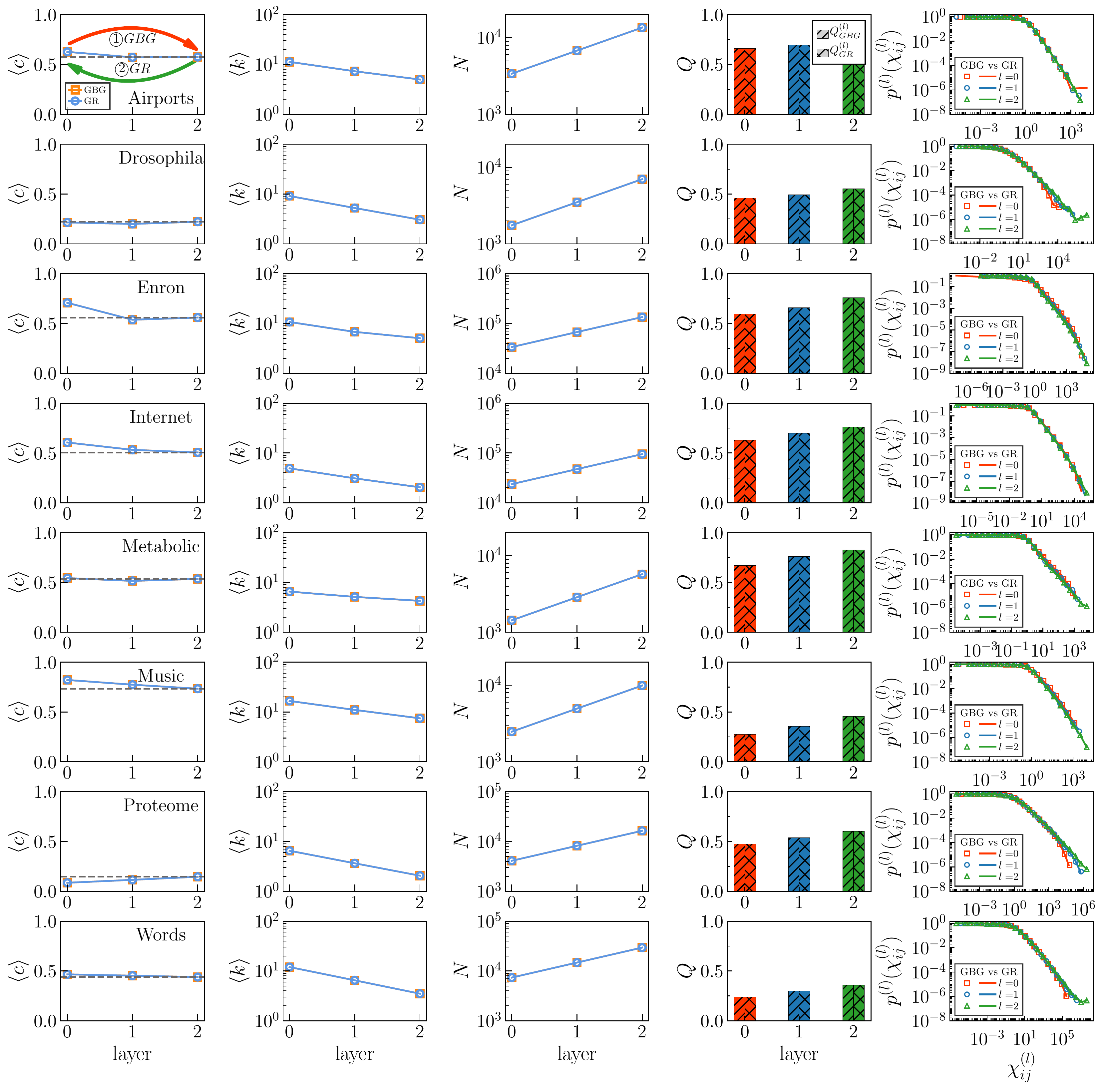}
	\end{tabular}
	\caption{\textbf{GBG vs GR on real networks with $b=2$.} Each column from left to right represents the mean clustering coefficient $\langle c \rangle$, average degree $\langle k \rangle$, network size $N$, modularity $Q$ and the connection probability on the GBG and GR flows. Each row shows the results for a real network.}
\end{figure}

\begin{figure}[htb]
	\centering
	\begin{tabular}{@{}c@{}}
		\includegraphics[width=0.9\linewidth]{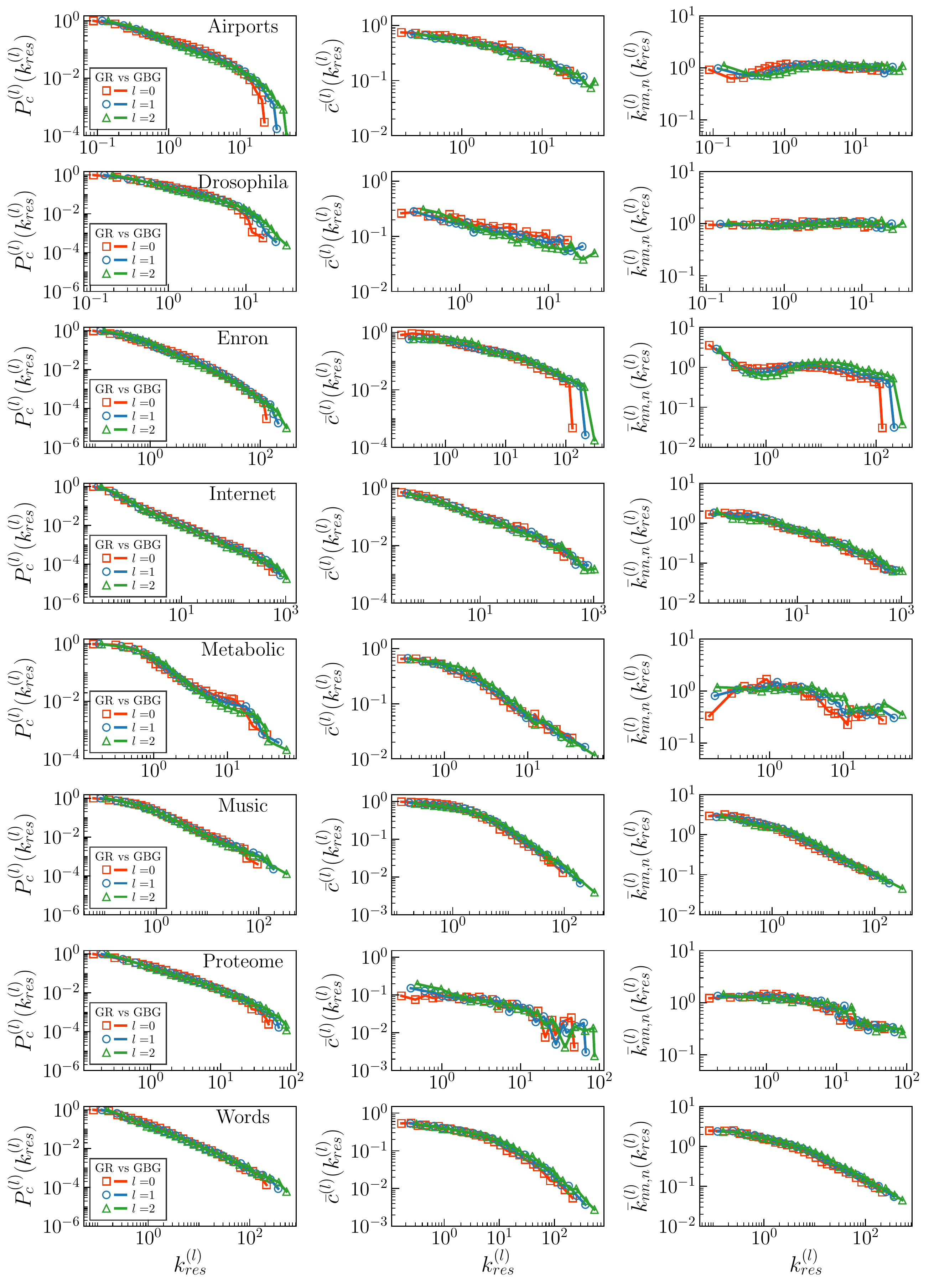}
	\end{tabular}
	\caption{\textbf{GBG vs GR on real networks with $b=2$.} Each column from left to right shows the complementary cumulative degree distribution, degree dependent clustering coefficient and degree-degree correlations of rescaled degrees $k^{(l)}_{res} = k^{(l)}/ \langle k^{(l)}\rangle$. Each row shows the results for a real network. }
\end{figure}

%=~=~=~=~=~=~=~=~=~=~=~=~=~=~=~=~=~=~=~=~=~=~=~=~=~=~=~=~=~=~=~=~=~=~=~=~=~=~=~=~=~=~=~=~=~=~=~=~=
\clearpage
\newpage
\subsection{Semigroup structure properties of GBG and GR models}

\begin{figure}[h]
	\centering
	\includegraphics[width=0.9\linewidth]{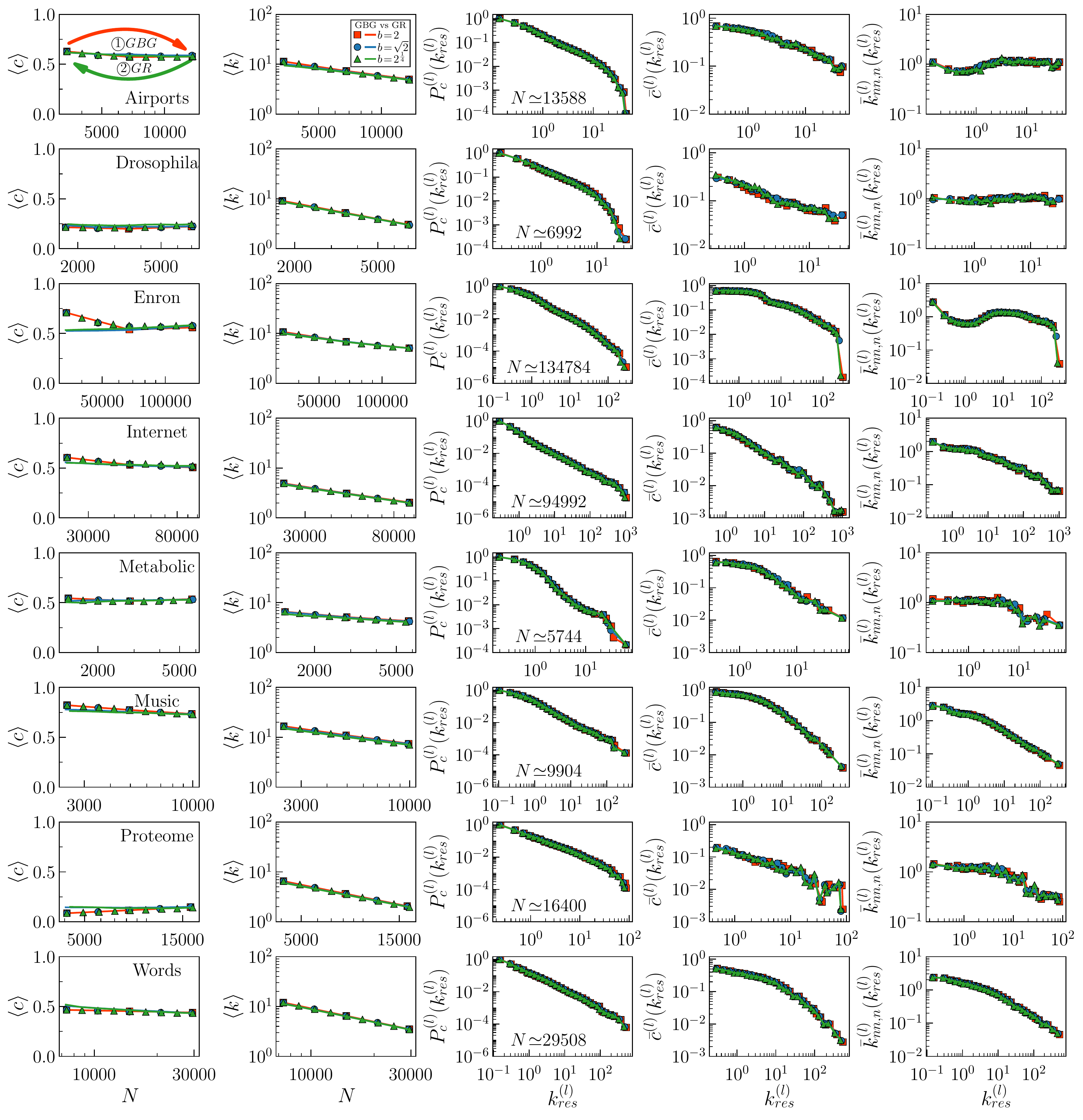}
	\caption{ \textbf{Semigroup structure properties of non-inflationary GBG and GR models for real networks.} Each row shows the results for a real network. Each column from left to right represents the mean clustering coefficient $\langle c \rangle$, average degree $\langle k \rangle$, versus network size $N$; The complementary cumulative degree distribution, degree dependent clustering coefficient and degree-degree correlations of rescaled degrees $k^{(l)}_{res} = k^{(l)}/ \langle k^{(l)}\rangle$. We perform non-inflationary GBG first with $b=2$, $b=\sqrt 2$, $b=2^{\frac{1}{4}}$ respectively. Then we go back with GR model correspondingly. When $N$ reach the same size, we compare network properties for different values of $b$ in the last three columns.
	}	
\end{figure}

%\begin{figure}[htb]
%	\centering
%	\includegraphics[width=1\linewidth]{fig_SI/GBG2GR_b0.pdf}	
%	\caption{\textbf{GBG vs GR on real networks with $b=2$.} Each column from left to right represents the complementary cumulative degree distribution, degree dependent clustering coefficient and degree-degree correlations of rescaled degrees $k^{(l)}_{res} = k^{(l)}/ \langle k^{(l)}\rangle$;  modularity $Q$ and the connection probability on the GBG and GR flows. Each row shows the results for a real network.}
%\end{figure}
%\begin{figure}[htb]
%	\centering
%	\includegraphics[width=1\linewidth]{fig_SI/GBG2GR_b1.pdf}	
%	\caption{\textbf{GBG vs GR on real networks with $b=\sqrt 2$.} Each column from left to right represents the complementary cumulative degree distribution, degree dependent clustering coefficient and degree-degree correlations of rescaled degrees $k^{(l)}_{res} = k^{(l)}/ \langle k^{(l)}\rangle$;  modularity $Q$ and the connection probability on the GBG and GR flows. Each row shows the results for a real network.}
%\end{figure}
%\begin{figure}[htb]
%	\centering
%	\includegraphics[width=1\linewidth]{fig_SI/GBG2GR_b2.pdf}	
%	\caption{\textbf{GBG vs GR on real networks with $b=2^{\frac{1}{4}}$.} Each column from left to right represents the complementary cumulative degree distribution, degree dependent clustering coefficient and degree-degree correlations of rescaled degrees $k^{(l)}_{res} = k^{(l)}/ \langle k^{(l)}\rangle$;  modularity $Q$ and the connection probability on the GBG and GR flows. Each row shows the results for a real network.}
%\end{figure}
%===========================

\begin{figure}[h]
	\centering
	\includegraphics[width=0.9\linewidth]{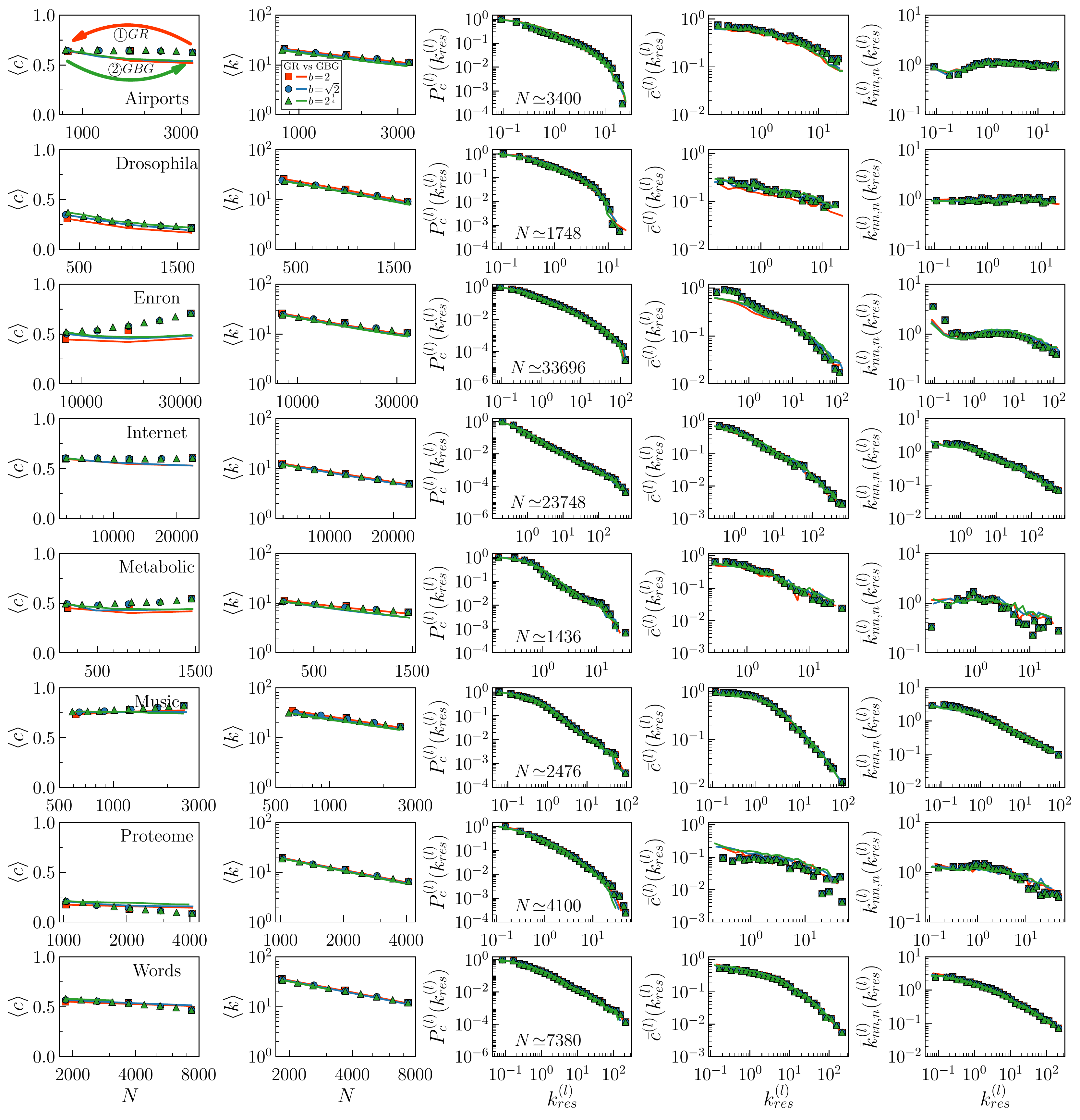}
	\caption{ \textbf{Semigroup structure properties of GR and non-inflationary GBG models for real networks.} Each row shows the results for a real network. Each column from left to right represents the mean clustering coefficient $\langle c \rangle$, average degree $\langle k \rangle$, versus network size $N$; The complementary cumulative degree distribution, degree dependent clustering coefficient and degree-degree correlations of rescaled degrees $k^{(l)}_{res} = k^{(l)}/ \langle k^{(l)}\rangle$. We perform GR model first with $b=2$, $b=\sqrt 2$, $b=2^{\frac{1}{4}}$ respectively. Then we increase the network with non-inflationary GBG model correspondingly. When $N$ reach the same size, we compare network properties for different values of $b$ in the last three columns.
	}	
\end{figure}
\clearpage
\newpage
%\section{Inflationary GBG is a statistical inverse of deflationary GR.}

\begin{figure}[h]
	\centering
	\includegraphics[width=1.0\linewidth]{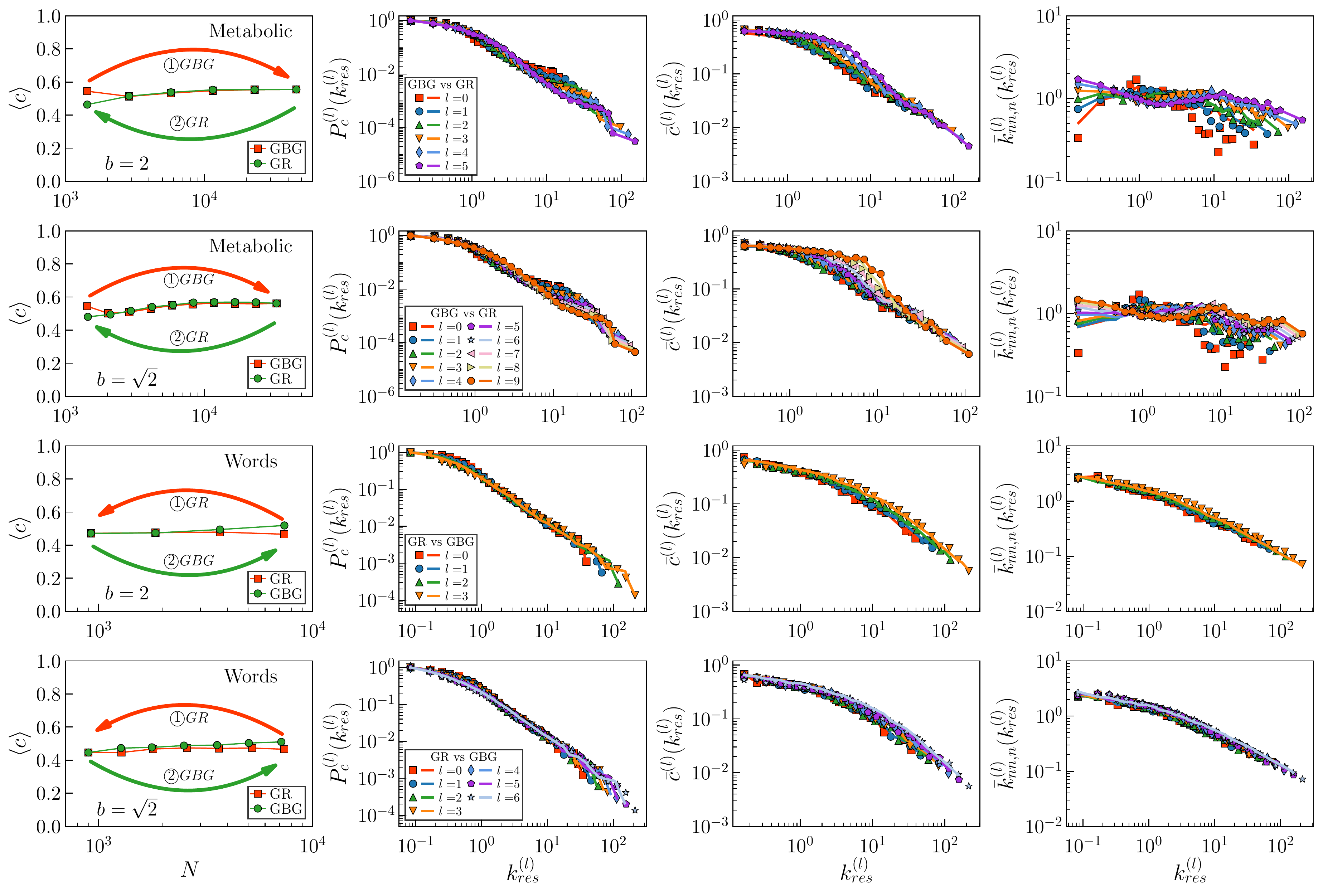}
	\caption{\textbf{Inflationary GBG is a statistical inverse of deflationary GR.} We first perform inflationary GBG (GBG+addition of links) and apply the deflationary GR (GR+pruning of links) back. The first two rows show the results for Metabolic network with $b=2$ and $b=\sqrt 2$, respectively. In the last two rows, we perform the deflationary GR (GR+pruning of links) first and apply the inflationary GBG (GBG+addition of links) back for Words network with $b=2$ and $b=\sqrt 2$, respectively. 
		Each column from left to right represents the mean clustering coefficient $\langle c \rangle$ versus network size $N$, the complementary cumulative degree distribution, degree dependent clustering coefficient and degree-degree correlations of rescaled degrees $k^{(l)}_{res} = k^{(l)}/ \langle k^{(l)}\rangle$.
	}	
\end{figure}

%
%==================================================================================================================
\clearpage
\newpage
\subsection{Predicting the evolution of the JCN and the WTW}
\begin{figure}[h]
	\centering
	\includegraphics[width=0.9\linewidth]{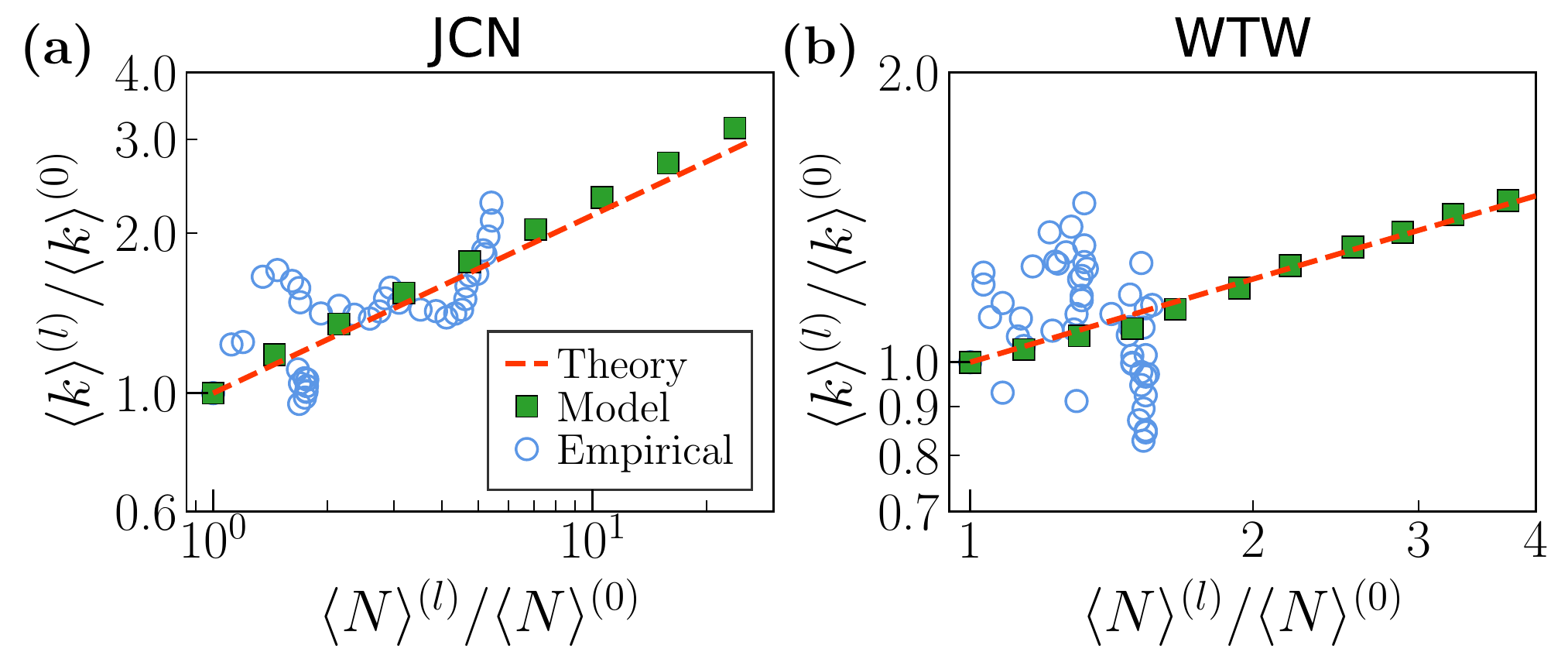}
	\caption{ {\bf Scaling of the average degree with network size.} Empirical data is compared with simulation results and the theoretical prediction. The slope in the theoretical line is $s=0.3344\pm0.0380$ for the JCN from 1950-1960 to 1965-1975, and $s=0.2874\pm0.1290$ for the WTW from $1950$ to $1965$. The inflationary GBG model is started from 1965-1975 with $b=1.5$, $a=1.415$ in JCN and 1965 with $b=1.15$, $a=1.090$ in WTW, respectively.}
	\label{fig1-1}
\end{figure}

\begin{figure}[h]
	\centering
	\includegraphics[width=0.9\linewidth]{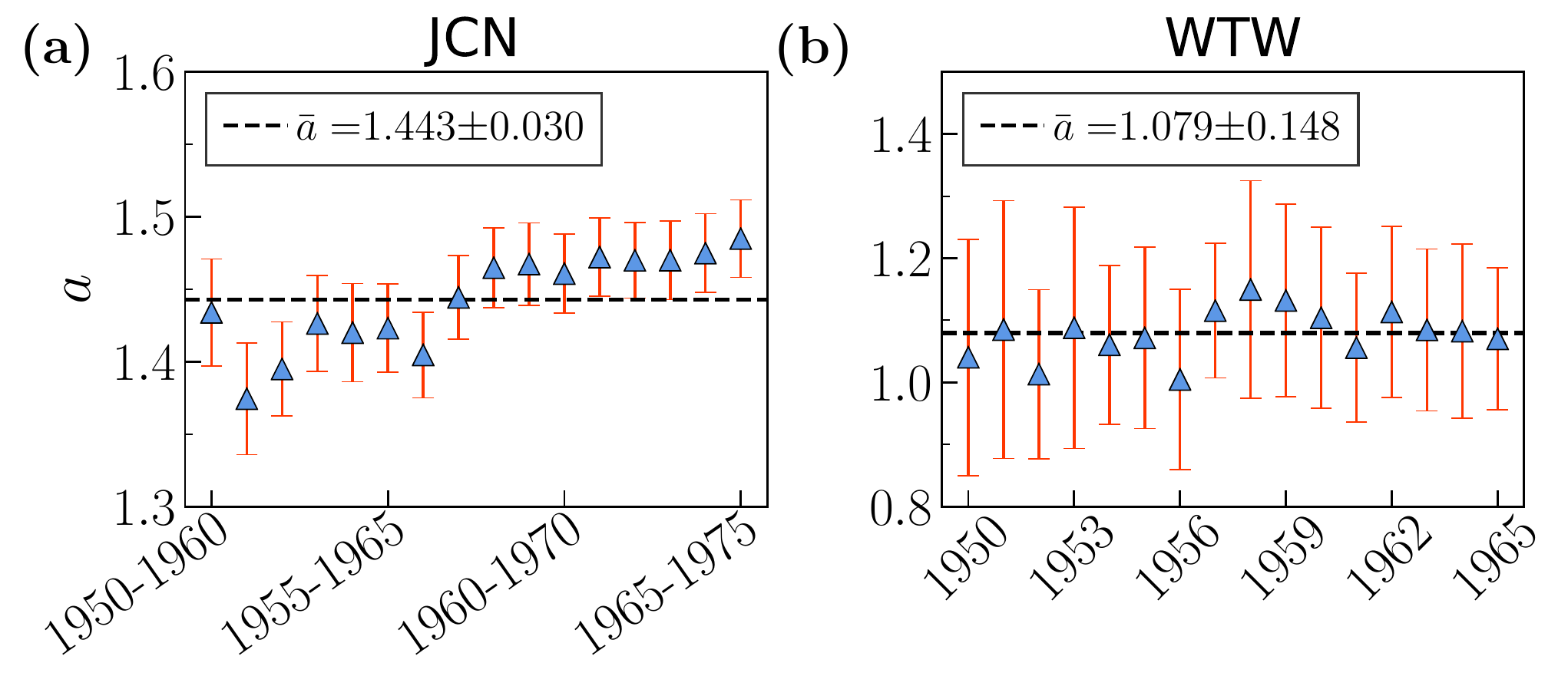}
	\caption{ {\bf Parameter $a$ estimated from empirical snapshots.}
		The slope $s=\frac{\ln b^{-\nu}+\ln a}{\ln b}=\frac{\ln a \varphi}{\ln b}$ of $\langle k\rangle$ as a function of $N$ in log-log scale is $s=0.3344\pm0.0380$ for the JCN from 1950-1960 to 1965-1975, and $s=0.2874\pm0.1290$ for the WTW from $1950$ to $1965$. More specificity, with the slope $s$ and standard deviation $\sigma_{s}$ on hand, we perform $100$ realizations for each snapshot in non-inflationary GBG model and get the average $\bar{\varphi}$ and standard deviation $\sigma_{\bar{\varphi}}$. Then we find the parameter $a$ for this snapshot and the corresponding propagation of error $\sigma_{a}=a\sqrt{s^2(\ln b)^2(\frac{\sigma_{s}}{s})^2+(\frac{\sigma_{\bar{\varphi}}}{\bar{\varphi}})^2}$.
	}
	\label{fig:estimate_a}
\end{figure}

\begin{figure}[h]
	\centering
	\includegraphics[width=1\linewidth]{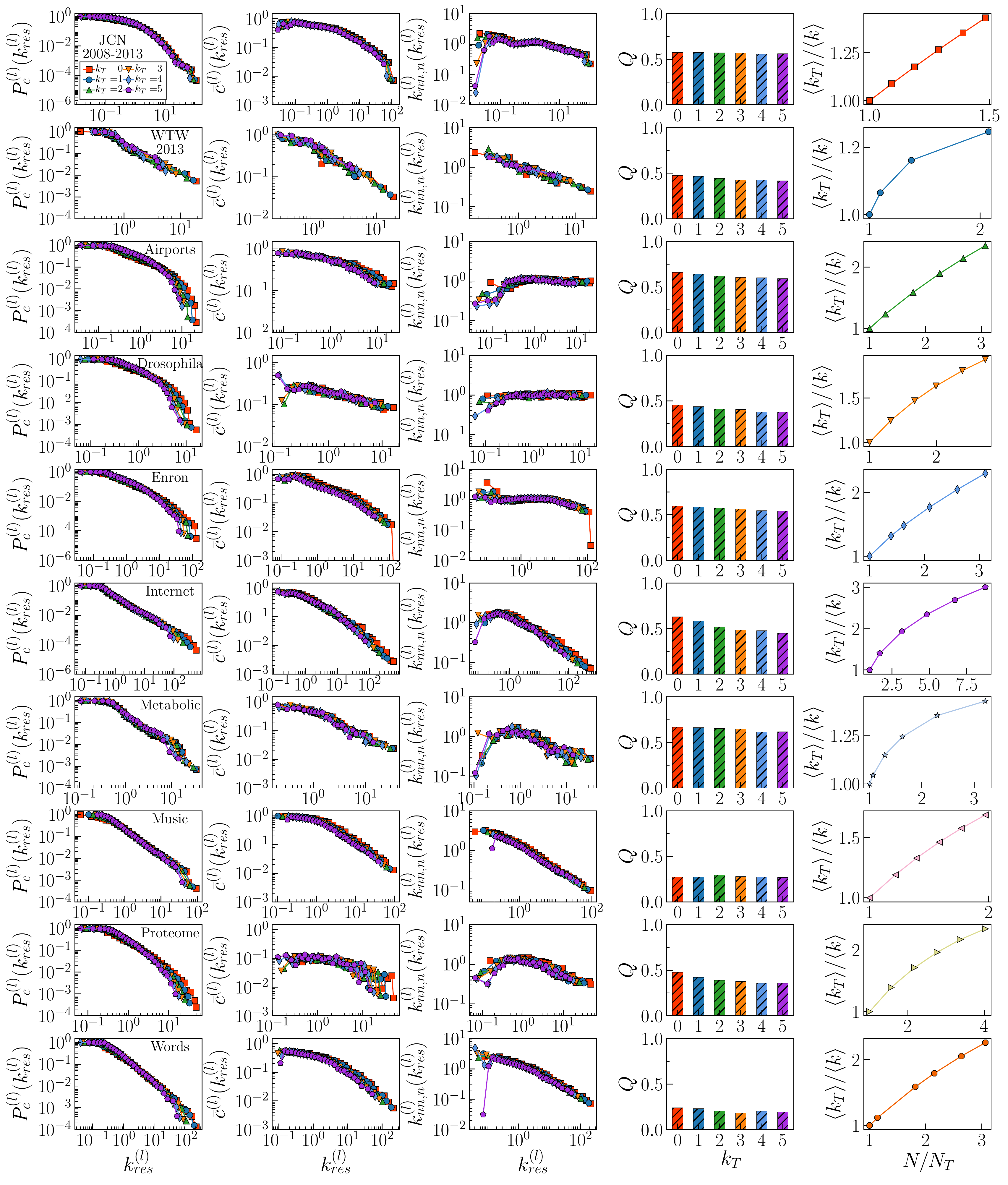}
	\caption{\textbf{Local rich-club and nested self-similarity effects in the nested hierarchy of subgraphs produced by degree thesholding.}
		Each column from left to right shows the complementary cumulative distribution $P_c^{(l)}(k_{res}^{(l)})$ of rescaled degrees $k^{(l)}_{res} = k^{(l)}/ \langle k^{(l)}\rangle$, degree dependent clustering coefficient over rescaled-degree classes, degree-degree correlations measured by the normalized average nearest-neighbour degree $\bar{k}_{nn,n}^{(l)} (k_{res}^{(l)}) = \bar{k}_{nn}^{(l)} (k_{res}^{(l)}) \langle  k^{(l)}\rangle/\langle(k^{(l)})^2\rangle$, the modularity $Q$, and the ratio of the subgraph average degree $ \langle k_T\rangle$ to the original average degree $\langle k\rangle$ as a function of the inverse relative subgraph size $N/N_T$. The subgraphs are obtained by removing nodes with degrees below threshold $k_T$ from the original network.	  }	
\end{figure}
\begin{figure}[htb]
	\centering
	\begin{tabular}{@{}c@{}}
		\includegraphics[width=1\linewidth]{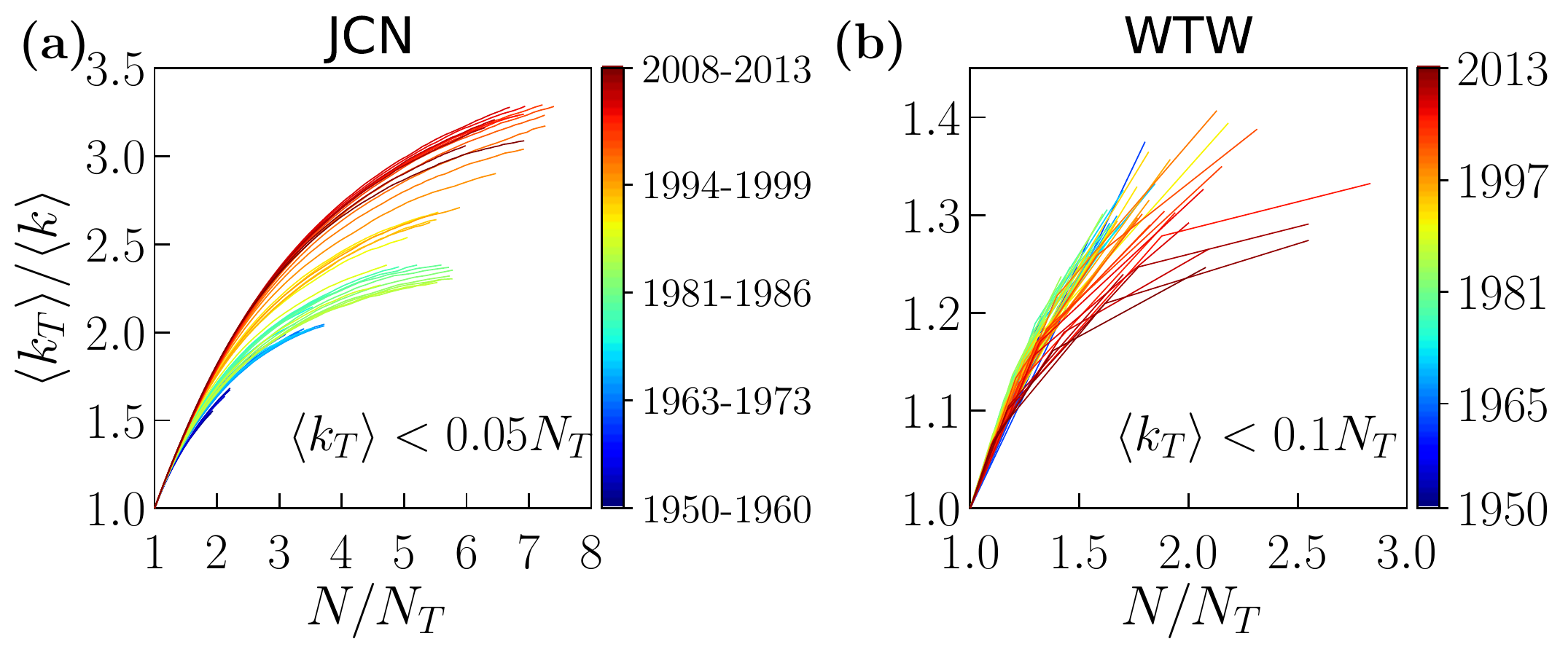}
	\end{tabular}
	\caption{\textbf{Ratio of the subgraph average degree  $ \langle k_T\rangle$ to the original average degree $\langle k\rangle$ as a function of
			the inverse relative subgraph size $N/N_T$ for the different snapshots of the JCN and the WTW.} The subgraphs are obtained by removing nodes with degrees below threshold $k_T$ from the original network. To reduce the effects of finite-size, the data are shown only for the subgraph average degree $\langle k_T\rangle$ smaller than $0.05N_T$ in JCN and $0.1N_T$ in WTW.} 
\end{figure}
\begin{figure}[h]
	\centering
	\includegraphics[width=0.9\linewidth]{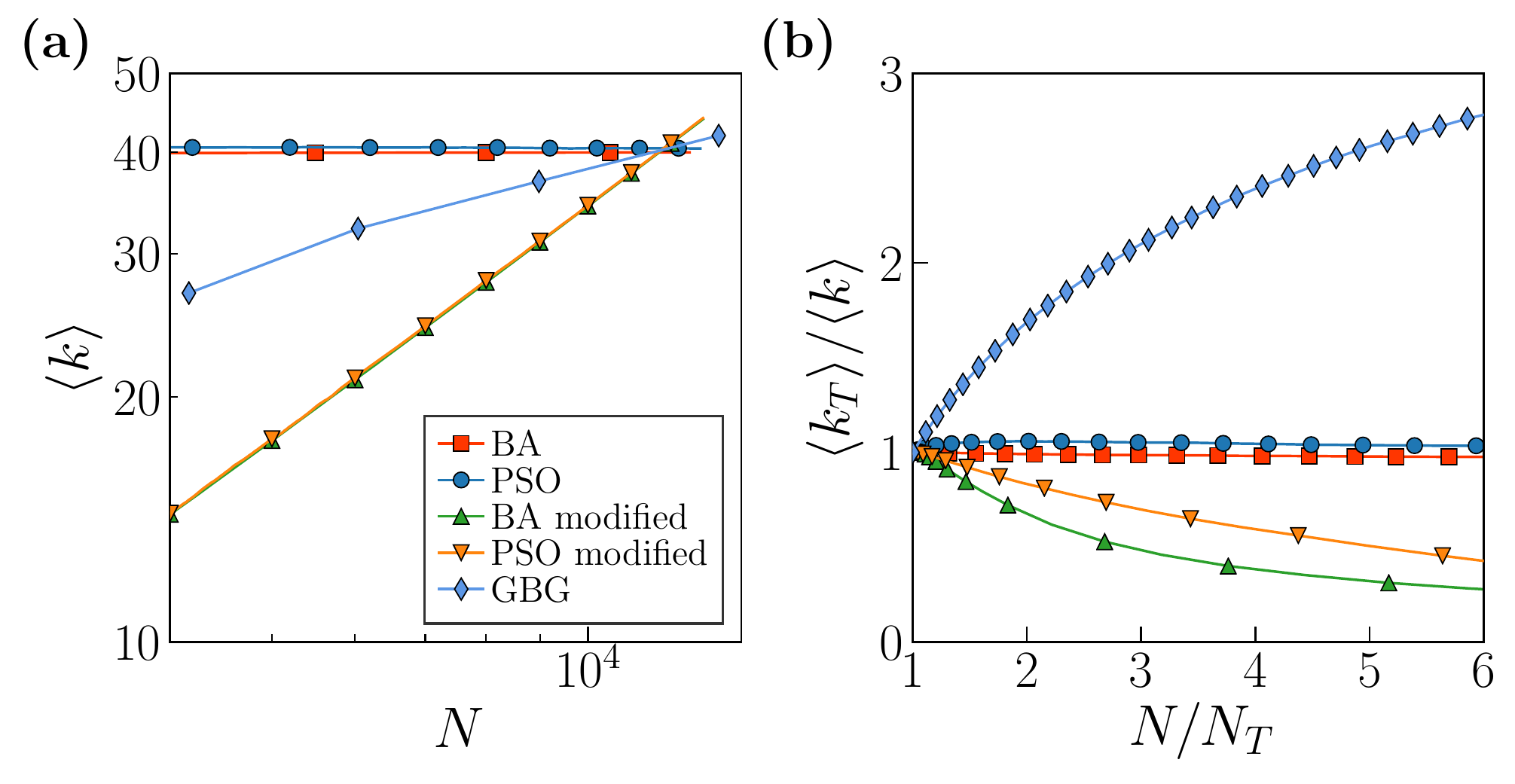}
	\caption{\textbf{Local rich-club effect in different growing network models.}
		(a) Behavior of the average degree $\langle k \rangle$ as a function of network size $N$. In particular, for the Barab{\'a}si-Albert model (BA)~\cite{Barabasi1999}, we set $m=20$, and for the Popularity Similarity Optimization model (PSO)~\cite{Papadopoulos2012} we use parameters $\gamma=2.76,T=0.4,m=20$ so that the generated networks eventually reach the same average degree, clustering and degree distribution than a targeted network, i.e., the empirical snapshot 1994-1999 in JCN. Notice that the average degree does not change with network size in the standard BA and PSO models. To make the average degree increasing with network size, we modify the standard models by changing  $m$ as $m=1+\text{Int}(t/t_m)$, where $t$ denotes the current network size in the node addition process so that $m$ only increases after remaining stable for periods of lengths $t_m$. We set $t_m=300$ such that the average degree of the targeted network can be reached. We name the new versions BA modified and PSO modified. (b) Ratio of subgraphs average degree $ \langle k_T\rangle$ to the original average degree $\langle k\rangle$ as a function of
		the inverse relative subgraph size $N/N_T$ in the final graphs reached by the different growing models. Even if the average degree in the BA modified and the PSO modified grows with the system size as shown in (a), the behavior of the relative average degree in the final graphs in (b) shows that the observed behavior in real networks cannot be reproduced. 
	}	
\end{figure}

\begin{figure}[!htp]
	\centering
	\includegraphics[width=1\linewidth]{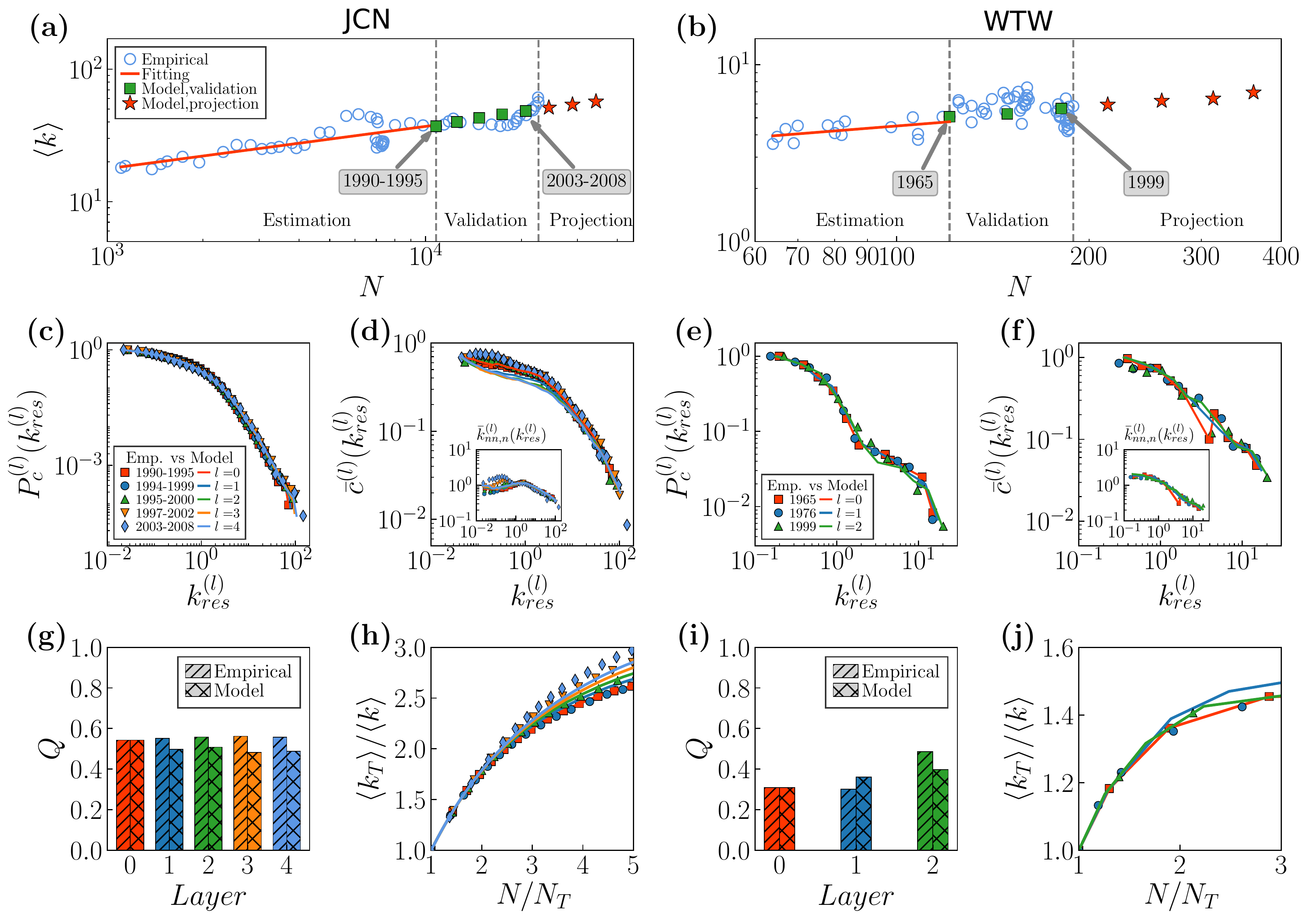}
	\caption{
		\textbf{The GBG model predicts the self-similar evolution of the JCN and the WTW with different $b$ and starting time points.} Similar to Fig.~5 in the main paper, but we take $b=1.2$ for the JCN and WTW (the corresponding values of $a$ are $1.185$ and $1.130$, respectively). We grow the networks with the GBG transformation with starting time points in 1990-1995 for the JCN and in 1965 for the WTW. The topological properties of simulated and empirical networks are compared. The results show that the prediction of the GBG method is robust, and that the evolution of the two systems is reproduced (at a statistical level) even when using different values of $b$ and different starting time points. }
	\label{fig:emp2model}
\end{figure}

\begin{figure}[h]
	\centering
	\includegraphics[width=0.9\linewidth]{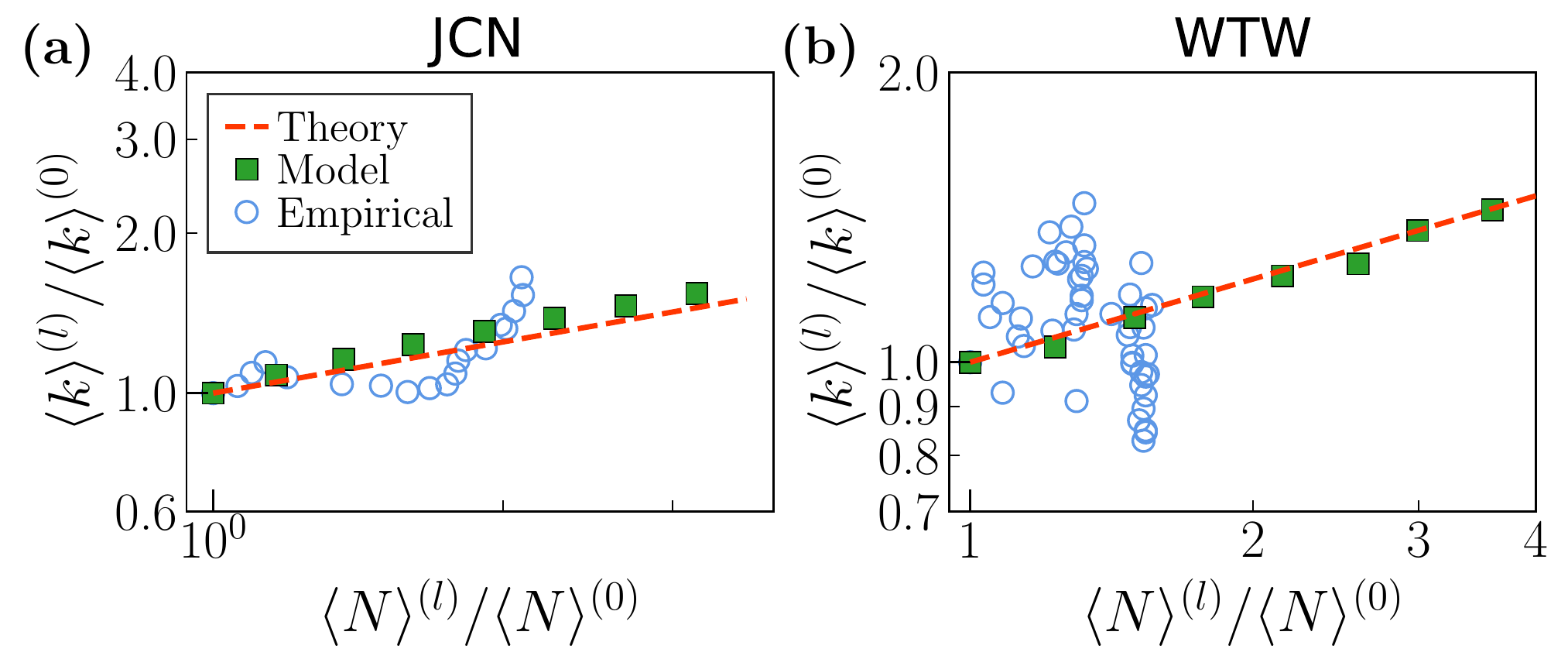}
	\caption{ {\bf Scaling of the average degree with network size.} Empirical data is compared with simulation results and the theoretical prediction. The slope in the theoretical line is $s=0.3195\pm0.0426$ for the JCN from 1950-1960 to 1990-1995, and $s=0.2874\pm0.1290$ for the WTW from $1950$ to $1965$. The inflationary GBG model is started from 1990-1995 with $b=1.2$, $a=1.185$ in JCN and 1965 with $b=1.2$, $a=1.130$ in WTW, respectively.}	
\end{figure}

\begin{figure}[h]
	\centering
	\includegraphics[width=0.9\linewidth]{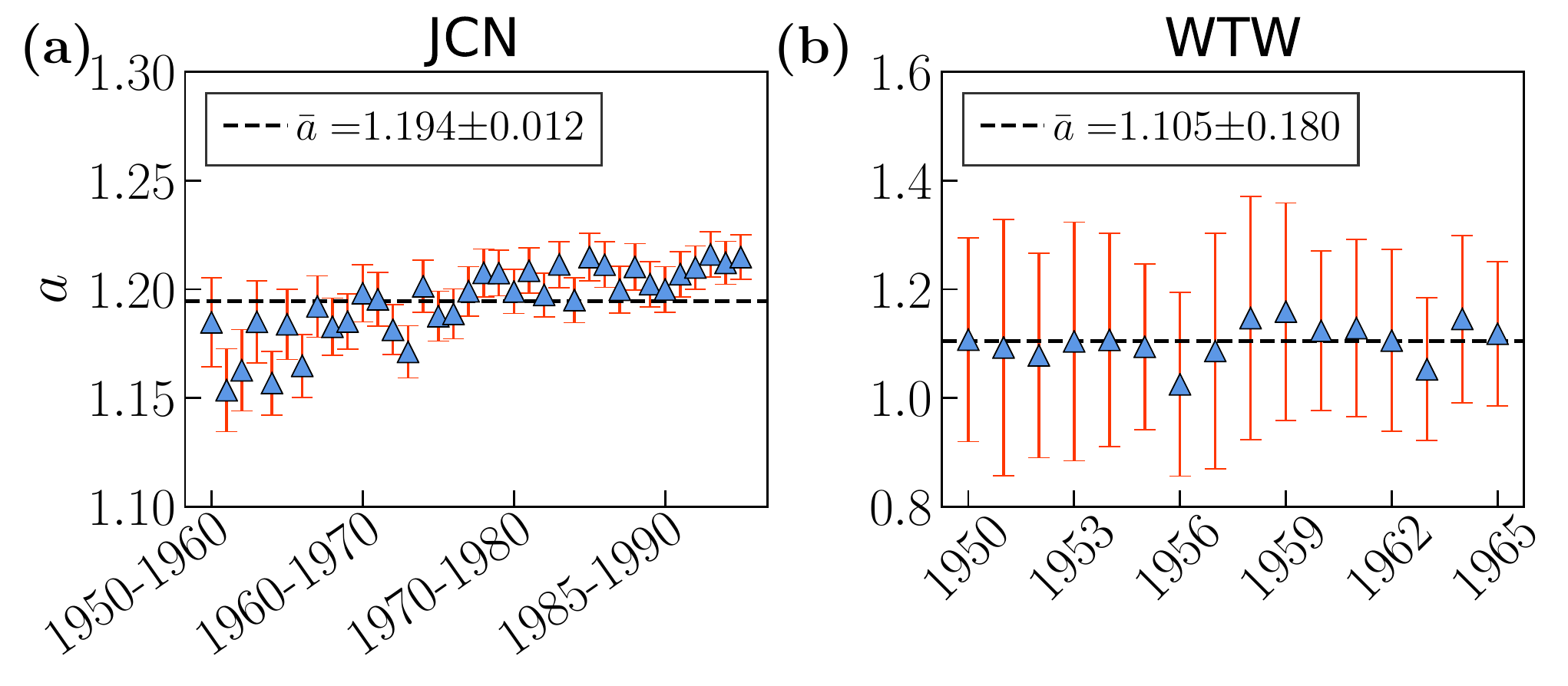}
	\caption{ {\bf Parameter $a$ estimated from empirical snapshots.}
		The slope $s=\frac{\ln b^{-\nu}+\ln a}{\ln b}=\frac{\ln a \varphi}{\ln b}$ of $\langle k\rangle$ as a function of $N$ in log-log scale is $s=0.3195\pm0.0426$ for the JCN from 1950-1960 to 1990-1995, and $s=0.2874\pm0.1290$ for the WTW from $1950$ to $1965$. More specificity, with the slope $s$ and standard deviation $\sigma_{s}$ on hand, we perform $100$ realizations for each snapshot in non-inflationary GBG model and get the average $\bar{\varphi}$ and standard deviation $\sigma_{\bar{\varphi}}$. Then we find the parameter $a$ for this snapshot and the corresponding propagation of error $\sigma_{a}=a\sqrt{s^2(\ln b)^2(\frac{\sigma_{s}}{s})^2+(\frac{\sigma_{\bar{\varphi}}}{\bar{\varphi}})^2}$.
	}	
\end{figure}

%==================================================================================================================
\newpage
\clearpage

\begin{figure}[htb]
	\centering
	\begin{tabular}{@{}c@{}}
		\includegraphics[width=0.85\linewidth]{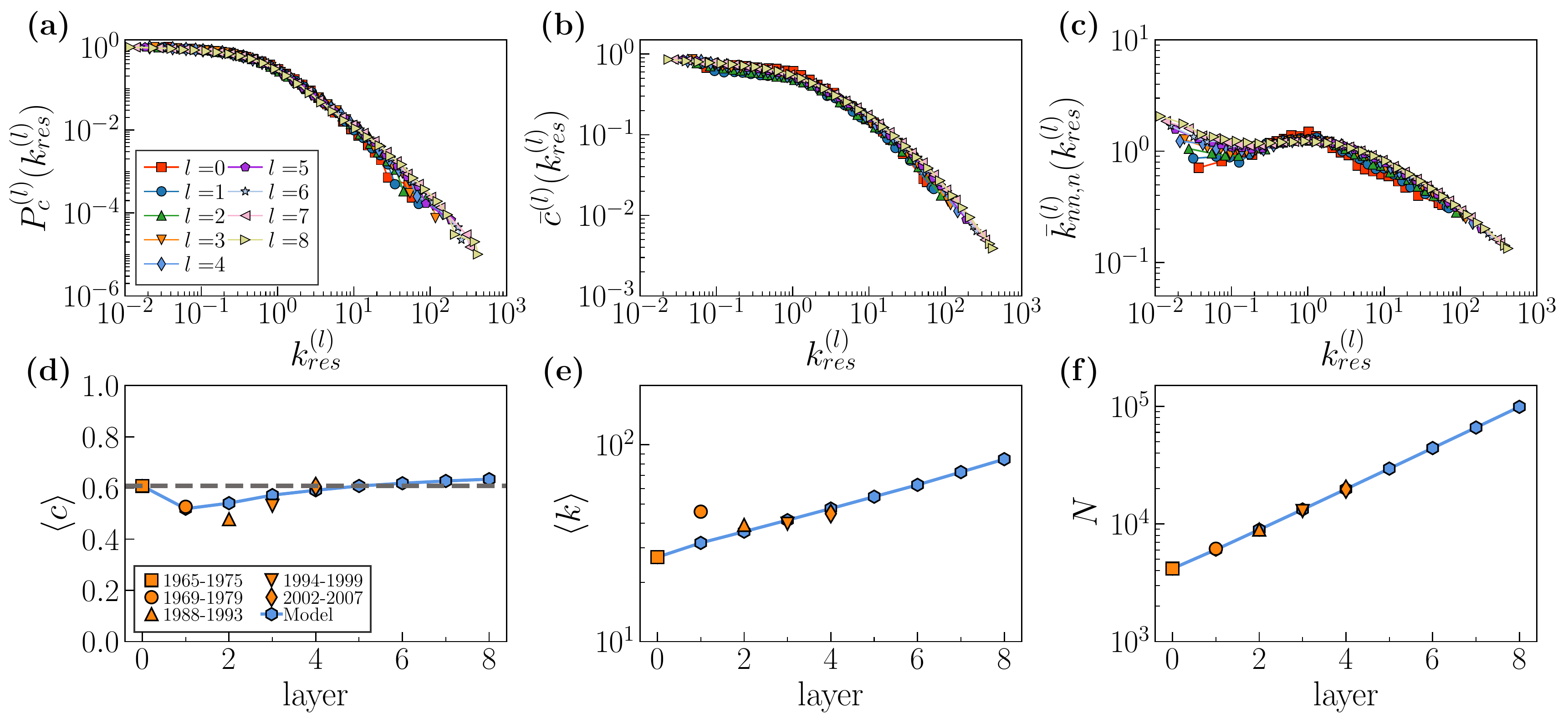}
	\end{tabular}
	\caption{\textbf{The self-similarity of inflationary GBG model in JCN.} (a)-(c) show the complementary cumulative degree distribution, degree dependent clustering coefficient and degree-degree correlations over rescaled degrees $k^{(l)}_{res} = k^{(l)}/ \langle k^{(l)}\rangle$, respectively. The self-similar behavior are still preserved up to $l=8$. Comparing average clustering (d), average degree (e), and network size (f) of the empirical JCN snapshots and of layers evolved by the GBG. Orange symbols represent empirical snapshots and blue pentagons correspond to simulated GBG networks with approximately the same size. The starting time point is 1965-1975, and $b=1.5$.}
\end{figure}

\begin{figure}[htb]
	\centering
	\begin{tabular}{@{}c@{}}
		\includegraphics[width=0.85\linewidth]{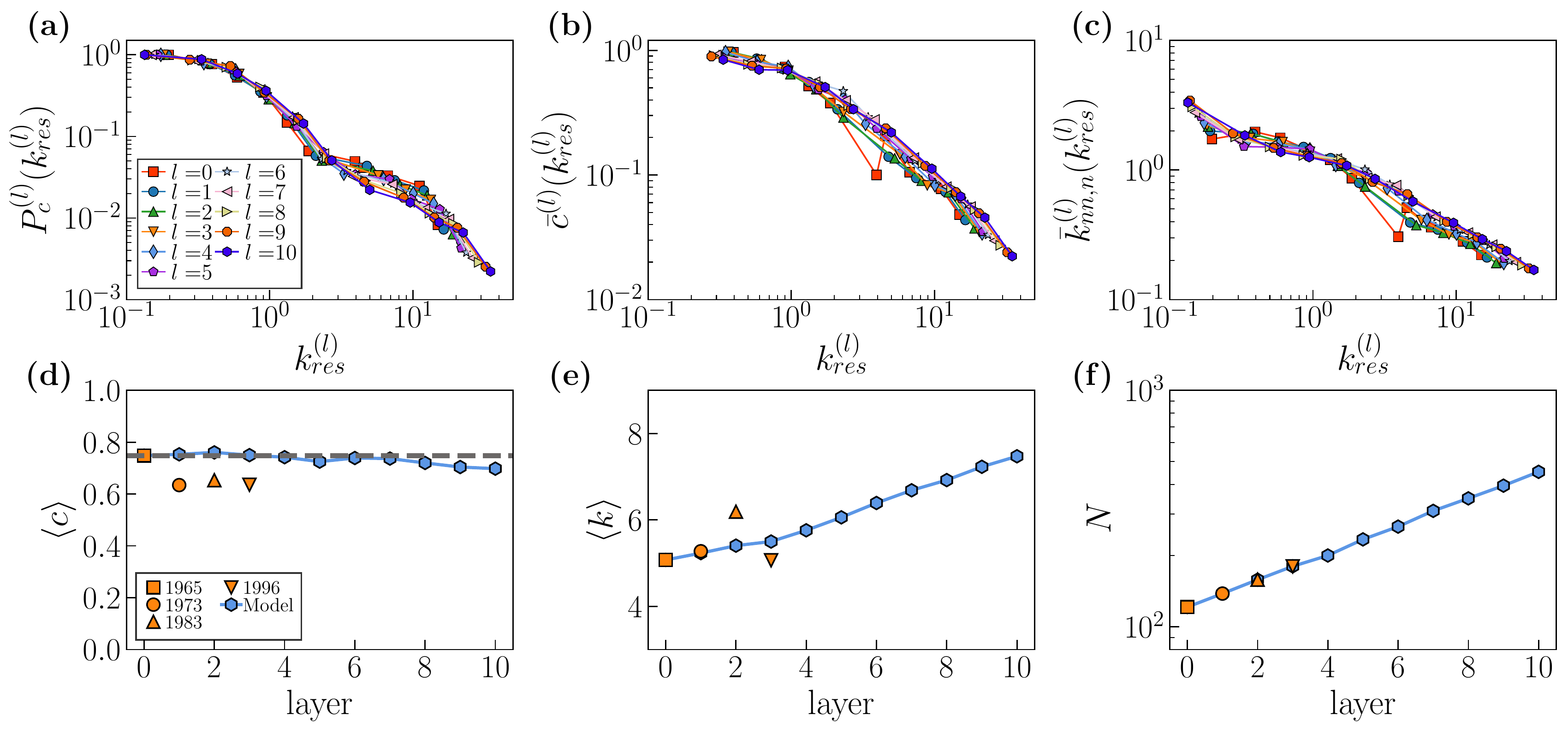}
	\end{tabular}
	\caption{\textbf{The self-similarity of inflationary GBG model in WTW.} (a)-(c) show the complementary cumulative degree distribution, degree dependent clustering coefficient and degree-degree correlations over rescaled degrees $k^{(l)}_{res} = k^{(l)}/ \langle k^{(l)}\rangle$, respectively. The self-similar behavior are still preserved up to $l=10$. Comparing average clustering (d), average degree (e), and network size (f) of the empirical WTW snapshots and of layers evolved by the GBG. Orange symbols represent empirical snapshots and blue pentagons correspond to simulated GBG networks with approximately the same size. The starting time point is $1965$, and $b=1.15$.}
\end{figure}

\begin{figure}[h]
	\centering
	\includegraphics[width=1\linewidth]{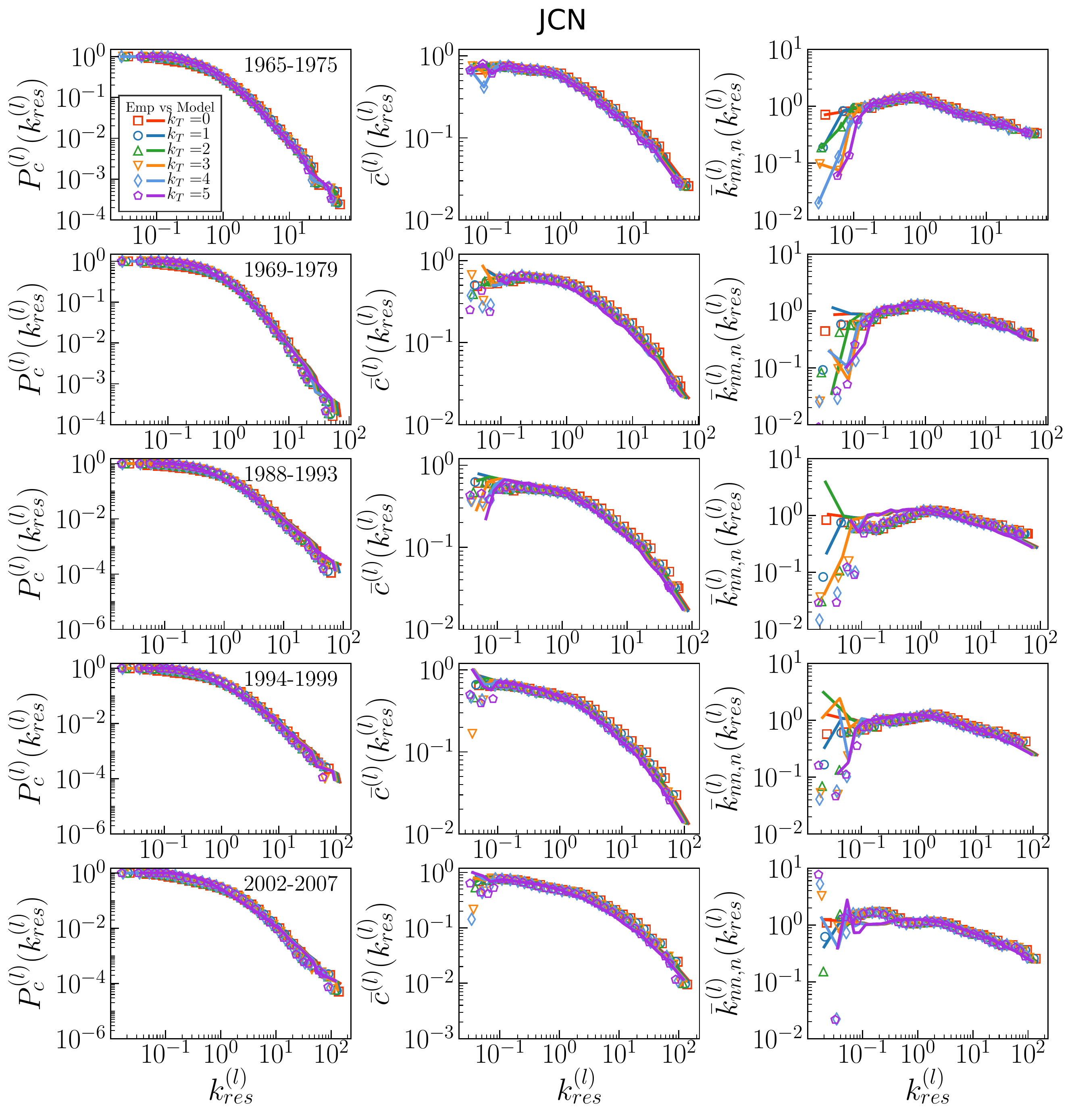}
	\caption{ {\bf Comparing self-similarity in the hierarchy of nested subgraphs in empirical JCN snapshots and in layers evolved by the GBG.}  From left to right, each column shows the complementary cumulative degree distribution $P_c^{(l)}(k_{res}^{(l)})$ of rescaled degrees $k^{(l)}_{res} = k^{(l)}/ \langle k^{(l)}\rangle$, the degree dependent clustering coefficient $\bar{c}^{(l)}(k_{res}^{(l)})$ over rescaled degrees, and the degree-degree correlations measured by the normalized average nearest-neighbour degree $\bar{k}_{nn,n}^{(l)} (k_{res}^{(l)}) = \bar{k}_{nn}^{(l)} (k_{res}^{(l)}) \langle  k^{(l)}\rangle/\langle(k^{(l)})^2\rangle$. Each row shows results for a different time snapshot. The subgraphs are obtained by removing nodes with degree below a threshold $k_T$.	
	}	
\end{figure}
\begin{figure}[h]
	\centering
	\includegraphics[width=0.9\linewidth]{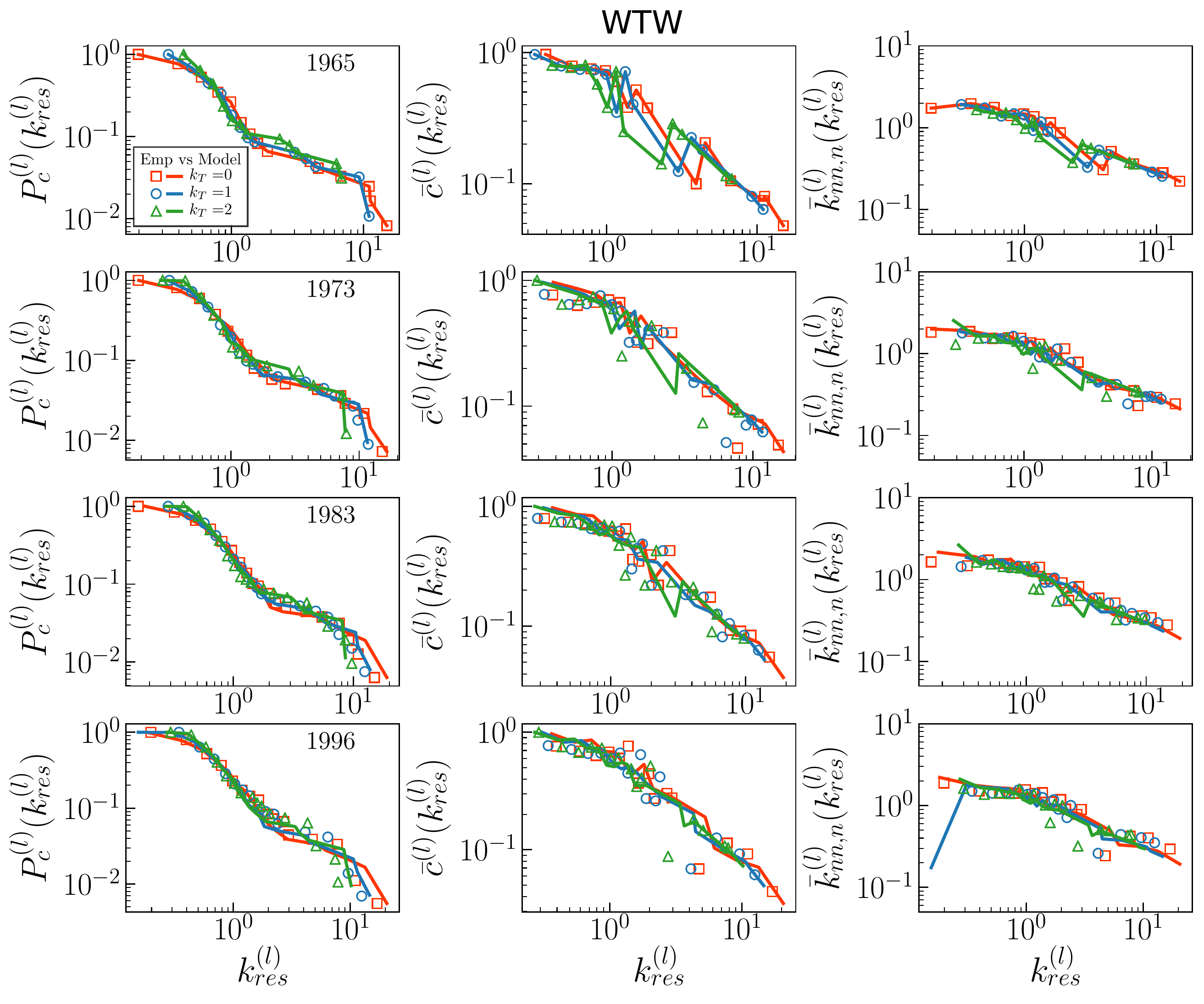}
	\caption{ {\bf Comparing self-similarity in the hierarchy of nested subgraphs in empirical WTW snapshots and in layers evolved by the GBG.}  From left to right, each column shows the complementary cumulative degree distribution $P_c^{(l)}(k_{res}^{(l)})$ of rescaled degrees $k^{(l)}_{res} = k^{(l)}/ \langle k^{(l)}\rangle$, the degree dependent clustering coefficient $\bar{c}^{(l)}(k_{res}^{(l)})$ over rescaled degrees, and the degree-degree correlations measured by the normalized average nearest-neighbour degree $\bar{k}_{nn,n}^{(l)} (k_{res}^{(l)}) = \bar{k}_{nn}^{(l)} (k_{res}^{(l)}) \langle  k^{(l)}\rangle/\langle(k^{(l)})^2\rangle$. Each row shows results for a different time snapshot. The subgraphs are obtained by removing nodes with degree below a threshold $k_T$.	
	}	
\end{figure}

%==================================================================================================================
%
%
%==================================================================================================================

\begin{figure}[htb]
	\centering
	\begin{tabular}{@{}c@{}}
		\includegraphics[width=0.9\linewidth]{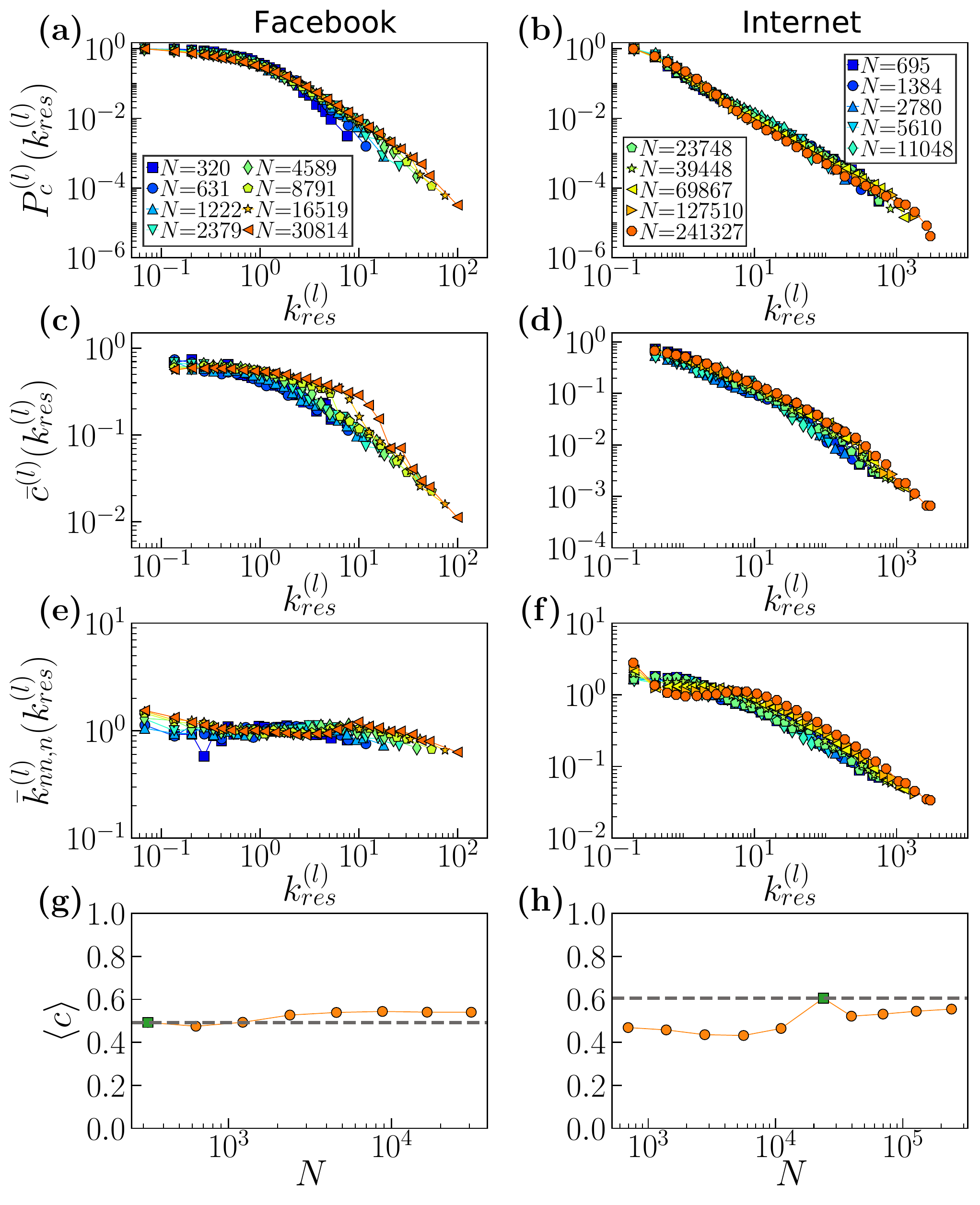}
	\end{tabular}
	\caption{\textbf{Network properties of upscaled and downscaled network replicas of real networks.} (a) and (b), the complementary cumulative degree distribution, (c) and (d),degree dependent clustering coefficient, (e) and (f),degree-degree correlations over rescaled degrees $k^{(l)}_{res} = k^{(l)}/ \langle k^{(l)}\rangle$, and (g) and (h) average clustering coefficient $\langle c\rangle$. Results for upscaled Facebook network replicas are shown on the left column, and results for downscaled Internet replicas are on the right column. The green symbol in (g) and (h) indicates the value for the real network.}
\end{figure}

\clearpage
\newpage
%%%%%%%%%%%%%%%%%%%%%%%%%%%%%%%%%%%
\section{Supplementary Tables S1 to S3}

\renewcommand{\arraystretch}{0.98} %Row height
\newcommand*{\thd}[1]{\multicolumn{1}{l}{#1}}
\begin{table*}[h] 
	\begin{ruledtabular}
		\centering		
		\caption{Year by year statistics for the Journal Citation Network. Columns are: the time windows of each network (Years), the number of nodes ($N$), the average degree ($\langle k \rangle$), and the average local clustering coefficient ($\langle c \rangle$).}
		\label{tab:CNS}		
		\begin{tabular}{*{5}{l}{c}|*{5}{l}} 
			%			\\[0.1cm]	%			
			\thd{No.}	&\thd{Years}  &\thd{$N$}  &\thd{$\langle k\rangle$}  &\thd{$\langle c\rangle$} && \thd{No.}	&\thd{Years}  &\thd{$N$}  &\thd{$\langle k\rangle$}  &\thd{$\langle c\rangle$}\\
			\hline 
			0      &1900-1910        &118    &4.068      &0.403      &&52     &1952-1962        &1381   &17.586     &0.632      \\ 
			1      &1901-1911        &127    &4.315      &0.409      &&53     &1953-1963        &1474   &19.228     &0.630      \\ 
			2      &1902-1912        &151    &5.682      &0.387      &&54     &1954-1964        &1543   &20.139     &0.629      \\ 
			3      &1903-1913        &147    &4.571      &0.433      &&55     &1955-1965        &1724   &21.831     &0.631      \\ 
			4      &1904-1914        &159    &5.761      &0.463      &&56     &1956-1966        &1945   &19.788     &0.645      \\ 
			5      &1905-1915        &179    &6.715      &0.461      &&57     &1957-1967        &2313   &23.739     &0.637      \\ 
			6      &1906-1916        &181    &7.348      &0.494      &&58     &1958-1968        &2549   &26.846     &0.622      \\ 
			7      &1907-1917        &176    &6.091      &0.489      &&59     &1959-1969        &2821   &26.747     &0.620      \\ 
			8      &1908-1918        &185    &6.151      &0.476      &&60     &1960-1970        &3061   &24.989     &0.625      \\ 
			9      &1909-1919        &178    &5.011      &0.501      &&61     &1961-1971        &3281   &25.367     &0.616      \\ 
			10     &1910-1920        &177    &4.757      &0.498      &&62     &1962-1972        &3468   &25.950     &0.615      \\ 
			11     &1911-1921        &204    &6.245      &0.525      &&63     &1963-1973        &3805   &27.710     &0.614      \\ 
			12     &1912-1922        &240    &7.083      &0.527      &&64     &1964-1974        &3925   &25.381     &0.615      \\ 
			13     &1913-1923        &253    &7.455      &0.522      &&65     &1965-1975        &4168   &26.844     &0.608      \\ 
			14     &1914-1924        &228    &6.167      &0.523      &&66     &1966-1976        &4657   &33.135     &0.589      \\ 
			15     &1915-1925        &236    &6.373      &0.526      &&67     &1967-1977        &4994   &33.486     &0.580      \\ 
			16     &1916-1926        &245    &6.580      &0.534      &&68     &1968-1978        &5634   &44.415     &0.540      \\ 
			17     &1917-1927        &250    &7.080      &0.523      &&69     &1969-1979        &6154   &45.717     &0.527      \\ 
			18     &1918-1928        &279    &7.290      &0.534      &&70     &1970-1980        &6727   &43.615     &0.528      \\ 
			19     &1919-1929        &304    &9.033      &0.539      &&71     &1972-1981        &7026   &42.295     &0.515      \\ 
			20     &1920-1930        &296    &7.797      &0.560      &&72     &1974-1982        &7071   &39.770     &0.503      \\ 
			21     &1921-1931        &287    &8.000      &0.562      &&73     &1976-1983        &6971   &29.752     &0.518      \\ 
			22     &1922-1932        &290    &8.234      &0.567      &&74     &1978-1984        &7060   &27.995     &0.505      \\ 
			23     &1923-1933        &332    &9.440      &0.545      &&75     &1980-1985        &7026   &25.625     &0.491      \\ 
			24     &1924-1934        &334    &9.246      &0.538      &&76     &1981-1986        &7275   &26.329     &0.490      \\ 
			25     &1925-1935        &339    &10.124     &0.559      &&77     &1982-1987        &7342   &26.996     &0.490      \\ 
			26     &1926-1936        &340    &8.506      &0.550      &&78     &1983-1988        &7360   &27.677     &0.487      \\ 
			27     &1927-1937        &362    &10.331     &0.554      &&79     &1984-1989        &7334   &27.087     &0.489      \\ 
			28     &1928-1938        &364    &9.610      &0.541      &&80     &1985-1990        &7379   &28.461     &0.484      \\ 
			29     &1929-1939        &372    &9.435      &0.538      &&81     &1986-1991        &7260   &28.621     &0.486      \\ 
			30     &1930-1940        &380    &10.053     &0.553      &&82     &1987-1992        &8016   &37.887     &0.468      \\ 
			31     &1931-1941        &399    &10.832     &0.559      &&83     &1988-1993        &8945   &39.171     &0.478      \\ 
			32     &1932-1942        &408    &11.770     &0.554      &&84     &1989-1994        &9816   &37.690     &0.505      \\ 
			33     &1933-1943        &399    &11.599     &0.555      &&85     &1990-1995        &10788  &37.057     &0.516      \\ 
			34     &1934-1944        &381    &11.470     &0.555      &&86     &1991-1996        &11433  &38.225     &0.521      \\ 
			35     &1935-1945        &400    &11.725     &0.546      &&87     &1992-1997        &11831  &40.454     &0.523      \\ 
			36     &1936-1946        &578    &11.626     &0.613      &&88     &1993-1998        &12225  &42.360     &0.520      \\ 
			37     &1937-1947        &596    &12.651     &0.608      &&89     &1994-1999        &12874  &39.697     &0.530      \\ 
			38     &1938-1948        &653    &11.525     &0.606      &&90     &1995-2000        &14675  &38.525     &0.542      \\ 
			39     &1939-1949        &701    &14.382     &0.598      &&91     &1996-2001        &16114  &38.276     &0.562      \\ 
			40     &1940-1950        &741    &16.159     &0.597      &&92     &1997-2002        &17174  &37.238     &0.582      \\ 
			41     &1941-1951        &777    &15.483     &0.591      &&93     &1998-2003        &18106  &37.874     &0.597      \\ 
			42     &1942-1952        &794    &15.657     &0.591      &&94     &1999-2004        &18879  &38.501     &0.607      \\ 
			43     &1943-1953        &800    &15.650     &0.586      &&95     &2000-2005        &19249  &40.333     &0.607      \\ 
			44     &1944-1954        &827    &14.989     &0.578      &&96     &2001-2006        &19389  &42.643     &0.607      \\ 
			45     &1945-1955        &781    &15.636     &0.586      &&97     &2002-2007        &19750  &44.656     &0.606      \\ 
			46     &1946-1956        &846    &18.234     &0.597      &&98     &2003-2008        &20703  &44.987     &0.609      \\ 
			47     &1947-1957        &943    &18.554     &0.590      &&99     &2004-2009        &21744  &48.983     &0.600      \\ 
			48     &1948-1958        &970    &16.097     &0.586      &&100    &2005-2010        &22151  &52.850     &0.594      \\ 
			49     &1949-1959        &1053   &17.673     &0.585      &&101    &2006-2011        &22627  &56.650     &0.586      \\ 
			50     &1950-1960        &1106   &18.029     &0.594      &&102    &2007-2012        &22556  &61.182     &0.576      \\ 
			51     &1951-1961        &1139   &18.586     &0.593      &&103    &2008-2013        &21460  &49.790     &0.594      \\ 		
		\end{tabular}
	\end{ruledtabular}
\end{table*}

\renewcommand{\arraystretch}{0.85} %Row height
\begin{table*}[h] 
	\begin{ruledtabular}
		\centering		
		\caption{Year by year statistics for the World Trade Web. Columns are: the time of each network (Year), the number of nodes ($N$), the average degree ($\langle k \rangle$), and the average local clustering coefficient ($\langle c \rangle$).}
		\label{tab:WTW}		
		\begin{tabular}{*{5}{l}{c}|*{5}{l}} 
			%			\\[0.1cm]	%			
			\thd{No.}	&\thd{Year}  &\thd{$N$}  &\thd{$\langle k\rangle$}  &\thd{$\langle c\rangle$} &&  \thd{No.}	&\thd{Year}  &\thd{$N$}  &\thd{$\langle k\rangle$}  &\thd{$\langle c\rangle$}\\
			\hline
			0      &1870   &24     &3.667      &0.709      &&72     &1942   &-    &-        &-        \\ 
			1      &1871   &24     &4.167      &0.704      &&73     &1943   &-    &-        &-        \\ 
			2      &1872   &23     &4.348      &0.709      &&74     &1944   &-    &-        &-        \\ 
			3      &1873   &26     &3.154      &0.700      &&75     &1945   &-    &-        &-        \\ 
			4      &1874   &26     &3.769      &0.713      &&76     &1946   &-    &-        &-        \\ 
			5      &1875   &25     &2.800      &0.420      &&77     &1947   &-    &-        &-        \\ 
			6      &1876   &27     &3.333      &0.759      &&78     &1948   &61     &2.426      &0.236      \\ 
			7      &1877   &27     &3.630      &0.709      &&79     &1949   &62     &3.097      &0.826      \\ 
			8      &1878   &31     &3.806      &0.774      &&80     &1950   &64     &3.562      &0.828      \\ 
			9      &1879   &28     &3.786      &0.664      &&81     &1951   &65     &3.908      &0.793      \\ 
			10     &1880   &29     &3.793      &0.707      &&82     &1952   &67     &4.328      &0.726      \\ 
			11     &1881   &28     &4.000      &0.700      &&83     &1953   &69     &3.594      &0.679      \\ 
			12     &1882   &27     &3.185      &0.614      &&84     &1954   &70     &4.514      &0.639      \\ 
			13     &1883   &29     &4.483      &0.693      &&85     &1955   &78     &4.538      &0.698      \\ 
			14     &1884   &28     &3.357      &0.690      &&86     &1956   &80     &4.100      &0.640      \\ 
			15     &1885   &28     &3.929      &0.718      &&87     &1957   &82     &4.000      &0.644      \\ 
			16     &1886   &29     &3.655      &0.719      &&88     &1958   &81     &4.469      &0.648      \\ 
			17     &1887   &29     &4.276      &0.761      &&89     &1959   &83     &4.771      &0.639      \\ 
			18     &1888   &29     &4.069      &0.710      &&90     &1960   &98     &4.429      &0.639      \\ 
			19     &1889   &29     &4.069      &0.702      &&91     &1961   &106    &5.038      &0.727      \\ 
			20     &1890   &31     &3.613      &0.677      &&92     &1962   &108    &3.759      &0.613      \\ 
			21     &1891   &31     &4.194      &0.732      &&93     &1963   &108    &4.056      &0.641      \\ 
			22     &1892   &30     &3.933      &0.673      &&94     &1964   &118    &5.068      &0.725      \\ 
			23     &1893   &31     &3.871      &0.682      &&95     &1965   &121    &5.074      &0.748      \\ 
			24     &1894   &25     &4.720      &0.732      &&96     &1966   &125    &6.112      &0.716      \\ 
			25     &1895   &28     &6.214      &0.766      &&97     &1967   &125    &6.288      &0.725      \\ 
			26     &1896   &31     &5.161      &0.698      &&98     &1968   &127    &5.654      &0.690      \\ 
			27     &1897   &32     &4.125      &0.665      &&99     &1969   &131    &5.847      &0.668      \\ 
			28     &1898   &34     &4.294      &0.662      &&100    &1970   &131    &4.718      &0.629      \\ 
			29     &1899   &34     &5.000      &0.695      &&101    &1971   &136    &5.397      &0.686      \\ 
			30     &1900   &29     &4.000      &0.740      &&102    &1972   &137    &5.635      &0.667      \\ 
			31     &1901   &28     &4.643      &0.722      &&103    &1973   &138    &5.275      &0.634      \\ 
			32     &1902   &28     &4.071      &0.721      &&104    &1974   &141    &6.383      &0.670      \\ 
			33     &1903   &29     &4.552      &0.707      &&105    &1975   &147    &6.925      &0.684      \\ 
			34     &1904   &29     &4.345      &0.700      &&106    &1976   &149    &6.456      &0.682      \\ 
			35     &1905   &34     &4.118      &0.621      &&107    &1977   &150    &6.427      &0.697      \\ 
			36     &1906   &32     &3.938      &0.707      &&108    &1978   &148    &5.473      &0.734      \\ 
			37     &1907   &31     &3.484      &0.785      &&109    &1979   &153    &6.601      &0.684      \\ 
			38     &1908   &31     &4.839      &0.741      &&110    &1980   &155    &7.019      &0.656      \\ 
			39     &1909   &32     &4.875      &0.801      &&111    &1981   &156    &5.487      &0.653      \\ 
			40     &1910   &37     &3.405      &0.711      &&112    &1982   &157    &5.694      &0.663      \\ 
			41     &1911   &38     &3.158      &0.641      &&113    &1983   &158    &6.190      &0.653      \\ 
			42     &1912   &41     &3.463      &0.685      &&114    &1984   &160    &6.713      &0.673      \\ 
			43     &1913   &41     &3.805      &0.772      &&115    &1985   &159    &5.887      &0.669      \\ 
			44     &1914   &-    &-        &-        &&116    &1986   &160    &7.425      &0.749      \\ 
			45     &1915   &-    &-        &-        &&117    &1987   &159    &6.239      &0.734      \\ 
			46     &1916   &-    &-        &-        &&118    &1988   &161    &6.348      &0.747      \\ 
			47     &1917   &-    &-        &-        &&119    &1989   &160    &6.450      &0.706      \\ 
			48     &1918   &-    &-        &-        &&120    &1990   &159    &5.950      &0.713      \\ 
			49     &1919   &-    &-        &-        &&121    &1991   &157    &4.624      &0.677      \\ 
			50     &1920   &50     &3.640      &0.737      &&122    &1992   &171    &5.696      &0.618      \\ 
			51     &1921   &52     &3.346      &0.597      &&123    &1993   &178    &5.416      &0.640      \\ 
			52     &1922   &53     &3.321      &0.704      &&124    &1994   &179    &5.966      &0.659      \\ 
			53     &1923   &54     &3.667      &0.686      &&125    &1995   &179    &5.520      &0.645      \\ 
			54     &1924   &54     &3.630      &0.819      &&126    &1996   &180    &5.067      &0.636      \\ 
			55     &1925   &51     &3.490      &0.767      &&127    &1997   &180    &5.056      &0.649      \\ 
			56     &1926   &53     &3.547      &0.661      &&128    &1998   &180    &5.156      &0.609      \\ 
			57     &1927   &55     &3.091      &0.723      &&129    &1999   &183    &4.415      &0.632      \\ 
			58     &1928   &56     &3.643      &0.798      &&130    &2000   &184    &6.435      &0.653      \\ 
			59     &1929   &56     &3.607      &0.758      &&131    &2001   &184    &4.957      &0.654      \\ 
			60     &1930   &46     &2.739      &0.618      &&132    &2002   &185    &5.514      &0.630      \\ 
			61     &1931   &57     &3.825      &0.832      &&133    &2003   &185    &4.205      &0.621      \\ 
			62     &1932   &57     &3.439      &0.723      &&134    &2004   &185    &4.541      &0.572      \\ 
			63     &1933   &56     &3.321      &0.701      &&135    &2005   &186    &4.688      &0.607      \\ 
			64     &1934   &57     &3.193      &0.642      &&136    &2006   &184    &4.804      &0.596      \\ 
			65     &1935   &53     &3.396      &0.642      &&137    &2007   &186    &4.290      &0.568      \\ 
			66     &1936   &59     &3.085      &0.618      &&138    &2008   &186    &4.312      &0.587      \\ 
			67     &1937   &60     &3.067      &0.313      &&139    &2009   &186    &5.774      &0.581      \\ 
			68     &1938   &56     &3.000      &0.248      &&140    &2010   &187    &4.930      &0.600      \\ 
			69     &1939   &-    &-        &-        &&141    &2011   &186    &4.903      &0.580      \\ 
			70     &1940   &-    &-        &-        &&142    &2012   &186    &5.161      &0.623      \\ 
			71     &1941   &-    &-        &-        &&143    &2013   &189    &5.820      &0.629      \\ 
			
		\end{tabular}
	\end{ruledtabular}
\end{table*}

\renewcommand{\arraystretch}{1} %Row height
\begin{table*}[t] 
	\begin{ruledtabular}
		\centering		
		\caption{Overview of the considered real-world networks.Columns are: the name  of each network (Name), the number of nodes ($N$), the average degree ($\langle k \rangle$), the average local clustering coefficient ($\langle c \rangle$), the exponent of the power-law degree distribution ($\gamma $), the hyperbolic embedding parameter $\beta$ and $\mu$, and the stable distribution fitting parameters $[\alpha,\eta,c,d]$.}
		\label{tab:networkproperties}		
		\begin{tabular}{*{13}{l}} 
			%			\\[0.1cm]	%			
			\thd{No.}	&\thd{Name}  &\thd{$N$} &\thd{$\langle k\rangle$}  &\thd{$\langle c\rangle$} &\thd{$\gamma$}  &\thd{$\beta$} &\thd{$\mu$}  &\thd{$[\alpha,\eta,c,d]$}\\
			\hline
			0      & Airports     	& 3397  	& 11.32       & 0.63  	&1.88    	& 1.955    &0.0274598  
			& [0.500, 1.000, 6.429  -4.295]  \\ 
			1      & Drosophila     & 1748  	& 9.13       & 0.22  	&1.93    	& 1.088    &0.0047457   
			& [0.781,  0.992, 3.670, -7.993]\\ 
			2      & Enron     		& 33696  	& 10.73       & 0.70  	&2.66    	& 2.663     &0.0365131              
			& [0.500, 1.000, 26.117, -23.839]     \\ 
			3      & Internet      	& 23748  	& 4.92        & 0.61  	&2.17   	& 1.979     &0.0640209              
			& [0.671, 0.988, 1.946, -1.684]\\ 
			4      & Metabolic     	& 1436 		& 6.57        & 0.54  	&2.60     	& 2.104     &0.0507968               
			& [0.752, 0.904, 9.510, -8.357]\\ 
			5      & Music     		& 2476  	& 16.66       & 0.82    &2.27     	& 2.192     &0.0207461        
			& [0.494, 0.999, 20.630, -19.403]     \\ 
			6      & Proteome     	& 4100   	& 6.52        & 0.09  	& 2.25    	& 1.005    &0.0008276             
			& [0.859, 0.999, 4.227, -16.070]   \\ 
			7      & Words      	& 7377    	& 11.99       & 0.47    & 2.25      & 1.006    &0.0005121            
			& [0.755, 1.000, 3.419, -6.055]     \\ 
			8      & Facebook      	& 320    	& 14.81       & 0.49    & 3.80      & 1.868         &0.0199591            
			& [0.670,   1.000,  45.000, -40.000]     \\
			9      & JCN(1965-1975)      	& 4168    	& 26.84        & 0.61    & 2.96      & 1.937       &0.0114668            
			& [0.677, 0.987, 108.053, -150.490]     \\ 
			10      & JCN(1990-1995)      	& 10788    	& 37.06        & 0.52    & 2.97      & 1.576     &0.0061764          
			& [0.843, 0.999,  92.088, -363.259]     \\ 
			11      & WTW(1965)      	& 121    	& 5.07        & 0.75    & 2.77      & 2.353    &0.1337139         
			& [0.500, 0.966, 1.323, 0.700]     \\ 		
		\end{tabular}
	\end{ruledtabular}
\end{table*}
%==================================================================================================================
\newpage
\clearpage
\bibliography{reference}

\end{document}